\documentclass[smallextended, final]{svjour3}
\smartqed
\makeatletter
\newcommand{\mylabel}[2]{#2\def\@currentlabel{#2}\label{#1}}
\makeatother

\usepackage{booktabs} 
\usepackage{dsfont}
\usepackage{amsmath}
\usepackage{amssymb}
\usepackage{stmaryrd}

\newtheorem{assumption}{Assumption}

\usepackage{tikz}
\usetikzlibrary{positioning}
\usetikzlibrary{calc,trees,positioning,arrows,chains,shapes.geometric,%
    decorations.pathreplacing,decorations.pathmorphing,shapes,%
    matrix,shapes.symbols,automata}

\usetikzlibrary{fit}
\usetikzlibrary{shapes}
\usetikzlibrary{decorations.markings}

\tikzstyle{randomVariable}=[circle,fill=white,draw=black,text=black,minimum size=0.8cm]
\tikzstyle{param}=[draw=none,fill=none,text=black,minimum size=0.2cm]

\DeclareMathOperator*{\argmin}{arg\,min}
\newcommand{\edge}[2]{(#1,#2)}

\newcommand{\ceil}[1]{\left\lceil #1 \right\rceil}
\newcommand{\floor}[1]{\left\lfloor #1 \right\rfloor}

\usepackage[toc,page]{appendix}
\usepackage{hyperref}

\begin{document}

\title{Optimal Control of Dynamic Bipartite Matching Models}

\author{Arnaud Cadas \and Josu Doncel \and Ana Bu\v{s}i\'c} 

\institute{Arnaud Cadas \and Ana Bu\v{s}i\'c \at INRIA, Paris, France.\\ DI ENS, CNRS, PSL Research University, Paris, France.\\
\email{arnaud.cadas@inria.fr}\\
\email{ana.busic@inria.fr}
\and Josu Doncel \at University of the Basque Country, UPV-EHU, Leioa, Spain.\\
\email{josu.doncel@ehu.eus}}

\maketitle

\begin{abstract}
A dynamic bipartite matching model is given by a bipartite matching graph which determines the possible matchings between the various types of supply and demand items. Both supply and demand items arrive to the system according to a stochastic process. Matched pairs leave the system and the others wait in the queues, which induces a holding cost. We model this problem as a Markov Decision Process and study the discounted cost and the average cost problem. 
We fully characterize the optimal matching policy for complete matching graphs and for the
$N$-shaped matching graph. In the former case, the optimal policy consists of matching everything and, in the latter case, it prioritizes the matchings in the extreme edges and is of threshold type for the diagonal edge. In addition, for the average cost problem, we compute the optimal threshold value. For more general graphs, we need to consider some assumptions
on the cost of the nodes. For complete graphs minus one edge, we provide conditions on the
cost of the nodes such that the optimal policy of the $N$-shaped matching graph extends 
to this case. For acyclic graphs, we show that, when the cost of the extreme edges is large,
the optimal matching policy prioritizes the matchings in the extreme edges. We also study the
$W$-shaped matching graph and, using simulations, we show that there are cases where 
it is not optimal to prioritize to matchings in the extreme edges.
\end{abstract}

\begin{acknowledgements}
 Funding from the French
National Research Agency grant ANR-16-CE05-0008, the Marie Sklodowska-Curie grant agreement No 777778, the Basque Government, Spain, Consolidated Research Group Grant IT649-13, 
and the Spanish Ministry of Economy and Competitiveness project MTM2016-76329-R. 
\end{acknowledgements}

\section{Introduction}

The theory of matching started with Petersen and K\"{o}nig and was under a lot of interests in graph theory with problems like maximum matchings. It was extended to online matching setting \cite{VSA,FMMM,JL} where one population is static and the other arrives according to a stochastic process. In the recent years, fully dynamic matching models have been considered where the whole population is random. The importance of matching models was shown through applications in many fields: health \cite{KidneySite,Kidney}, ridesharing \cite{BKQ}, power grid \cite{ZDC}, or pattern recognition \cite{SR}.

We study matching models from a queueing theory perspective, where a supply item and a demand item arrive to the system at each time step and can be matched or stay in buffers. \cite[Theorem 1]{Mairesse_Stability} proves that in a matching model where items arrive one by one, there exists no arrival distribution which verifies the necessary stability conditions for bipartite matching graphs. This result justifies why we assume arrivals by pairs as in \cite{BoundedRegret,Stability}. We consider that there is a holding cost that is a function of the buffer sizes. Our objective is to find the optimal matching policy in the discounted cost problem and in the average cost problem for general bipartite matching graphs. For this purpose, we model this problem as a Markov Decision Process.

The search for good policies and the performance analysis of various matching models have received great interest in the recent literature.  For example, the FCFS infinite bipartite matching model was introduced in \cite{FCFSModel} and further studied in \cite{ExactFCFS,ABMW18} that established the reversibility of the dynamics and the product form of stationary distribution. In \cite{Stability} the bipartite matching model was extended to other matching policies. In that paper, the authors established the necessary stability conditions and studied the stability region of various policies, including priorities and MaxWeigth. It is shown that MaxWeight has maximal stability region. For ride-sharing systems, state-dependent dispatch policies were identified in \cite{BKQ} which achieved exponential decay of the demand-dropping probability in heavy traffic regime. In \cite{GW}, the authors presented the imbalance process and derived a lower bound on the holding costs. 

Optimality results are scarce, and have been derived for some matching models in the asymptotic regimes. An extension of the greedy primal-dual algorithm was developed in \cite{NS} and was proved to be asymptotically optimal for the long-term average matching reward. However, they considered rewards on the edges, which differs from our model with holding costs. In \cite{BoundedRegret}, the authors considered the asymptotic heavy-traffic setting, and identified a policy that is approximately optimal with bounded regret using a workload relaxation approach. 

In this paper, we study the optimal matching policy of bipartite graphs in the non-asymptotic setting, i.e. we allow for any arrival rates under the stability conditions established in \cite{Stability}. We use arguments of structured policies to analyze the optimal matching control and we fully characterize the optimal policy for 
the following matching models: 

\begin{itemize}
\item
We first consider a matching model that is a complete graph. For this case, 
we show that the policy that matches all the possible items is optimal. We remark that this 
result is applied, in particular, to all the bipartite and connected graphs of less than 4 
nodes (except for the $N$-shaped graph). 
\item
We then consider the $N$-shaped graph, which is formed by two supply and two demand classes. For this system, we show that the optimal matching policy is of threshold type for the diagonal edge and with priority to the end edges of the matching graph. We also compute the optimal threshold for the average cost problem. This part of the paper was already published in a conference \cite{CDB}.
\end{itemize}

If we aim to extend the optimal matching policy identified in the case $N$ to more general bipartite matching graphs, we require certain assumptions on the cost of the nodes. 
For instance, in a complete graph minus one node, we provide conditions on the cost of the nodes such that the cost associated to any state coincides with the cost of its projection on the $N$-shaped graph. 
As a consequence of this, there is a threshold-type policy that is optimal in a matching model that consists of a complete graph minus one edge. 
Moreover, we show that, in an arbitrary acyclic graph, if the cost of the extreme nodes is larger or equal than that of their neighbors nodes, the optimal matching policy prioritizes the matchings of the extreme edges. 
Finally, we explore
the $W$-shaped graph and we differentiate two cases. 
On one hand, we consider that
the cost of the extreme nodes is large, in which case we conjecture that the optimal policy is of threshold type with priority to the extreme edges. We also 
present the set of properties that are needed to prove our conjecture, how they can be used to show that
this policy is optimal and what are the difficulties in their preservation. On the other hand, 
we consider that the cost of the middle edge is large and, according to our numerical experiments, the matching policy that prioritizes the extreme edges is not optimal. 

The remainder of the article is organized as follows. In Section~\ref{sec:Model-description}, we describe the optimal control
problem we investigate in this paper. We characterize the optimal matching policy for 
complete matching graphs in Section~\ref{sec:CompleteGraphs} and for $N$-shaped graphs in Section~\ref{sec:N}. 
Then, in Section~\ref{sec:general}, we consider general matching graphs and, in Section~\ref{sec:W}, the $W$-shaped
matching graph. We provide the main conclusions of our work in Section~\ref{sec:Conclusion}.

\section{Model Description}\label{sec:Model-description}

\begin{figure}[htbp]
     \centering
    \begin{tikzpicture}[scale=0.6]
     \node[style={circle,draw}] at (2,0) (s1) {$s_1$};
     \node[style={circle,draw}] at (4,0) (s2) {$s_2$};
     \node[style={circle,draw}] at (6,0) (s3) {$s_3$};
     \node[style={circle,draw}] at (2,2) (d1) {$d_1$};
     \node[style={circle,draw}] at (4,2) (d2) {$d_2$};
     \node[style={circle,draw}] at (6,2) (d3) {$d_3$};
     \draw[<-] (s1) -- (2,-1)  node[below] {$\beta_1$} ;
     \draw[<-] (s2) -- (4,-1)  node[below] {$\beta_2$} ;
     \draw[<-] (s3) -- (6,-1)  node[below] {$\beta_3$} ;
     \draw[<-] (d1) -- (2,3)  node[above] {$\alpha_1$} ;
     \draw[<-] (d2) -- (4,3)  node[above] {$\alpha_2$} ;
     \draw[<-] (d3) -- (6,3)  node[above] {$\alpha_3$} ;
     \draw (s1) -- (d1);
     \draw (s2) -- (d1);
     \draw (s2) -- (d2);
     \draw (s3) -- (d2);
     \draw (s3) -- (d3);
    \end{tikzpicture}  
   \caption{A matching graph with three supply classes and three demand classes.}
   \label{fig:NN}
  \end{figure}
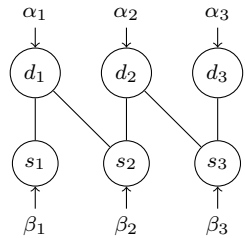

We consider a bipartite matching graph $(\mathcal D \cup \mathcal S,\mathcal E)$ where $\mathcal{D}=\{d_1,d_2,\dots,d_{n_D}\}$ and $\mathcal{S} =\{s_1,s_2,\dots,s_{n_S}\}$ are, respectively, 
the set of demand nodes (or queues) and the set of supply nodes. $\mathcal{E}\subset \mathcal{D}\times\mathcal{S}$ is the set of allowed matching pairs. In Figure~\ref{fig:NN} it is depicted an example of a matching graph with three demand nodes and three supply nodes. In each time slot $n$, a demand item and a supply item arrive to the system according to the i.i.d. arrival process $A(n)$. We assume independence between demand and supply arrivals.
The demand item arrives to the queue $d_i$ with probability $\alpha_i$ and the supply item arrives to the queue $s_j$ with probability $\beta_j$, i.e:
\[\forall \edge{i}{j}\in\mathcal{A}\quad \mathbb{P}(A(n)=e_{\edge{i}{j}})=\alpha_i \beta_j >0 \]
with $\sum_{i=1}^{n_\mathcal{D}}\alpha_i =1$, $\sum_{j=1}^{n_\mathcal{S}}\beta_j =1$ and where $\mathcal{A}=\mathcal{D}\times\mathcal{S}$ is the set of allowed arrival pairs, $e_{\edge{i}{j}}=e_{d_i}+e_{s_j}$ and $e_k\in\mathbb{N}^{n_\mathcal{D}+n_\mathcal{S}}$ is the vector of all zeros except in the $k$-th coordinate where it is equal to one, $k\in\mathcal{D}\cup\mathcal{S}$.  We assume that the $\alpha_i$ and $\beta_j$ are chosen such that the arrival distribution satisfies the necessary and sufficient conditions for stabilizability of the MDP model: \texttt{Ncond} given in \cite{Stability}, i.e $\forall D \subsetneq \mathcal{D}$, $\forall S \subsetneq \mathcal{S}$:
\begin{equation}\label{eq:stability} 
\sum_{d_i \in D}\alpha_i < \sum_{s_j \in \mathcal{S}(D)}\beta_j \text{  and }
 \sum_{s_j \in S}\beta_j < \sum_{d_i \in \mathcal{D}(S)}\alpha_i 
 \end{equation}
where $\mathcal{D}(j)=\{i\in\mathcal{D}: (i,j)\in\mathcal{E}\}$ is the set of demand classes that can be matched with a class $j$ supply and $\mathcal{S}(i)=\{j\in\mathcal{S}: (i,j)\in\mathcal{E}\}$ is the set of supply classes that can be matched with a class $i$ demand. The extension to subsets $S\subset\mathcal{S}$ and $D\subset\mathcal{D}$ is $\mathcal{D}(S)=\bigcup_{j\in S}\mathcal{D}(j)$ and $\mathcal{S}(D)=\bigcup_{i\in D}\mathcal{S}(i)$.

We denote by $q_k(n)$ the queue length of node $k$ at time slot $n$, where $k\in\mathcal{D}\cup\mathcal{S}$.
Let $Q(n)=(q_k(n))_{k\in\mathcal{D}\cup\mathcal{S}}$ be the vector of the queue length of all the nodes and $\mathcal{Q}=\{q\in\mathbb{N}^{n_{\mathcal{D}}+n_{\mathcal{S}}}:\sum_{k\in\mathcal{D}}q_k=\sum_{k\in\mathcal{S}}q_k$ be the set of all the possible queues length. We must have $q(n)\in\mathcal{Q}$ for all $n$. Matchings at time $n$ are carried out after the arrivals at time $n$. Hence, $Q(n)$ evolves over time according to the following expression:
\begin{equation}
Q(n+1)=Q(n)+A(n)-u(Q(n),A(n)),
\label{eq:x-evolution}
\end{equation}
where $u$ is a deterministic Markovian decision rule which maps the current state $Y(n)=\left(Q(n),A(n)\right)$ to the vector of the items that are matched at time $n$. Thus, $Y$ is a Markov Decision Process where the control is denoted by $u$. It is sufficient to consider only deterministic Markovian decision rules and not all history-dependent randomized decision rules as proved in \cite[Theorem 5.5.3]{puterman2005markov} and \cite[Proposition 6.2.1]{puterman2005markov}. Let us define $X(n)=Q(n)+A(n)$ as the vector of the queue length of all the nodes just after the arrivals. In order to ease the notations and because the matchings only depends on the queues length after the arrivals, we will note $u(Q(n),A(n))=u(X(n))$ for the reminder of the paper. When the state of the system is $Y(n)=(q,a)$, $x=q+a$, $u(x)$ must belong to the set of admissible matchings which is defined as:
\begin{multline*}
U_x=\left\{u=\sum_{\edge{i}{j}\in\mathcal E}u_{\edge{i}{j}}e_{\edge{i}{j}}\in\mathbb{N}^{n_{\mathcal{D}}+n_{\mathcal{S}}}: (a)\; \forall i\in\mathcal D, \sum_{k\in S(i)}u_{\edge{i}{k}}\leq x_{d_i},\right.\\
\left. (b)\; \forall j\in\mathcal S, \sum_{k\in D(j)}u_{\edge{k}{j}}\leq   x_{s_j}\right\}
\end{multline*}
where $u_{\edge{i}{j}}$ is the number of matchings in the edge $\edge{i}{j}$. $U_x$ is defined for all $x\in\mathcal{Q}$.
We consider a linear cost function on the buffer size of the nodes: $ c(Q(n),A(n))=c(X(n))=\sum_{k\in\mathcal{D}\cup\mathcal{S}}c_k x_k(n)$. Our analysis presented in the following sections holds for more general cost functions as long as they satisfy Assumption~\ref{ass:cost} and the assumptions of Theorem~\ref{thm:puterman_discounted} and Theorem~\ref{thm:puterman_average}. We  chose a linear cost function because it satisfies these assumptions (see Lemma~\ref{lem:technical-req} and Lemma~\ref{lem:technical-req-average} in Section~\ref{sec:preliminaries}) and allow us to give an analytical form for the optimal threshold in the $N$-shaped graph. The buffer size of the nodes is infinite, thus we are in the unbounded costs setting.

A matching policy $\pi$ is a sequence of deterministic Markovian decision rules, i.e. $\pi=\left(u(X(n))\right)_{n\geq 1}$.
The goal is to obtain the optimal matching policy for two optimization problems:
\begin{itemize}
\item
The average cost problem: 
\[
g^\ast=\inf_{\pi}g^\pi \quad \text{ with }g^\pi(y)=\lim_{N\to\infty}\dfrac1N\sum_{n=0}^{N-1}\mathbb E_y^\pi[c(Y(n))]
\]
\item
The discounted cost problem: 
\[
v^\ast_\theta=\inf_{\pi}v^\pi_\theta \quad \text{ with }v^\pi_\theta(y)=\lim_{N\to\infty}\sum_{n=0}^{N-1}\theta^t\mathbb E_y^\pi[c(Y(n))]
\]
\end{itemize}
where $\theta\in[0,1)$ is the discount factor and $y\in\mathcal{Y}=\mathcal{Q}\times\mathcal{A}$ is the starting state. Both problems admit an optimal stationary policy, i.e. the decision rule depend only on the state of the system and not on the time \cite{puterman2005markov}. 
The notation $\mathbb E_y^\pi$ indicates that the expectation is over the arrival process, given that $Y(0)=y$ and using the matching policy $\pi$ to determine the matched items $u(X(n))$ for all $n$.

As $A(n)$ are i.i.d., to ease the notation from now on, we denote by $A$ a random variable with the same distribution as $A(1)$. 
For a given function $v$, $Y(n)=(q,a)$, $x=q+a$, $u\in U_x$, we define for all $0\leq \theta\leq 1$:
\begin{align*}
L_u^\theta v(q,a)&=c(q,a)+\theta\mathbb E[v(q+a-u,A)]=c(x)+\theta\mathbb E[v(x-u,A)]
\\
L^\theta v(q,a)&=c(q,a)+\min_{u\in U_x}\theta\mathbb E[v(q+a-u,A)]=c(x)+\min_{u\in U_x}\theta\mathbb E[v(x-u,A)]
\end{align*}
and in particular, we define $T_u = L_u^1$ and $T = L^1$. A solution of the discounted cost problem can be obtained as a solution of the Bellman fixed point equation $v=L^\theta v$. In the average cost problem, the Bellman equation is given by $g^\ast + v = T v$.

We say that a value function $v$ or a decision rule $u$ is structured if it satisfies a special property, such as being increasing, decreasing or
convex. Throughout the article, by increasing we mean nondecreasing and we will use strictly increasing for increasing. A policy is called structured when it only uses structured decision rules. 

The framework of this work is that of property preservation when we apply the Dynamic Programming operator. First, we identify a set of structured value functions $V^\sigma$ and a set of structured deterministic Markovian decision rules $D^\sigma$ such that if the value function belongs to $V^\sigma$ an optimal decision rule belongs to $D^\sigma$. Then, we show that the properties of $V^\sigma$ are preserved by the Dynamic Programming operator and that they hold in the limit. 

\begin{assumption}\label{ass:cost}
The cost function $c$ is a nonnegative function with the same structured properties as the 
value function, i.e $c\in V^\sigma$.
\end{assumption}

Theorem~\ref{thm:puterman_discounted} \cite[Theorem 1]{HJM12} lets us conclude that there exists an optimal policy which can be chosen in the set of structured stationary matching policies $\Pi^\sigma =\{\pi=\left(u(X(n))\right)_{n\geq 1} : u\in D^\sigma\}$.

\begin{theorem}{\cite[Theorem 1]{HJM12}}
Assume that the following properties hold: there exists  positive function $w$ on $\mathcal{Y}$ such that
\begin{align}
&\sup_{y}\frac{c(y)}{w(y)}<+\infty,\label{eq:A_discounted-1}\\
&\sup_{(y,u)}\frac{1}{w(y)}\sum_{y^\prime}\mathbb P(y^\prime|y,u)w(y^\prime)<+\infty,\label{eq:A_discounted-2}
\end{align}
and for every $\mu$, $0\leq\mu<1$, there exists $\eta$, $0\leq\eta<1$ and some integer $J$ such that
for every $J$-tuple of Markov deterministic decision rules $\pi=(u_1,\dots,u_J)$ and every $y$
\begin{align}
\mu^J\sum_{y^\prime}P_\pi(y^\prime|y)w(y^\prime)<\eta w(y),\label{eq:A_discounted-3}
\end{align}
where $P_\pi$ denotes the $J$-step transition matrix under policy $\pi$. 
Let $0\leq \theta < 1$. Let $V_w$ the set of functions in the state space which have a finite $w$-weighted supremum norm, i.e., 
$\sup_y|v(y)/w(y)|<+\infty$. Assume that 
\begin{itemize}
\item[($\ast$)] for each $v\in V_w$, there exists a deterministic Markov decision rule $u$ such that $L^\theta v=L_u^\theta v$.
\end{itemize}
Let $V^\sigma$ and $D^\sigma$ be such that
\begin{itemize}
\item[(a)] $v\in V^\sigma$ implies that $L^\theta v\in V^\sigma$;
\item[(b)] $v\in V^\sigma$ implies that there exists a decision $u^\prime\in D^\sigma$ such that 
$u^\prime\in \argmin_{u} L_u^\theta v$;
\item[(c)] $V^\sigma$ is a closed subset of the set of value functions under pointwise convergence.
\end{itemize} 
Then, there exists an optimal stationary policy $\pi^\ast=\left(u^\ast (X(n))\right)_{n\geq 1}$ that belongs to $\Pi^\sigma$ with $u^\ast\in \argmin_{u} L_u^\theta v$.
\label{thm:puterman_discounted}
\end{theorem}

This result is an adapted version of \cite[Theorem 6.11.3]{puterman2005markov}. The former removes the need to verify that $V^\sigma\subset V_w$ (assumption made in the latter) and its statement separates the structural requirements (\textit{(a)}, \textit{(b)} and \textit{(c)}) from the technical requirements related to the unboundedness of the cost function (\eqref{eq:A_discounted-1}, \eqref{eq:A_discounted-2}, \eqref{eq:A_discounted-3} and \textit{($\ast$)}).

In the case of the average cost problem, we will use the results
of the discounted cost problem. We consider the average cost problem as a limit when $\theta$ tends to one and we show that the properties still hold for this limit. In order to prove the optimality in the average cost case, we will use \cite[Theorem 8.11.1]{puterman2005markov}:

\begin{theorem}{\cite[Theorem 8.11.1]{puterman2005markov}}
Suppose that the following properties hold: 
\begin{itemize}
\item[\mylabel{ass:average-1}{(A1)}]
$
\exists\: C\in\mathbb{R}, \forall y=(q,a)\in \mathcal{Y}, x=q+a,\; -\infty<C\leq c(x) < +\infty,
$
\item[\mylabel{ass:average-2}{(A2)}]
$
\forall y\in \mathcal{Y}, \forall\: 0\leq \theta < 1,\; v_\theta^\ast (y) < +\infty
$
\item[\mylabel{ass:average-3}{(A3)}]
$
\exists\: H\in\mathbb{R}, \forall y\in \mathcal{Y}, \forall\: 0\leq \theta < 1,\; -\infty<H\leq v_{\theta}^\ast(y) - v_{\theta}^\ast(0) 
$ 
\item[\mylabel{ass:average-4}{(A4)}]
There exists a nonnegative function $M(y)$ such that
\begin{itemize}
\item[\mylabel{itm:A4-a}{(a)}] $\forall y\in \mathcal{Y},\: M(y)<+\infty$
\item[\mylabel{itm:A4-b}{(b)}] $\forall y\in \mathcal{Y}, \forall\: 0\leq \theta < 1,\; v_{\theta}^\ast(y) - v_{\theta}^\ast(0) \leq M(y) $
\item[\mylabel{itm:A4-c}{(c)}] There exists $u\in U_0$ for which  $\sum_{y}\mathbb P(y|0,u)M(y)<+\infty $
\end{itemize}
\end{itemize}

Let $H$ and $M$ be defined in Assumptions~\ref{ass:average-3} and~\ref{ass:average-4}. We define a subset $V_H^\sigma$ of $V^\sigma$ which contains all the value functions $v\in V^\sigma$ such that $H\leq v(y) - v(0)\leq M(y)$ for all $y\in\mathcal{Y}$. Then, if 
\begin{itemize}
\item[(a)]\label{itm:average-a} for any sequence $(\theta_n)_{n\geq 0}$, $0\leq \theta_n < 1$, for which $\underset{n\to +\infty}{lim} \theta_n = 1$,
$$
\underset{n\to +\infty}{lim} [v_{\theta_n}^\ast - v_{\theta_n}^\ast(0)e] \in V_H^\sigma \quad\text{ with $e(y)=1$ for all $y\in\mathcal{Y}$} 
$$ and
\item[(b)]\label{itm:average-b} $v\in V_H^\sigma$ implies that there exists a  decision $u^\prime\in D^\sigma$ such that 
$u^\prime\in \argmin_{u} T_uv$;
\end{itemize}
Then $D^\sigma \cap \argmin_{u} T_uv \neq \emptyset$ and $u^\ast\in D^\sigma \cap  \argmin_{u} T_uv$ implies that the stationary matching policy which uses $u^\ast$ is lim sup average optimal.
\label{thm:puterman_average}
\end{theorem}
\begin{remark}
When $v_\theta (0)$ is used in this theorem, we mean $v_\theta (y)$ with $y=(0,a)$ for any $a\in\mathcal{A}$.
\end{remark}

\subsection{Preliminaries}\label{sec:preliminaries}

We first show that the technical requirements of Theorem~\ref{thm:puterman_discounted} 
due to the unboundness of the costs are verified when the cost function is linear.

\begin{lemma}\label{lem:technical-req}
Assuming that the cost function is linear, the conditions (\eqref{eq:A_discounted-1}, \eqref{eq:A_discounted-2}, \eqref{eq:A_discounted-3} and \textit{($\ast$)}) are verified. 
\end{lemma}

\begin{proof}
We show that the technical details given in \eqref{eq:A_discounted-1}-\eqref{eq:A_discounted-3} are verified. Let $y=(q,a)\in\mathcal{Q}\times\mathcal{A}$, $x=q+a$, we 
choose $w(y)=\sum_{i}x_i$. In our case, the cost is a linear function of $x$ therefore, 
$c(y)/w(y)\leq \max_{i\in\mathcal{D}\cup\mathcal{S}}c_i$. This shows \eqref{eq:A_discounted-1}. In addition,
\begin{multline*}
\frac{1}{w(y)}\sum_{y^\prime}\mathbb P(y^\prime|y,u)w(y^\prime)=\mathbb E\left[\left.\frac{w(x-u+a^\prime)}{w(y)}\right\vert A=a^\prime\right]\\
\leq E\left[\left.\frac{w(x+a^\prime)}{w(y)}\right\vert A=a^\prime\right]=\frac{\sum_{i}x_i+2}{\sum_{i}x_i}\leq 2
\end{multline*}
since $w(y)$ is increasing and two items arrive to the system in each step following a process which is independent of the state of the system. This shows
\eqref{eq:A_discounted-2}. Finally, we can repeat the previous argument to show that for all $J$-step matching policy $\pi$
$$
\sum_{y^\prime} P_\pi(y^\prime|y)w(y^\prime)\leq \sum_{y^\prime} P_{\pi_0}(y^\prime|y)w(y^\prime)=w(y) +2J.
$$
where $\pi_0 =(0,\cdots,0)$ is the policy which does not match any items.
Therefore, \eqref{eq:A_discounted-3} is satisfied if there exist $J$ integer and $\eta<1$ such that
$$
\mu^J (w(y) +2J)\leq \eta w(y) \iff \eta>\frac{\mu^J(w(y)+2J)}{w(y)}.
$$
Since it is decreasing with $J$ and when $J\to\infty$ it tends to zero, there exists a $J$ integer such that $\eta$ is less than one and, therefore, \eqref{eq:A_discounted-3} is also verified.

Since for each state of the system, the set of admissible matching policies is finite, it follows that (*) holds.
\end{proof}

From this result, we conclude that, when we study the optimal matching policy 
in the discounted cost problem with linear 
costs, we only need to check that conditions (a), (b) and (c) of 
Theorem~\ref{thm:puterman_discounted} are verified.

We now show that Assumptions~\textit{(A1)} to \textit{(A4)} of 
Theorem~\ref{thm:puterman_average} hold when the cost function is a linear function.

\begin{lemma}\label{lem:technical-req-average}
If the cost function is linear, then Assumptions~\textit{(A1)} to \textit{(A4)} hold. 
\end{lemma}

\begin{proof}
First, by Assumption~\ref{ass:cost}, we know that the cost function is nonnegative and therefore
Assumption~\textit{(A1)} holds using $C=0$. 
Let $\pi_{ML}$ be the policy Match the Longest as defined in \cite[Definition 2.6]{Stability}. This policy is stable for any bipartite matching graph as proved in \cite[Theorem 7.1]{Stability}, which means that the derived Markov chain is positive recurrent and $g^{\pi_{ML}}<\infty$. Moreover, the set $\{ y\in\mathcal Y : c(y)<g^{\pi_{ML}}\}$ is nonempty because $g^{\pi_{ML}}>\min_{\edge{i}{j}\in\mathcal{A}}c_{d_i}+c_{s_j}$ almost surely and $c(0,a)=c_{d_i}+c_{s_j}$ (for any $a=e_{\edge{i}{j}}\in\mathcal{A}$). It is also finite because $g^{\pi_{ML}}<\infty$ and $c$ is increasing in $y$ (because $c$ is linear). Therefore, we can use \cite[Theorem 8.10.9]{puterman2005markov} and Assumptions~\textit{(A2)} to \textit{(A4)} hold.
\end{proof}

From this result, we conclude that, when we study the optimal matching policy 
in the average cost problem with linear 
costs, we only need to check that conditions (a) and (b) of 
Theorem~\ref{thm:puterman_average} are verified.

\section{Complete Graphs}\label{sec:CompleteGraphs}

\begin{figure}[htbp]
     \centering
    \begin{tikzpicture}[scale=0.6]
     \node[style={circle,draw}] at (2,0) (s1) {$s_1$};
     \node[style={circle,draw}] at (2,2) (d1) {$d_1$};
     \draw[<-] (s1) -- (2,-1)  node[below] {$1$} ;
     \draw[<-] (d1) -- (2,3)  node[above] {$1$} ;
     \draw (s1) -- (d1) node[pos=0.5,left] {};
    
     \node[style={circle,draw}] at (6,0) (2s1) {$s_1$};
     \node[style={circle,draw}] at (5,2) (2d1) {$d_1$};
     \node[style={circle,draw}] at (7,2) (2d2) {$d_2$};
     \draw[<-] (2d1) -- (5,3)  node[above] {$\alpha_1$} ;
     \draw[<-] (2d2) -- (7,3)  node[above] {$\alpha_2$} ;
     \draw[<-] (2s1) -- (6,-1)  node[below] {$1$} ;
     \draw (2s1) -- (2d1) node[pos=0.5,left] {};
     \draw (2s1) -- (2d2) node[pos=0.5,left] {};
     
     \node[style={circle,draw}] at (12,0) (3s1) {$s_1$};
     \node[style={circle,draw}] at (10,2) (3d1) {$d_1$};
     \node[style={circle,draw}] at (12,2) (3d2) {$d_2$};
     \node[style={circle,draw}] at (14,2) (3d3) {$d_3$};
     \draw[<-] (3d1) -- (10,3)  node[above] {$\alpha_1$} ;
     \draw[<-] (3d2) -- (12,3)  node[above] {$\alpha_2$} ;
     \draw[<-] (3d3) -- (14,3)  node[above] {$\alpha_3$} ;
     \draw[<-] (3s1) -- (12,-1)  node[below] {$1$} ;
     \draw (3s1) -- (3d1) node[pos=0.5,left] {};
     \draw (3s1) -- (3d2) node[pos=0.5,left] {};
     \draw (3s1) -- (3d3) node[pos=0.5,left] {};

     \node[style={circle,draw}] at (17,0) (4s1) {$s_1$};
     \node[style={circle,draw}] at (19,0) (4s2) {$s_2$};
     \node[style={circle,draw}] at (17,2) (4d1) {$d_1$};
     \node[style={circle,draw}] at (19,2) (4d2) {$d_2$};
     \draw[<-] (4s1) -- (17,-1)  node[below] {$\beta_1$} ;
     \draw[<-] (4s2) -- (19,-1)  node[below] {$\beta_2$} ;
     \draw[<-] (4d1) -- (17,3)  node[above] {$\alpha_1$} ;
     \draw[<-] (4d2) -- (19,3)  node[above] {$\alpha_2$} ;
     \draw (4s1) -- (4d1) node[pos=0.5,left] {};
     \draw (4s2) -- (4d1) node[pos=0.75,right] {};;
     \draw (4s2) -- (4d2) node[pos=0.5,right] {};
     \draw (4s1) -- (4d2) node[pos=0.5,right] {};
    \end{tikzpicture}  
   \caption{Every bipartite and connected matching graphs of less than $4$ nodes (except the $N$-shaped matching graph).}
   \label{fig:small-matching-graph}
  \end{figure}
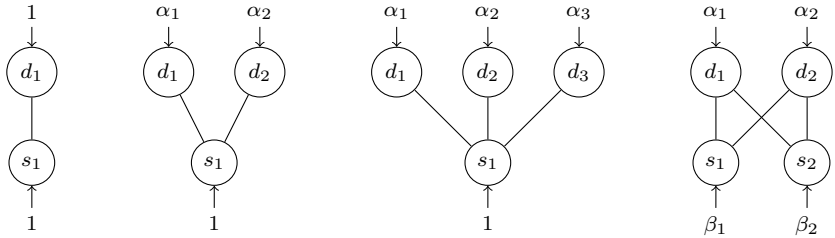

We start our study with the smallest bipartite graphs.
Immediately, we can see that all bipartite and connected graphs of less than $4$ nodes (except the $N$-shaped graph that will be studied in Section~\ref{sec:N}) are complete, i.e $\mathcal{E}=\mathcal{D}\times\mathcal{S}$. Without loss of generality, we can focus on the matching graphs such that there are more demand nodes than supply nodes, i.e $n_{\mathcal{D}}\geq n_{\mathcal{S}}$ (see Figure~\ref{fig:small-matching-graph}).

For these matching graphs, the stabilizability conditions come down to $0< \alpha_i <1$ for $i\in\{1,2,3\}$ and $0< \beta_j <1$ for $j\in\{1,2\}$.
The optimal policy is very intuitive. 
We can empty the system from any state and because we only consider costs on the number of items at each node and not rewards on which edge we match, it does not matter how do we empty the system.
However, leaving some items in the system will only increase the total cost in the long run. 
Thus, we can easily convince ourselves that matching everything is optimal.
\medbreak
Let us now consider the more general case when there are any number of demand nodes and any number of supply nodes, i.e $n_{\mathcal{D}}\in\mathbb{N}^{\ast}$ and $n_{\mathcal{S}}\in\mathbb{N}^{\ast}$.
We will show that the optimal policy for the complete matching graph ($(\mathcal{D}\cup\mathcal{S},\mathcal{E})$ with $\mathcal{E}=\mathcal{D}\times\mathcal{S}$) is to match everything. 
For this purpose, we first present the properties of the value function. 
Then, we show how these properties characterize the optimal decision rule and how they are preserved by the dynamic programming operator. 
Finally, we prove the desired results in Theorem~\ref{thm:CG-opt-control} and Theorem~\ref{thm:CG-opt-control-average}.

\subsection{Value Function Properties}
\label{sec:CG-definitions}

We define the only property that is needed to prove that matching everything is optimal:

\begin{definition}[Increasing property]\label{def:CG-increasing}
Let $\edge{i}{j}\in\mathcal{E}$. 
We say that a function $v$ is increasing in $\edge{i}{j}$ or $v\in \mathcal{I}_{\edge{i}{j}}$ if
$$
\forall a\in\mathcal{A},\forall q\in\mathcal Q, \quad v(q+e_{\edge{i}{j}},a)\geq v(q,a).
$$ 
We also note $\mathcal{I}_a =\cup_{\edge{i}{j}\in\mathcal{E}}\mathcal{I}_{\edge{i}{j}}$.
\end{definition}

We aim to characterize the optimal matching policy using Theorem~\ref{thm:puterman_discounted} and Theorem~\ref{thm:puterman_average}. 
Thus, in the remainder of the article, we will consider the following set of structured value functions
$V^\sigma=\mathcal{I}_{a}$.

\subsection{Optimal decision rule}

In this section, we show that, for any $v\in V^\sigma$, there is a control which matches everything and minimizes $L_u^\theta v$.

\begin{definition}[Full decision rule]\label{def:CG-optimal-decision-rule}
A decision rule $u_x$ is said to be full if it matches everything, i.e $x - u_x = 0$.
We define $D^\sigma$ as the set of decision rules that are full.
\end{definition}

In the next proposition, we establish that there exists an optimal decision rule which is full.

\begin{proposition}
Let $v\in \mathcal I_{a}$, let $0\leq \theta\leq 1$. For any $q\in\mathcal Q$ and any $a\in\mathcal{A}$, $x=q+a$, there exists $u^\ast\in U_x$ such that $u^\ast\in \argmin_{u\in U_x} L_u^\theta v(q,a)$ and $x - u^\ast = 0$. In particular, this result holds for the average operator: $T_u$.
\label{prop:CG-optimal-control}
\end{proposition}

\begin{proof}
Let $v\in \mathcal I_{a}$, $0\leq\theta\leq 1$, $q\in\mathcal Q$, $a\in\mathcal{A}$, $x=q+a$ and $u\in U_x$. We are going to match every items that are left after the matching $u$.
\medbreak
For $i=1,\cdots,n_{\mathcal{D}}$ and for $j=1,\cdots,n_{\mathcal{S}}$ ($\edge{i}{j}\in\mathcal{E}$), let $m_{\edge{i}{j}}=\min(x_{d_i}-\sum_{k\in\mathcal{S}(i)} u^{i,j-1}_{\edge{i}{k}}, x_{s_j}-\sum_{k\in\mathcal{D}(j)} u^{i,j-1}_{\edge{k}{j}})$ (where $u^{i,0}=u^{i-1,n_{\mathcal{S}}}$ and $u^{0,n_{\mathcal{S}}}=u$) be the number of possible matchings that can be made in $\edge{i}{j}$.
We define $u^{i,j}$ that matches all the the possible $\edge{i}{j}$ of $x-u^{i,j-1}$, that is, of the remaining items when we apply $u^{i,j-1}$: $u^{i,j} = u^{i,j-1}+m_{\edge{i}{j}}e_{\edge{i}{j}}$. 
We also verify that this policy is admissible, i.e $u^{i,j}\in U_x$: $u^{i,j}\in\mathbb{N}^{n_{\mathcal{D}}+n_{\mathcal{S}}}$ because $u^{i,j-1}\in U_x$.  
\textit{(a)} is true because 
$\sum_{k\in\mathcal{S}(i)} u^{i,j}_{\edge{i}{k}}=\sum_{k\in\mathcal{S}(i)} u^{i,j-1}_{\edge{i}{k}}+m_{\edge{i}{j}}\leq x_{d_i}$.
\textit{(b)} is true because 
$\sum_{k\in\mathcal{D}(j)} u^{i,j}_{\edge{k}{j}}=\sum_{k\in\mathcal{D}(j)} u^{i,j-1}_{\edge{k}{j}}+m_{\edge{i}{j}}\leq x_{s_j}$.
Then, we can use the fact that $v\in\mathcal{I}_{\edge{i}{j}}\subset \mathcal{I}_a$ to show that $L^\theta_{u^{i,j}} v(q,a)\leq L^\theta_{u^{i,j-1}} v(q,a)$.
\medbreak
In the end, we get $u^{n_{\mathcal{D}},n_{\mathcal{S}}}$, that we note as $u^\ast$, such that $L^\theta_{u^\ast} v(q,a)\leq L^\theta_{u} v(q,a)$ and every edge is emptied, i.e $x-u^\ast=0$. 
Indeed, if for all $i\in\mathcal{D}$, there exist $j\in\mathcal{S}$ such that $m_{\edge{i}{j}}=x_{d_i}-\sum_{k\in\mathcal{S}(i)} u^{i,j-1}_{\edge{i}{k}}$, then $u^\ast_{d_i}=\sum_{k\in\mathcal{S}(i)}u^\ast_{\edge{i}{k}}=\sum_{k\in\mathcal{S}(i)}u^{i,j}_{\edge{i}{k}}=\sum_{k\in\mathcal{S}(i)}u^{i,j-1}_{\edge{i}{k}}+m_{\edge{i}{j}}=x_{d_i}$. Thus, for all $i\in\mathcal{D}$, $x_{d_i}-u^\ast_{d_i}=0$. So $\sum_{i\in\mathcal{D}}\left( x_{d_i} -\sum_{j\in\mathcal{S}}u^\ast_{\edge{i}{j}}\right)=0$ which is equivalent to $\sum_{j\in\mathcal{S}}x_{s_j} -\sum_{j\in\mathcal{S}}\sum_{i\in\mathcal{D}}u^\ast_{\edge{i}{j}}=0$ (because $x\in\mathcal{Q}$), i.e $\sum_{j\in\mathcal{S}}x_{s_j} -u^\ast_{s_j}=0$ which means that for all $j\in\mathcal{S}$ we have $x_{s_j} -u^\ast_{s_j}=0$ (because $x_{s_j}-u^\ast_{s_j}\geq 0$ as $u^\ast \in U_x$). Otherwise, there exist $i\in\mathcal{D}$ such that for all $j\in\mathcal{S}$, we have $m_{\edge{i}{j}}=x_{s_j}-\sum_{k\in\mathcal{D}(j)} u^{i,j-1}_{\edge{k}{j}}$, then $u^\ast_{s_j}=\sum_{k\in\mathcal{D}(j)}u^\ast_{\edge{k}{j}}=\sum_{k\in\mathcal{D}(j)}u^{i,j}_{\edge{k}{j}}=\sum_{k\in\mathcal{D}(j)}u^{i,j-1}_{\edge{k}{j}}+m_{\edge{i}{j}}=x_{s_j}$. Thus, for all $j\in\mathcal{S}$, $x_{s_j}-u^\ast_{s_j}=0$. So $\sum_{j\in\mathcal{S}}\left( x_{s_j} -\sum_{i\in\mathcal{D}}u^\ast_{\edge{i}{j}}\right) =0$ which is equivalent to $\sum_{i\in\mathcal{D}}x_{d_i} -\sum_{i\in\mathcal{D}}\sum_{j\in\mathcal{S}}u^\ast_{\edge{i}{j}}=0$ (because $x\in\mathcal{Q}$), i.e $\sum_{i\in\mathcal{D}}x_{d_i} -u^\ast_{d_i}=0$ which means that for all $i\in\mathcal{D}$ we have $x_{d_i} -u^\ast_{d_i}=0$ (because $x_{d_i}-u^\ast_{d_i}\geq 0$ as $u^\ast \in U_x$).
\\
This was done for any $u\in U_x$ and because $U_x$ is finite for every $x\in \mathcal Q$, we can choose $u$ such that it belongs to $\argmin_{u^\prime\in U_x} L_{u^\prime}^\theta v(q,a)$ giving the final result.
\end{proof}

\subsection{Value Function Property Preservation}
\label{sec:CG-propagation}

In this section, we show that the properties of the value function defined in Section~\ref{sec:CG-definitions} are preserved by the dynamic programming operator. 
In other words, we show that if $v\in V^\sigma$, then $L^\theta v\in V^\sigma$. We recall that $ V^\sigma= \mathcal{I}_{a}$. 

\begin{lemma}
If a function $v\in \mathcal I_{a}$, then $L^\theta v\in \mathcal I_{a}$.
\label{lem:CG-increasing}
\end{lemma}

\begin{proof}

Let $\underline{q}\in\mathcal Q$ and $a\in\mathcal{A}$, $\underline{x}=\underline{q}+a$. Let $\edge{i}{j}\in\mathcal{E}$, we define $\overline{q}=\underline{q}+e_{\edge{i}{j}}$, $\overline{x}=\overline{q}+a$. Since $v$ is increasing with $\edge{i}{j}$, we have that $v(\overline{q},a)\geq v(\underline{q},a)$. We aim to show that 
$L^\theta v(\overline{q},a)\geq L^\theta v(\underline{q},a)$.
\medbreak
Let $u_{\underline{x}}\in\argmin_{u\in U_{\underline{x}}} L^\theta v(\underline{q},a)$ and $u_{\overline{x}}\in\argmin_{u\in U_{\overline{x}}} L^\theta v(\overline{q},a)$. Using Proposition~\ref{prop:CG-optimal-control}, we can choose $u_{\underline{x}}$ such that $\underline{x}-u_{\underline{x}}=0$ and $u_{\overline{x}}$ such that $\overline{x}-u_{\overline{x}}=0$. Hence, 
\begin{align*}
L^\theta v(\underline{q},a) & =c(\underline{x})+\theta\mathbb E[v(\underline{x}-u_{\underline{x}},A)]\\
& =c(\underline{x})+\theta\mathbb E[v(0,A)]\\
& \leq c(\overline{x})+\theta\mathbb E[v(0,A)]\\
& =c(\overline{x})+\theta\mathbb E[v(\overline{x}-u_{\overline{x}},A)]\\
& = L^\theta v(\overline{q},a).
\end{align*}
since $c\in\mathcal I_{\edge{i}{j}}\subset\mathcal{I}_a$ from 
Assumption~\ref{ass:cost}.
This inequality remains valid for any $\edge{i}{j}\in\mathcal{E}$, giving us the final result.
\end{proof}

\subsection{Structure of the Optimal Policy}

We now present that, using the result of Theorem~\ref{thm:puterman_discounted}, there exists an optimal matching policy which is formed of a sequence of decision rules that belong to $D^\sigma$.

\begin{theorem}
The optimal control for the discounted cost problem is to match everything.
\label{thm:CG-opt-control}
\end{theorem}

\begin{proof}
We apply Theorem~\ref{thm:puterman_discounted} where $V^\sigma$ is the set of functions satisfying Definition~\ref{def:CG-increasing} and 
$D^\sigma$ the set defined in Definition~\ref{def:CG-optimal-decision-rule}.
\eqref{eq:A_discounted-1}-\eqref{eq:A_discounted-3} are verified because of Lemma~\ref{lem:technical-req}.

We now focus on the structural conditions of the theorem. From Lemma~\ref{lem:CG-increasing}, it follows $(a)$ since they show that if $v\in V^\sigma$, then $L^\theta v\in V^\sigma$. The result of 
Proposition~\ref{prop:CG-optimal-control} shows $(b)$ because the policy that belong to $D^\sigma$ minimize $L_u^\theta v$ 
if $v\in V^\sigma$. Finally, since limits preserve inequalities, the point-wise
convergence of functions of $V^\sigma$ belong to this set, which shows $(c)$.
\end{proof}

The following theorem shows that the previous result is also proven for the average cost problem.

\begin{theorem}
The optimal control for the average cost problem is to match everything. 
\label{thm:CG-opt-control-average}
\end{theorem}
\begin{proof}
We want to apply Theorem~\ref{thm:puterman_average} using the same value function set $V^\sigma$ and the same decision rule set $D^\sigma$ as in the proof of the previous theorem. Assumptions~\textit{(A1)} to \textit{(A4)} hold because of Lemma~\ref{lem:technical-req-average}. 

Let $(\theta_n)_{n\in\mathbb N}$ be a sequence such that $0\leq \theta_n < 1$ for all $n\in\mathbb N$ and $\underset{n\to +\infty}{lim} \theta_n = 1$. Let $n\in\mathbb N$.  We know that $v_{\theta_n}^\ast \in V^\sigma$ (see the proof of Theorem~\ref{thm:CG-opt-control}). The inequalities in the definitions of the properties used in $V^\sigma$ still hold if we add a constant to $v$, thus $v_{\theta_n}^\ast - v_{\theta_n}^\ast(0)e \in V^\sigma$. Using Assumption~\textit{(A3)} and Assumption~\textit{(A4)}, we have $H\leq v_{\theta_n}^\ast - v_{\theta_n}^\ast(0)e\leq M$, so $v_{\theta_n}^\ast - v_{\theta_n}^\ast(0)e \in V_H^\sigma$. This last result holds for each $n\in\mathbb N$ and since limits preserve inequalities $ V_H^\sigma$ is a closed set, $\underset{n\to +\infty}{lim} [v_{\theta_n}^\ast - v_{\theta_n}^\ast(0)e] \in V_H^\sigma$ which shows $(a)$. The result of 
Proposition~\ref{prop:CG-optimal-control} shows $(b)$ because the policy that belong to $D^\sigma$ minimize $L_u^1 v=T_u v$ 
if $v\in V_H^\sigma \subset V^\sigma$.
\end{proof}

\section{$N$-shaped graph}\label{sec:N} 

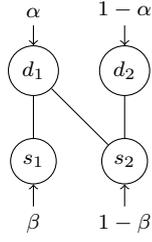
\begin{figure}[htbp]
     \centering
    \begin{tikzpicture}[scale=0.6]
     \node[style={circle,draw}] at (2,0) (s1) {$s_1$};
     \node[style={circle,draw}] at (4,0) (s2) {$s_2$};
     \node[style={circle,draw}] at (2,2) (d1) {$d_1$};
     \node[style={circle,draw}] at (4,2) (d2) {$d_2$};
     \draw[<-] (s1) -- (2,-1)  node[below] {$\beta$} ;
     \draw[<-] (s2) -- (4,-1)  node[below] {$1-\beta$} ;
     \draw[<-] (d1) -- (2,3)  node[above] {$\alpha$} ;
     \draw[<-] (d2) -- (4,3)  node[above] {$1-\alpha$} ;
     \draw (s1) -- (d1) node[pos=0.5,left] {};
     \draw (s2) -- (d1) node[pos=0.75,right] {};;
     \draw (s2) -- (d2) node[pos=0.5,right] {};
    \end{tikzpicture}  
   \caption{The $N$-shaped matching graph.}
   \label{fig:matching2}
  \end{figure}

The first model where difficulties arise due to the fact that we can not always empty the system is the $N$-shaped graph. It is formed by two supply nodes and two demand nodes with a $N$-shaped set of edges (see Figure~\ref{fig:matching2}). 
For this graph, we have $\mathcal{D}=\{d_1, d_2\}$, $\mathcal{S}=\{s_1, s_2\}$ and $\mathcal{E}=\{\edge{1}{1},\edge{1}{2},\edge{2}{2}\}$. Let us also define $\edge{2}{1}$ as the imaginary edge between $d_2$ and $s_1$ (imaginary because $\edge{2}{1} \notin \mathcal E$) that we introduce to ease the notations. To ensure stability, we assume that $\alpha>\beta$. 

We will show that the optimal policy for this case has a specific structure. For this purpose, we first present 
the properties of the value function. Then, we show how these properties characterize the optimal decision rule and how they are preserved by the dynamic programming operator. Finally, we prove the desired results in Theorem~\ref{thm:opt-control-thr-priority} and Theorem~\ref{thm:opt-control-thr-priority-average}.

\subsection{Value Function Properties}
\label{sec:definitions}

We start by using the increasing property in $\edge{1}{1}$ and $\edge{2}{2}$ as defined in Definition~\ref{def:CG-increasing} that we will note $\mathcal{I}_{\edge{1}{1}}$ and $\mathcal{I}_{\edge{2}{2}}$.
We also define an increasing property in the imaginary edge $\edge{2}{1}$ using the same definition as in Definition~\ref{def:CG-increasing} (even if $\edge{2}{1}\notin \mathcal{E}$) that we will note $\mathcal{I}_{\edge{2}{1}}$.

\begin{remark}
The increasing property in $\edge{2}{1}$ can be interpreted as the fact that we prefer to match $\edge{1}{1}$ and $\edge{2}{2}$ rather than to match $\edge{1}{2}$. 
Indeed, $v(q+e_{\edge{1}{1}}+e_{\edge{2}{2}}-e_{\edge{1}{2}},a)=v(q+e_{\edge{2}{1}},a)\geq v(q,a)$.
\end{remark}

We also define the convexity in $\edge{1}{2}$ and $\edge{2}{1}$ as follows:

\begin{definition}[Convexity property]\label{def:convex}
A function $v$ is convex in $\edge{1}{2}$ or $v\in\mathcal{C}_{\edge{1}{2}}$ if $v(q+e_{\edge{1}{2}},a)-v(q,a)$ is increasing in $\edge{1}{2}$, i.e., $\forall a\in\mathcal{A}$, $\forall q\in\mathcal Q$ such that $q_{d_1}\geq q_{s_1}$, we have
$$ v(q+2e_{\edge{1}{2}},a)-v(q+e_{\edge{1}{2}},a)\geq v(q+e_{\edge{1}{2}},a)-v(q,a).$$
Likewise, $v$ is convex in $\edge{2}{1}$ or $v\in\mathcal{C}_{\edge{2}{1}}$ if $v(q+e_{\edge{2}{1}},a)-v(q,a)$ is increasing in $\edge{2}{1}$, i.e., $\forall a\in\mathcal{A}$, $\forall q\in\mathcal Q$ such that $q_{s_1}\geq q_{d_1}$, we have 
$$v(q+2e_{\edge{2}{1}},a)-v(q+e_{\edge{2}{1}},a)\geq v(q+e_{\edge{2}{1}},a)-v(q,a).$$
\end{definition}

\begin{definition}[Boundary property]\label{def:boundary}
A function $v\in\mathcal B$ if 
$$
\forall a\in\mathcal{A}, \quad v(0,a)-v(e_{\edge{2}{1}},a)\leq v(e_{\edge{1}{2}},a)-v(0,a).
$$ 
\end{definition}

As we will show in Proposition~\ref{prop:u-min-Lv}, the properties $\mathcal I_{\edge{1}{1}}$, $\mathcal I_{\edge{2}{2}}$, $\mathcal I_{\edge{2}{1}}$ and $\mathcal C_{\edge{1}{2}}$ characterize the optimal decision rule. 
On the other hand, $\mathcal C_{\edge{2}{1}}$ and $\mathcal B$ are required to show that $\mathcal C_{\edge{1}{2}}$ is preserved by the operator $L^\theta$.

We aim to characterize the optimal matching policy using Theorem~\ref{thm:puterman_discounted} and Theorem~\ref{thm:puterman_average}. 
Thus, in the remainder of the section, we will consider the following set of structured value functions
\begin{equation}\label{eq:N-V-sigma}
V^\sigma=\mathcal{I}_{\edge{1}{1}}\cap \mathcal{I}_{\edge{2}{2}} \cap \mathcal{I}_{\edge{2}{1}} \cap \mathcal{C}_{\edge{1}{2}}  \cap \mathcal{ C}_{\edge{2}{1}}\cap\mathcal{B}.
\end{equation}

\subsection{Optimal decision rule}

In this section, we show that, for any $v\in V^\sigma$, there is a control of threshold-type in $\edge{1}{2}$ with priority to $\edge{1}{1}$ and $\edge{2}{2}$ that minimizes $L_u^\theta v$.

\begin{definition}[Threshold-type decision rule]\label{def:threshold_policy}
A decision rule $u_x$ is said to be of threshold type in $\edge{1}{2}$ with priority to $\edge{1}{1}$ and $\edge{2}{2}$ if:
\begin{enumerate}
\item it matches all of $\edge{1}{1}$ and $\edge{2}{2}$.
\item it matches $\edge{1}{2}$ only if the remaining items (in $d_1$ and $s_2$) are above a specific threshold, denoted by $t$ (with $t\in \mathbb{N}\cup\infty$).
\end{enumerate}
i.e, $u_x = \min(x_{d_1},x_{s_1})e_{\edge{1}{1}}+\min(x_{d_2},x_{s_2})e_{\edge{2}{2}}+k_t (x)e_{\edge{1}{2}}$ where 
$ k_t (x)=
\left\{\begin{array}{lr}
0 & if\; x_{d_1}-x_{s_1}\leq t \\
x_{d_1}-x_{s_1}-t & otherwise
\end{array}\right.
$.

We define $D^\sigma$ as the set of decision rules that are of threshold type in $\edge{1}{2}$ with priority to $\edge{1}{1}$ and $\edge{2}{2}$ for any $t\in \mathbb{N}\cup\infty$.

\end{definition}

\begin{remark}
If $t=\infty$, the decision rule will never match $\edge{1}{2}$. Otherwise, the decision rule will match $\edge{1}{2}$ until the remaining items in $d_1$ and $s_2$ are below or equal to the threshold $t$.
\end{remark}

In the next proposition, we establish that there exists an optimal decision rule with priority to $\edge{1}{1}$ and $\edge{2}{2}$.

\begin{proposition}
Let $v\in \mathcal I_{\edge{1}{1}}\cap \mathcal I_{\edge{2}{2}}\cap \mathcal I_{\edge{2}{1}}$, let $0\leq \theta\leq 1$. For any $q\in\mathcal Q$ and $a\in\mathcal{A}$, $x=q+a$, there exists $u^\ast\in U_x$ such that $u^\ast\in \argmin_{u\in U_x} L_u^\theta v(q,a)$, $u^\ast_{\edge{1}{1}}=\min(x_{d_1},x_{s_1})$ and $u^\ast_{\edge{2}{2}}=\min(x_{d_2},x_{s_2})$. In particular, this result holds for the average operator: $T_u$.
\label{prop:s1-d2-optimal-control}
\end{proposition}

\begin{proof}

Let $v\in \mathcal I_{\edge{1}{1}}\cap \mathcal I_{\edge{2}{2}}\cap\mathcal I_{\edge{2}{1}}$, $0\leq\theta\leq 1$, $q\in\mathcal Q$, $a\in\mathcal{A}$, $x=q+a$ and $u\in U_x$. The maximum number of matchings in $\edge{1}{1}$ is denoted by $m_{\edge{1}{1}}=\min(x_{d_1},x_{s_1})$ and in $\edge{2}{2}$ by $m_{\edge{2}{2}}=\min(x_{d_2},x_{s_2})$. 
\medbreak
Let $p_{\edge{1}{2}}=\min(u_{\edge{1}{2}},x_{s_1}-u_{\edge{1}{1}},x_{d_2}-u_{\edge{2}{2}})$ be the number of possible matchings that can be
transformed from $\edge{1}{2}$ to $\edge{1}{1}$ and $\edge{2}{2}$ matchings. We define a policy $u^{p_{\edge{1}{2}}}$ that removes the $p_{\edge{1}{2}}$ 
matchings in $\edge{1}{2}$ and matches $p_{\edge{1}{2}}$ times $\edge{1}{1}$ and $\edge{2}{2}$, that is, 
$u^{p_{\edge{1}{2}}}=u+p_{\edge{1}{2}}(e_{\edge{1}{1}}+e_{\edge{2}{2}}-e_{\edge{1}{2}})$. We verify that this policy is admissible, i.e $u^{p_{\edge{1}{2}}}\in U_x$: $u^{p_{\edge{1}{2}}}\in\mathbb{N}^{4}$ is true because $u\in U_x$. \textit{(a)} is true because $u^{p_{\edge{1}{2}}}_{\edge{2}{2}}=u_{\edge{2}{2}}+p_{\edge{1}{2}}\leq x_{d_2}$. \textit{(b)} is true because $u^{p_{\edge{1}{2}}}_{\edge{1}{1}}=u_{\edge{1}{1}}+p_{\edge{1}{2}}\leq x_{s_1}$. Then, we can use the fact that $v\in\mathcal I_{\edge{2}{1}}$ to show that $L^\theta_{u^{p_{\edge{1}{2}}}} v(q,a)\leq L^\theta_{u} v(q,a)$. 
\medbreak
Moreover, we define $u^\ast$ that matches all the the possible $\edge{1}{1}$ and $\edge{2}{2}$ of $x-u^{p_{\edge{1}{2}}}$, that is, of the remaining items when we apply $u^{p_{\edge{1}{2}}}$: $u^\ast=u^{p_{\edge{1}{2}}} + e_{\edge{1}{1}}(m_{\edge{1}{1}}-u^{p_{\edge{1}{2}}}_{\edge{1}{1}})+e_{\edge{2}{2}}(m_{\edge{2}{2}}-u^{p_{\edge{1}{2}}}_{\edge{2}{2}})$. We also verify that this policy is admissible, i.e $u^\ast\in U_x$: $u^\ast\in\mathbb{N}^4$ is true because $u^{p_{\edge{1}{2}}}\in U_x$, $m_{\edge{1}{1}}\geq 0$ and $m_{\edge{2}{2}}\geq 0$. We immediately get that $u^\ast_{\edge{2}{2}}=m_{\edge{2}{2}}\leq x_{d_2}$ and $u^\ast_{\edge{1}{1}}=m_{\edge{1}{1}}\leq x_{s_1}$. For $x_{d_1}$ and $x_{s_2}$, we need to discuss the value of $p_{\edge{1}{2}}$:
\begin{enumerate}
\item If $p_{\edge{1}{2}}=u_{\edge{1}{2}}$. Then we have $u^\ast_{\edge{1}{2}}=u^{p_{\edge{1}{2}}}_{\edge{1}{2}}=0$. Thus,
\begin{align*}
u^\ast_{\edge{1}{1}}+u^\ast_{\edge{1}{2}}&=m_{\edge{1}{1}}\leq x_{d_1}\\ 
u^\ast_{\edge{2}{2}}+u^\ast_{\edge{1}{2}}&=m_{\edge{2}{2}}\leq x_{s_2}
\end{align*}
\item If $p_{\edge{1}{2}}=x_{s_1}-u_{\edge{1}{1}}$. Then,
\begin{multline*}
u^\ast_{\edge{1}{1}}+u^\ast_{\edge{1}{2}}=m_{\edge{1}{1}}+u_{\edge{1}{2}}-x_{s_1}+u_{\edge{1}{1}}\leq u_{\edge{1}{2}}+u_{\edge{1}{1}}\leq x_{d_1}\\ 
u^\ast_{\edge{2}{2}}+u^\ast_{\edge{1}{2}}=m_{\edge{2}{2}}+u_{\edge{1}{2}}-x_{s_1}+u_{\edge{1}{1}}\leq x_{d_2}+u_{\edge{1}{2}}-x_{s_1}+u_{\edge{1}{1}}\\
=x_{s_2}+u_{\edge{1}{2}}-x_{d_1}+u_{\edge{1}{1}}\leq x_{s_2}
\end{multline*}
\item If $p_{\edge{1}{2}}=x_{d_2}-u_{\edge{2}{2}}$. Then,
\begin{multline*}
u^\ast_{\edge{1}{1}}+u^\ast_{\edge{1}{2}}=m_{\edge{1}{1}}+u_{\edge{1}{2}}-x_{d_2}+u_{\edge{2}{2}}\leq x_{s_1}+u_{\edge{1}{2}}-x_{d_2}+u_{\edge{2}{2}}\\
=x_{d_1}+u_{\edge{1}{2}}-x_{s_2}+u_{\edge{2}{2}}\leq x_{d_1}\\ 
u^\ast_{\edge{2}{2}}+u^\ast_{\edge{1}{2}}=m_{\edge{2}{2}}+u_{\edge{1}{2}}-x_{d_2}+u_{\edge{2}{2}}\leq u_{\edge{1}{2}}+u_{\edge{2}{2}}\leq x_{s_2}
\end{multline*}
\end{enumerate}
In every case, \textit{(a)} and \textit{(b)} are true. Hence, since $v\in \mathcal I_{\edge{1}{1}}\cap \mathcal I_{\edge{2}{2}}$, it results that $L^\theta_{u^\ast}v(q,a)\leq L^\theta_{u^{p_{\edge{1}{2}}}} v(q,a)$. 
\medbreak
As a result, we have $L^\theta_{u^\ast}v(q,a)\leq L^\theta_{u} v(q,a)$, $u^\ast_{\edge{1}{1}}=m_{\edge{1}{1}}=\min(x_{s_1},x_{d_1})$ and $u^\ast_{\edge{2}{2}}=m_{\edge{2}{2}}=\min(x_{s_2},x_{d_2})$. This was done for any $u\in U_x$ and because $U_x$ is finite for every $x\in \mathcal Q$, we can choose $u$ such that it belongs to $\argmin_{u^\prime\in U_x} L_{u^\prime}^\theta v(q,a)$ giving the final result.
\end{proof}

From this result, it follows that there exists an optimal decision rule that matches all possible $\edge{1}{1}$ and $\edge{2}{2}$. Our goal now is to find the optimal number of matchings in $\edge{1}{2}$. We introduce first some notation: 
\begin{definition}\label{def:optimal_l3}
Let $0\leq \theta \leq 1$, $x\in\mathcal{Q}$. We define:

$ K_x =\left\{\begin{array}{ll}
\{0\} &if \; x_{d_1}\leq x_{s_1} \\
\{0,\cdots,\min(x_{d_1}-x_{s_1},x_{s_2}-x_{d_2})\} &otherwise
\end{array}\right.$\\ the set of possible matching in $\edge{1}{2}$ after having matched all possible $\edge{1}{1}$ and $\edge{2}{2}$.

\end{definition}

\begin{remark}
The state of the system after having matched all possible $\edge{1}{1}$ and $\edge{2}{2}$ is of the form $(0,l,l,0)$ if $x_{d_1}\leq x_{s_1}$ and of the form $(l,0,0,l)$ otherwise (because of the definition of $\mathcal Q$ and $U_x$).
\end{remark}

Finally, we prove that a decision rule of threshold type in $\edge{1}{2}$ with priority to $\edge{1}{1}$ and $\edge{2}{2}$ is optimal. This is done by choosing the right $t$ for different cases such that $k_t (x)$ is the optimal number of matchings in $\edge{1}{2}$ for a given $x$.

\begin{proposition}
Let $v\in \mathcal{I}_{\edge{1}{1}}\cap \mathcal{I}_{\edge{2}{2}} \cap \mathcal I_{\edge{2}{1}} \cap \mathcal{C}_{\edge{1}{2}}$. Let $0\leq \theta\leq 1$. For any $q\in\mathcal{Q}$ and for any $a\in\mathcal{A}$, $x=q+a$, there exists $u^\ast\in D^\sigma$ (see Definition~\ref{def:threshold_policy}) such that $u^\ast\in\argmin_{u\in U_x} L_u^\theta v(q,a)$. In particular, this result holds for the average operator: $T_u$.
\label{prop:u-min-Lv}
\end{proposition}

\begin{proof}
Let $q\in\mathcal Q$, $a\in\mathcal{A}$, $x=q+a$ and $u\in U_x$. We note $m_{\edge{1}{1}}=\min(x_{s_1},x_{d_1})$ and $m_{\edge{2}{2}}=\min(x_{s_2},x_{d_2})$. We assumed that $v\in \mathcal{I}_{\edge{1}{1}}\cap \mathcal{I}_{\edge{2}{2}} \cap \mathcal I_{\edge{2}{1}}$, so we can use Proposition~\ref{prop:s1-d2-optimal-control} : $\exists u^\prime\in U_x$ such that $L_{u^\prime}^\theta v(q,a)\leq L_u^\theta v(q,a)$ and $u^\prime = m_{\edge{1}{1}}e_{\edge{1}{1}}+m_{\edge{2}{2}}e_{\edge{2}{2}}+ke_{\edge{1}{2}}$ with $k\in K_x$. We now have to prove that there exists $t\in\mathbb{N}\cup\infty$ such that 
\begin{equation}\label{eq:optimal_pol}L_{u^\ast}^\theta v(q,a)\leq L_{u^\prime}^\theta v(q,a),\quad \forall k\in K_x
\end{equation}
where $u^\ast \in D^\sigma$ (see Definition~\ref{def:threshold_policy}). If $x_{s_1}\geq x_{d_1}$, then $K_x={0}$ and we have $u^\ast = u^\prime$ which satisfies \eqref{eq:optimal_pol}. Otherwise, $x_{s_1}< x_{d_1}$ and $K_x\neq\{0\}$. Therefore, the state of the system after having matched $u^\prime$ (or $u^\ast$), i.e $x-u^\prime$ (or $x-u^\ast$), is of the form $(l,0,0,l)$. So when we compare $L_{u^\ast}^\theta v(q,a)$ with $L_{u^\prime}^\theta v(q,a)$, this comes down to comparing $\mathbb{E}[v(j^\ast e_{\edge{1}{2}},A)]$ with $\mathbb{E}[v(j^\prime e_{\edge{1}{2}},A)]$ ($j^\ast,j^\prime \in K_x$).
\smallbreak
First of all, suppose that $\forall j\in\mathbb{N},\; \mathbb{E}[v((j+1)e_{\edge{1}{2}},A)]-\mathbb{E}[v(je_{\edge{1}{2}},A)]\leq 0$. We choose $t=\infty$, so $k_t (x)=0$ and $u^\ast = m_{\edge{1}{1}}e_{\edge{1}{1}}+m_{\edge{2}{2}}e_{\edge{2}{2}}$. By assumption, we have $L_{u^\ast}^\theta v(q,a)\leq L_{u^\ast +e_{\edge{1}{2}}}^\theta v(q,a)\leq \cdots \leq L_{u^\ast +ke_{\edge{1}{2}}}^\theta v(q,a)$ for all $k\in K_x$ and because $u^\ast +ke_{\edge{1}{2}}=u^\prime$, we have proven \eqref{eq:optimal_pol}.
\smallbreak
Then, suppose that $\mathbb{E}[v(e_{\edge{1}{2}},A)]-\mathbb{E}[v(0,A)]\geq 0$. We choose $t=0$, so $k_t (x)=x_{d_1}-x_{s_1}$ and $u^\ast = m_{\edge{1}{1}}e_{\edge{1}{1}}+m_{\edge{2}{2}}e_{\edge{2}{2}}+(x_{d_1}-x_{s_1})e_{\edge{1}{2}}$. By assumption and because $v$ is convex in $\edge{1}{2}$, we have 
\[L_{u^\ast}^\theta v(q,a)\leq L_{u^\ast -e_{\edge{1}{2}}}^\theta v(q,a)\leq \cdots \leq L_{u^\ast -ke_{\edge{1}{2}}}^\theta v(q,a)\] for all $k\in K_x$ and because $u^\ast -(x_{d_1}-x_{s_1}-k)e_{\edge{1}{2}}=u^\prime$ (with $x_{d_1}-x_{s_1}-k\in K_x$ for all $k\in K_x$), we have proven \eqref{eq:optimal_pol}.
\smallbreak
Finally, suppose that $\exists j\in\mathbb{N}^\ast,\; \mathbb{E}[v((j+1)e_{\edge{1}{2}},A)]-\mathbb{E}[v(je_{\edge{1}{2}},A)]\geq 0$. Let $\underline{j}=\min\{j\in\mathbb{N}^\ast:\; \mathbb{E}[v((j+1)e_{\edge{1}{2}},A)]-\mathbb{E}[v(je_{\edge{1}{2}},A)]\geq 0 \}$. By definition of $\underline{j}$ and by convexity of $v$ in $\edge{1}{2}$, we have 
\begin{equation}
\mathbb{E}[v((\underline{j}-l)e_{\edge{1}{2}},A)]-\mathbb{E}[v((\underline{j}-l-1)e_{\edge{1}{2}},A)]\leq 0\quad \forall l\in \llbracket 0~;~ \underline{j}-1 \rrbracket \label{eq:threshold-dec}
\end{equation}
and
\begin{equation}
\mathbb{E}[v((\underline{j}+1+l)e_{\edge{1}{2}},A)]-\mathbb{E}[v((\underline{j}+l)e_{\edge{1}{2}},A)]\geq 0\quad \forall l\in\mathbb{N}\label{eq:threshold-inc}
\end{equation}
We choose $t=\underline{j}$. If $x_{d_1}-x_{s_1} \leq \underline{j}$, then we have $k_t(x)=0$ and $L_{u^\ast}^\theta v(q,a)\leq L_{u^\prime}^\theta v(q,a)$ for all $k\in K_x$ by \eqref{eq:threshold-dec} ($0\leq k\leq x_{d_1}-x_{s_1}\leq \underline{j}$). Otherwise $x_{d_1}-x_{s_1} > \underline{j}$, then $k_t(x)=x_{d_1}-x_{s_1}-\underline{j}$ and $L_{u^\ast}^\theta v(q,a)=c(x)+\theta\mathbb{E}[v(\underline{j}e_{\edge{1}{2}},A)]$. Therefore, for all $k\in K_x$, $L_{u^\ast}^\theta v(q,a)\leq L_{u^\prime}^\theta v(q,a)$ by \eqref{eq:threshold-dec} if $k\geq \underline{j}$ or by \eqref{eq:threshold-inc} if $k\leq \underline{j}$, which proves \eqref{eq:optimal_pol}.

\end{proof}

\subsection{Value Function Property Preservation}
\label{sec:propagation}

In this section, we show that the properties of the value function defined in Section~\ref{sec:definitions} are preserved by the dynamic programming operator. 
In other words, we show that if $v\in V^\sigma$, then $L^\theta v\in V^\sigma$. We recall that $ V^\sigma= \mathcal{I}_{\edge{1}{1}}\cap \mathcal{I}_{\edge{2}{2}} \cap \mathcal I_{\edge{2}{1}} \cap \mathcal{C}_{\edge{1}{2}}  \cap \mathcal{C}_{\edge{2}{1}}\cap\mathcal B$. 

We first show that the monotonicity on $\edge{1}{1}$, $\edge{2}{2}$ and $\edge{2}{1}$ is also preserved by the dynamic programming operator.

\begin{lemma}
If a function $v\in \mathcal I_{\edge{1}{1}}\cap \mathcal I_{\edge{2}{2}}\cap\mathcal I_{\edge{2}{1}}$, then $L^\theta v\in \mathcal I_{\edge{1}{1}}\cap\mathcal I_{\edge{2}{2}}\cap\mathcal I_{\edge{2}{1}}$.
\label{lem:increasing}
\end{lemma}

\begin{proof}

Let $\underline{q}\in\mathcal Q$ and $a\in\mathcal{A}$, $\underline{x}=\underline{q}+a$. We define $\overline{q}=\underline{q}+e_{\edge{1}{1}}$, $\overline{x}=\overline{q}+a$. Since $v$ is increasing with $\edge{1}{1}$, we have that $v(\overline{q},a)\geq v(\underline{q},a)$. We aim to show that 
$L^\theta v(\overline{q},a)\geq L^\theta v(\underline{q},a)$.
\medbreak
Let $u_{\overline{x}}\in\argmin_{u\in U_{\overline{x}}} L^\theta v(\overline{q},a)$. Since $(\overline{x})_{s_1}\geq 1$ and 
$(\overline{x})_{d_1}\geq 1$, using Proposition~\ref{prop:s1-d2-optimal-control}, we can choose $u_{\overline{x}}$ such that $(u_{\overline{x}})_{\edge{1}{1}}=\min(\overline{x}_{d_1},\overline{x}_{s_1})\geq 1$. 
Therefore, we can define $u_{\underline{x}}=u_{\overline{x}}-e_{\edge{1}{1}}$. $u_{\underline{x}}\in U_{\underline{x}}$ because 
$u_{\overline{x}}\in U_{\overline{x}}$, $(u_{\underline{x}})_{\edge{1}{1}}=(u_{\overline{x}})_{\edge{1}{1}}-1\geq 0$, $(u_{\underline{x}})_{\edge{1}{1}}+(u_{\underline{x}})_{\edge{1}{2}}=(u_{\overline{x}})_{\edge{1}{1}}-1+(u_{\overline{x}})_{\edge{1}{2}}\leq \overline{x}_{d_1}-1=\underline{x}_{d_1}$ and $(u_{\underline{x}})_{\edge{1}{1}}=(u_{\overline{x}})_{\edge{1}{1}}-1\leq \overline{x}_{s_1}-1=\underline{x}_{s_1}$. Besides, $\overline{x}-u_{\overline{x}}=\underline{x}-u_{\underline{x}}$ and $c(\overline{x})\geq c(\underline{x})$ since $c\in\mathcal I_{\edge{1}{1}}$ from 
Assumption~\ref{ass:cost}. Hence, 
\begin{align*}
L_{u_{\underline{x}}}^\theta v(\underline{q},a) & =c(\underline{x})+\theta\mathbb E[v(\underline{x}-u_{\underline{x}},A)]\\
& =c(\underline{x})-c(\overline{x})+c(\overline{x})+\theta\mathbb E[v(\overline{x}-u_{\overline{x}},A)]\\
& =c(\underline{x})-c(\overline{x})+L^\theta v(\overline{q},a)\\
& \leq L^\theta v(\overline{q},a).
\end{align*}
And, since $u_{\underline{x}}\in U_{\underline{x}}$, then by definition $L^\theta v(\underline{q},a)\leq L_{u_{\underline{x}}}^\theta v(\underline{q},a)$ and, 
as a result, $L^\theta v(\underline{q},a)\leq L^\theta v(\overline{q},a)$. The same arguments with $\overline{q}=\underline{q}+e_{\edge{2}{2}}$ can be made to show that $L^\theta v(\underline{q},a)\leq L^\theta v(\overline{q},a)$.

The proof is similar for $\mathcal{I}_{\edge{2}{1}}$ but also requires to handle the case when no matching can be made in $\edge{1}{2}$. Let $\underline{q}\in\mathcal Q$ and $a\in\mathcal{A}$, $\underline{x}=\underline{q}+a$. We denote $\overline{q}=\underline{q}+e_{\edge{1}{1}}+e_{\edge{2}{2}}-e_{\edge{1}{2}}$, $\overline{x}=\overline{q}+a$. Since $v\in \mathcal{I}_{\edge{2}{1}}$, we know that $v(\underline{q},a)\leq v(\overline{q},a)$. $c(\underline{x})\leq c(\overline{x})$ holds because of Assumption~\ref{ass:cost}. We aim to show that 
$L^\theta v(\underline{q},a)\leq L^\theta v(\overline{q},a)$.
\medbreak
Using Proposition~\ref{prop:s1-d2-optimal-control}, let $u_{\overline{x}}\in\argmin_{u\in U_{\overline{x}}} L^\theta_{v} v(\overline{q},a)$ such that $(u_{\overline{x}})_{\edge{2}{2}}=\min(\overline{x}_{d_2},\overline{x}_{s_2})$ and $(u_{\overline{x}})_{\edge{1}{1}}=\min(\overline{x}_{d_1},\overline{x}_{s_1})$. Suppose that $\underline{x}_{d_1}\geq 1$ and $\underline{x}_{s_2}\geq 1$. We define $u_{\underline{x}}=u_{\overline{x}}-e_{\edge{1}{1}}-e_{\edge{2}{2}}+e_{\edge{1}{2}}$, thus $\underline{x} - u_{\underline{x}}=\overline{x} - u_{\overline{x}}$ and $u_{\underline{x}}\in U_{\underline{x}}$ because $u_{\overline{x}}\in U_{\overline{x}}$, $\underline{x}_{s_1}=\overline{x}_{s_1}-1\geq (u_{\overline{x}})_{\edge{1}{1}}-1=\min(\overline{x}_{d_1},\overline{x}_{s_1})-1=\min(\underline{x}_{d_1},\underline{x}_{s_1}+1)-1\geq 0$, $\underline{x}_{d_2}=\overline{x}_{d_2}-1\geq (u_{\overline{x}})_{\edge{2}{2}}-1=\min(\overline{x}_{d_2},\overline{x}_{s_2})-1=\min(\underline{x}_{d_2}+1,\underline{x}_{s_2})-1\geq 0$. 
Thus,
\begin{align*}
L^\theta_{u_{\underline{x}}}v(\underline{q},a) & = c(\underline{x})+\theta\mathbb E[v(\underline{x}-u_{\underline{x}},A)]\\
& = c(\underline{x})-c(\overline{x})+c(\overline{x})+\theta\mathbb E[v(\overline{x}-u_{\overline{x}},A)]\\
& = c(\underline{x})-c(\overline{x})+L^\theta v(\overline{q},a)\\
& \leq L^\theta v(\overline{q},a),
\end{align*}
and $L^\theta v(\underline{q},a)\leq L^\theta_{u_{\underline{x}}}v(\underline{q},a)$ as $u_{\underline{x}}\in U_{\underline{x}}$. Suppose now that $\underline{x}_{d_1}=0$ or $\underline{x}_{s_2}=0$. In that case, we cannot do more matchings in state $\overline{x}$ than we can do in state $\underline{x}$: $u_{\overline{x}}\in U_{\underline{x}}$. Thus,
\begin{align*}
L^\theta_{u_{\overline{x}}}v(\underline{q},a) & = c(\underline{x})+\theta\mathbb E[v(\underline{x}-u_{\overline{x}},A)]\\
& \leq c(\underline{x})+\theta\mathbb E[v(\overline{x}-u_{\overline{x}},A)]\text{ since }v\in \mathcal{I}_{\edge{2}{1}}\\
& = c(\underline{x})-c(\overline{x})+c(\overline{x})+\theta\mathbb E[v(\overline{x}-u_{\overline{x}},A)]\\
& = c(\underline{x})-c(\overline{x})+L^\theta v(\overline{q},a)\\
& \leq L^\theta v(\overline{q},a),
\end{align*}
and $L^\theta v(\underline{q},a)\leq L^\theta_{u_{\overline{x}}}v(\underline{q},a)$ as $u_{\overline{x}}\in U_{\underline{x}}$. In both cases we get the desired result $L^\theta v(\underline{q},a)\leq L^\theta v(\overline{q},a)$.

\end{proof}

The proof that the dynamic programming operator preserves the convexity in $\edge{1}{2}$ was more difficult than anticipated. We had to introduce the boundary property $\mathcal B$ to show that $L^\theta v\in \mathcal C_{\edge{1}{2}}$ for a specific case.

\begin{lemma}
If $v\in \mathcal I_{\edge{1}{1}}\cap \mathcal I_{\edge{2}{2}}\cap \mathcal I_{\edge{2}{1}} \cap \mathcal C_{\edge{1}{2}}\cap \mathcal B$, 
then $L^\theta v\in \mathcal C_{\edge{1}{2}}$.
\label{lem:convex}
\end{lemma}
\begin{proof}
Let $\underline{q}\in\mathcal Q$, $\underline{q}_{d_1}\geq \underline{q}_{s_1}$, $a\in\mathcal{A}$, $\underline{x}=\underline{q}+a$, $\overline{q}=\underline{q}+e_{\edge{1}{2}}$, $\overline{x}=\overline{q}+a$, $\overline{\overline{q}}=\overline{q}+e_{\edge{1}{2}}$ and $\overline{\overline{x}}=\overline{\overline{q}}+a$. Since $v$ is convex in $\edge{1}{2}$, we have $v(\overline{q},a)-v(\underline{q},a)\leq v(\overline{\overline{q}},a)-v(\overline{q},a)$ (this inequality also holds for the cost function $c$ because of Assumption~\ref{ass:cost}). We aim to show that $L^\theta v(\overline{q},a)-L^\theta v(\underline{q},a)\leq L^\theta v(\overline{\overline{q}},a)-L^\theta v(\overline{q},a)$. For $y\in\left\{\underline{x},\overline{x},\overline{\overline{x}}\right\}$, let $u_{y}\in \argmin_u L_u^\theta v(y)$. From Proposition~\ref{prop:u-min-Lv}, we can choose $u_y$ such that $ u_{y}=\min(y_{d_1},y_{s_1})e_{\edge{1}{1}}+\min(y_{d_2},y_{s_2})e_{\edge{2}{2}}+k_t(y)e_{\edge{1}{2}}$.

Let us also define $m=\underline{x}-u_{\underline{x}}$. Suppose that $\underline{q}_{d_1}\geq \underline{q}_{s_1}+1$ or $a\in\mathcal{A}\setminus\{e_{\edge{2}{1}}\}$, we can distinguish 3 cases: (a) $k_t(\overline{x})>0$, (b) $k_t(\overline{x})=0$ and $k_t(\overline{\overline{x}})>0$ and (c) $k_t(\overline{x})=0$ and $k_t(\overline{\overline{x}})=0$:
\begin{enumerate}
\item[(a)] If $k_t(\overline{x})>0$. Then,
\begin{align*}
L^\theta v(\overline{q},a)-L^\theta v(\underline{q},a)&=c(\overline{x})-c(\underline{x})+\theta\mathbb E[v(m,A)-v(m,A)] \\
&\leq c(\overline{\overline{x}})-c(\overline{x})+\theta\mathbb E[v(m,A)-v(m,A)]\\
&=L^\theta v(\overline{\overline{q}},a)-L^\theta v(\overline{q},a)
\end{align*}
because $c\in\mathcal{C}_{\edge{1}{2}}$.
\item[(b)] If $k_t(\overline{x})=0$ and $k_t(\overline{\overline{x}})>0$. Then, 
\begin{align*}
L^\theta v(\overline{q},a)-L^\theta v(\underline{q},a)&=c(\overline{x})-c(\underline{x})+\theta\mathbb E[v(m+e_{\edge{1}{2}},A)-v(m,A)] \\
&=c(\overline{x})-c(\underline{x})+L^\theta v(\overline{q},a)-L_{u_{\overline{x}}+e_{\edge{1}{2}}}^\theta v(\overline{q},a) \\
&\leq c(\overline{\overline{x}})-c(\overline{x})\\
&=c(\overline{\overline{x}})-c(\overline{x})+\theta\mathbb E[v(m+e_{\edge{1}{2}},A)-v(m+e_{\edge{1}{2}},A)]\\
&=L^\theta v(\overline{\overline{q}},a)-L^\theta v(\overline{q},a)
\end{align*}
because $c\in\mathcal{C}_{\edge{1}{2}}$ and because $k_t(\underline{x})=k_t(\overline{x})=0$ and $1\in K_{\overline{x}}$.
\item[(c)] If $k_t(\overline{x})=0$ and $k_t(\overline{\overline{x}})=0$. Then,
\begin{align*}
L^\theta v(\overline{q},a)-L^\theta v(\underline{q},a)&=c(\overline{x})-c(\underline{x})+\theta\mathbb E[v(m+e_{\edge{1}{2}},A)-v(m,A)] \\
&\leq c(\overline{\overline{x}})-c(\overline{x})+\theta\mathbb E[v(m+2e_{\edge{1}{2}},A)-v(m+e_{\edge{1}{2}},A)]\\
&=L^\theta v(\overline{\overline{q}},a)-L^\theta v(\overline{q},a)
\end{align*}
because $c\in\mathcal{C}_{\edge{1}{2}}$ and $v\in\mathcal{C}_{\edge{1}{2}}$.
\end{enumerate}
Suppose now that 
$\underline{q}_{d_1}
=\underline{q}_{s_1}$ and $a=e_{\edge{2}{1}}$, we can distinguish 2 cases: $k_t(\overline{\overline{x}})>0$ 
and $k_t(\overline{\overline{x}})=0$:
\begin{itemize}
\item If $k_t(\overline{\overline{x}})>0$. Then, 
\begin{align*}
L^\theta v(\overline{q},a)-L^\theta v(\underline{q},a)&=c(\overline{x})-c(\underline{x})+\theta\mathbb E[v(m-e_{\edge{2}{1}},A)-v(m,A)] \\
&\leq c(\overline{x})-c(\underline{x})+\theta\mathbb E[v(m,A)-v(m,A)] \\
&\leq c(\overline{\overline{x}})-c(\overline{x})+\theta\mathbb E[v(m-e_{\edge{2}{1}},A)-v(m-e_{\edge{2}{1}},A)]\\
&=L^\theta v(\overline{\overline{q}},a)-L^\theta v(\overline{q},a)
\end{align*}
because $v\in\mathcal I_{\edge{2}{1}}$ and $c\in\mathcal{C}_{\edge{1}{2}}$.
\item If $k_t(\overline{\overline{x}})=0$. Then,
\begin{align*}
L^\theta v(\overline{q},a)-L^\theta v(\underline{q},a)&=c(\overline{x})-c(\underline{x})+\theta\mathbb E[v(m-e_{\edge{2}{1}},A)-v(m,A)] \\
&\leq c(\overline{\overline{x}})-c(\overline{x})+\theta\mathbb E[v(m-e_{\edge{2}{1}}+e_{\edge{1}{2}},A)\\
&\quad -v(m-e_{\edge{2}{1}},A)]\\
&=L^\theta v(\overline{\overline{q}},a)-L^\theta v(\overline{q},a)
\end{align*}
because $v\in\mathcal{B}$ and $c\in\mathcal{C}_{\edge{1}{2}}$.
\end{itemize}
\end{proof}

Then, to show that the dynamic programming operator preserves $\mathcal B$, we had to introduce an other property: $\mathcal C_{\edge{2}{1}}$ (convexity in $\edge{2}{1}$). The preservation of these two last properties are combined in the next lemma (see proof in Appendix~\ref{app:convex}).

\begin{lemma}
If $v\in \mathcal I_{\edge{1}{1}}\cap \mathcal I_{\edge{2}{2}}\cap \mathcal I_{\edge{2}{1}} \cap \mathcal C_{\edge{1}{2}}\cap\mathcal C_{\edge{2}{1}}\cap \mathcal B$, 
then $L^\theta v\in \mathcal C_{\edge{2}{1}}\cap \mathcal B$.
\label{lem:B-convex}
\end{lemma}

In this section, we have shown that the structural properties of the value function presented in Section~\ref{sec:definitions} are preserved by the dynamic programming operator. That is, if 
$v\in V^\sigma$, then $L^\theta v$ also belongs to this set. Using this result, we give the structure of the optimal policy in the next section.

\subsection{Structure of the Optimal Policy}\label{sec:N-optimal-structure}

We now present that, using the result of Theorem~\ref{thm:puterman_discounted}, there exists an optimal matching policy which is formed of a sequence of decision rules that belong to $D^\sigma$ (with a fixed threshold).

\begin{theorem}
The optimal control for the discounted cost problem is of threshold type in $\edge{1}{2}$ with priority to $\edge{1}{1}$ and $\edge{2}{2}$.
\label{thm:opt-control-thr-priority}
\end{theorem}

\begin{proof}
We apply Theorem~\ref{thm:puterman_discounted} where $V^\sigma$ is the set of functions defined in \eqref{eq:N-V-sigma} and 
$D^\sigma$ the set defined in Definition~\ref{def:threshold_policy}.
\eqref{eq:A_discounted-1}-\eqref{eq:A_discounted-3} are verified because of Lemma~\ref{lem:technical-req}.

We now focus on the structural conditions of the theorem. From Lemma~\ref{lem:increasing}, Lemma~\ref{lem:convex} and Lemma~\ref{lem:B-convex} of Section~\ref{sec:propagation}, it follows $(a)$ since they show that if $v\in V^\sigma$, then $L^\theta v\in V^\sigma$. The result of 
Proposition~\ref{prop:u-min-Lv} shows $(b)$ because the policy that belong to $D^\sigma$ minimize $L_u^\theta v$ 
if $v\in V^\sigma$. Finally, since limits preserve inequalities, the point-wise
convergence of functions of $V^\sigma$ belong to this set, which shows $(c)$.
\end{proof}

The following theorem shows that the previous result is also proven for the average cost problem.

\begin{theorem}
The optimal control for the average cost problem is of threshold type in $\edge{1}{2}$ with priority to $\edge{1}{1}$ and $\edge{2}{2}$. 
\label{thm:opt-control-thr-priority-average}
\end{theorem}
\begin{proof}
We want to apply Theorem~\ref{thm:puterman_average} using the same value function set $V^\sigma$ and the same decision rule set $D^\sigma$ as in the proof of the previous theorem. Assumptions~\textit{(A1)} to \textit{(A4)} hold because of Lemma~\ref{lem:technical-req-average}. 

Let $(\theta_n)_{n\in\mathbb N}$ be a sequence such that $0\leq \theta_n < 1$ for all $n\in\mathbb N$ and $\underset{n\to +\infty}{lim} \theta_n = 1$. Let $n\in\mathbb N$.  We know that $v_{\theta_n}^\ast \in V^\sigma$ (see the proof of Theorem~\ref{thm:opt-control-thr-priority}). The inequalities in the definitions of the properties used in $V^\sigma$ still hold if we add a constant to $v$, thus $v_{\theta_n}^\ast - v_{\theta_n}^\ast(0)e \in V^\sigma$. Using Assumption~\textit{(A3)} and Assumption~\textit{(A4)}, we have $H\leq v_{\theta_n}^\ast - v_{\theta_n}^\ast(0)e\leq M$, so $v_{\theta_n}^\ast - v_{\theta_n}^\ast(0)e \in V_H^\sigma$. This last result holds for each $n\in\mathbb N$ and since limits preserve inequalities $ V_H^\sigma$ is a closed set, $\underset{n\to +\infty}{lim} [v_{\theta_n}^\ast - v_{\theta_n}^\ast(0)e] \in V_H^\sigma$ which shows (a). The result of 
Proposition~\ref{prop:u-min-Lv} shows (b) because the policy that belong to $D^\sigma$ minimize $L_u^1 v=T_u v$ 
if $v\in V_H^\sigma \subset V^\sigma$.
\end{proof}

\subsection{Computing the Optimal Threshold}

We consider the matching policy of threshold type in $\edge{1}{2}$ with priority to $\edge{1}{1}$ and $\edge{2}{2}$ in the average cost case.

\begin{proposition}
Let $\rho=\frac{\beta(1-\alpha)}{\alpha(1-\beta)}\in(0,1)$, $R=\frac{c_{s_1}+c_{d_2}}{c_{d_1}+c_{s_2}}$ and $\Pi^{T_{\edge{1}{2}}}$ be the set of matching policy of threshold type in $\edge{1}{2}$ with priority to $\edge{1}{1}$ and $\edge{2}{2}$. Assume that the cost function is a linear function. The optimal threshold $t^\ast$, which minimizes the average cost on $\Pi^{T_{\edge{1}{2}}}$, is 
$$
t^\ast=\left\{\begin{array}{ll}
\ceil{k} & if\; f(\ceil{k})\leq f(\floor{k}) \\
\floor{k} & otherwise 
\end{array}\right.
$$
where $k=\frac{\log{\frac{\rho-1}{(R+1)\log{\rho}}}}{\log\rho}-1$ and
$
f(x)=(c_{d_1}+c_{s_2})x+(c_{d_1}+c_{d_2}+c_{s_1}+c_{s_2})\frac{\rho^{x+1}}{1-\rho}
-(c_{d_1}+c_{s_2})\frac{\rho}{1-\rho}+((c_{d_1}+c_{s_1})\alpha\beta+(c_{d_2}+c_{s_2})(1-\alpha)(1-\beta)+
(c_{d_2}+c_{s_1})(1-\alpha)\beta+(c_{d_1}+c_{s_2})\alpha(1-\beta)).
$
\label{prop:opt-threshold}

The threshold $t^*$ is positive.
\end{proposition}
\begin{proof}
The idea of the proof is to look at the Markov chain derived from the policy $u_t^\infty \in\Pi^{T_{\edge{1}{2}}}$. We show that the Markov chain is positive recurrent and we compute the stationary distribution. Using the strong law of large numbers for Markov chains, we show that the average cost $g^{u_t^\infty}$ is equal to the expected cost of the system under the stationary distribution. Then, we find an analytical form for the expected cost which depends on the threshold on $\edge{1}{2}$, i.e, $t$. Finally, we minimize the function over $t$. See details in Appendix~\ref{app:opt-threshold}.
\end{proof}

\section{General Matching Graphs}
\label{sec:general}

\subsection{Complete Graphs Minus One Edge}

\begin{figure}[htbp]
     \centering
    \begin{tikzpicture}[]

     \node[style={circle,draw}] at (0,0) (s1) {$s_1$};
     \node[style={circle,draw}] at (3,0) (s2) {$s_2$};
     \node[style={circle,draw}] at (5,0) (s3) {$s_3$};
     \node[style={circle,draw}] at (0,2) (d1) {$d_1$};
     \node[style={circle,draw}] at (2,2) (d2) {$d_2$};
     \node[style={circle,draw}] at (5,2) (d3) {$d_3$};
     \draw[<-] (s1) -- (0,-1)  node[below] {$\beta_1$} ;
     \draw[<-] (s2) -- (3,-1)  node[below] {$\beta_2$} ;
     \draw[<-] (s3) -- (5,-1)  node[below] {$\beta_3$} ;
     \draw[<-] (d1) -- (0,3)  node[above] {$\alpha_1$} ;
     \draw[<-] (d2) -- (2,3)  node[above] {$\alpha_2$} ;
     \draw[<-] (d3) -- (5,3)  node[above] {$\alpha_3$} ;
     \draw (s1) -- (d1);
     \draw (s1) -- (d2);
     \draw (s2) -- (d2);
     \draw (s2) -- (d3);
     \draw (s2) -- (d1);
     \draw (s3) -- (d2);
     \draw (s3) -- (d3);
     \draw (s3) -- (d1);
    \end{tikzpicture}  
   \caption{A complete matching graph minus the edge $\edge{3}{1}$.}
   \label{fig:complete_minus_one_graph}
  \end{figure}
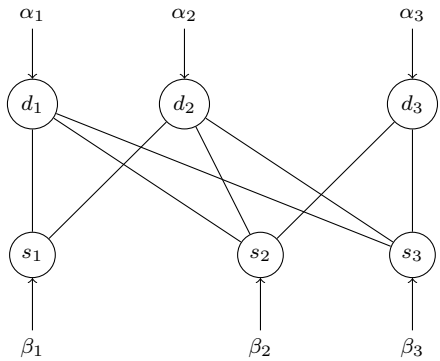

In this section, we are going to generalize the results of the $N$-shaped graph to any complete graph minus one edge with specific costs. 
Let us note this latter matching graph $\mathcal{G}=(\mathcal{D}\cup \mathcal{S},\mathcal{E})$ with $\mathcal{E}=\left( \mathcal{D}\times\mathcal{S}\right)\setminus\{\edge{i^\ast}{j^\ast}\}$, where $\edge{i^\ast}{j^\ast}$ ($i^\ast \in\mathcal{D}$ and $j^\ast\in\mathcal{S}$) is the missing edge (see Figure~\ref{fig:complete_minus_one_graph} for an example). 
We note $\mathcal{G}^N=(\mathcal{D}^N\cup \mathcal{S}^N,\mathcal{E}^N)$ the $N$-shaped matching graph with $\mathcal{D}^N$, $\mathcal{S}^N$ and $\mathcal{E}^N$ defined as in Section~\ref{sec:N}.

First, let us define some transformations, to move from $Y$ (the Markov Decision Process defined on $\mathcal{G}$) to $Y^N$ (the Markov Decision Process defined on $\mathcal{G}^N$).
\begin{definition}\label{def:complete-to-N}
Let $\mathcal{Q}$ and $\mathcal{A}$ be the sets of possible state and arrival of $\mathcal{G}$.
Let $\mathcal{Q}^N$ and $\mathcal{A}^N$ be the sets of possible state and arrival of $\mathcal{G}^N$.
We define the projection from $\mathcal{Q}$ to $\mathcal{Q}^N$ and the projection from $\mathcal{A}$ to $\mathcal{A}^N$ as 
\begin{align*} p^N_{\mathcal{Q}}(q)&=\left(\sum_{i\in\mathcal{D}(j^\ast)}q_{d_i},q_{d_{i^\ast}},q_{s_{j^\ast}},\sum_{j\in\mathcal{S}(i^\ast)}q_{s_j}\right)\text{ and }\\ 
p^N_{\mathcal{A}}(a)&=\left\{\begin{array}{lr}
e_{\edge{1}{1}} & if\; a=e_{\edge{i}{j^\ast}}, i\in\mathcal{D}(j^\ast) \\
e_{\edge{2}{2}} & if\; a=e_{\edge{i^\ast}{j}}, j\in\mathcal{S}(i^\ast) \\
e_{\edge{2}{1}} & if\; a=e_{\edge{i^\ast}{j^\ast}} \\
e_{\edge{1}{2}} & otherwise
\end{array}\right. .
\end{align*}
We can easily show that $p^N_{\mathcal{Q}}(x)=p^N_{\mathcal{Q}}(q)+p^N_{\mathcal{A}}(a)$ where $x=q+a$.
Let $a^N\in\mathcal{A}^N$, we also define $(p^{N}_{\mathcal{A}})^{-1}(a^N)=\{a\in\mathcal{A}:p^N_{\mathcal{A}}(a)=a^N\}$.
\end{definition}

Let $A$ be the arrival process associated to $\mathcal{G}$, we construct an arrival process $A^N$ for $\mathcal{G}^N$ in the following way:
\[\forall a^N\in\mathcal{A}^N,\quad \mathbb{P}(A^N = a^N)=\sum_{a\in (p_{\mathcal{A}}^N)^{-1}(a^N)}\mathbb{P}(A = a)\]

We formalize what we mean by specific costs in the following assumption:

\begin{assumption}\label{ass:equal-costs}
Let $c$ and $c^N$ be the cost functions associated to $Y$ and $Y^N$. We assume that $c$ and $c^N$ are cost functions such that $c(x)=c^N (p^N_{\mathcal{Q}}(x))$ for all $x\in\mathcal{Q}$.
\end{assumption}

For example, Assumption~\ref{ass:equal-costs} is satisfied for any linear cost function $c$ which have the same coefficient $c_{d_i}=c_1$ for all $i\in\mathcal{D}(j^\ast)$ and the same coefficient $c_{s_j}=c_2$ for all $j\in\mathcal{S}(i^\ast)$. We just have to construct a linear cost function $c^N$ with the following coefficients $c^N_{d_1}=c_1$, $c^N_{d_2}=c_{d_{i^\ast}}$, $c^N_{s_1}=c_{s_{j^\ast}}$ and $c^N_{d_2}=c_2$.

We now translates what is the decision rule of threshold-type in $\edge{1}{2}$ with priority in $\edge{1}{1}$ and $\edge{2}{2}$ for $\mathcal{G}$.

\begin{definition}[Threshold-type decision rule]\label{def:CMO-threshold-policy}
A decision rule $u_x$ is said to be of threshold type with priority to $i^\ast$ and $j^\ast$ if:
\begin{enumerate}
\item it matches all of $\edge{i}{j^\ast}$ for all $i\in\mathcal{D}(j^\ast)$ and $\edge{i^\ast}{j}$ for all $j\in\mathcal{S}(i^\ast)$.
\item it matches $\edge{i}{j}$ ($i\in\mathcal{D}(j^\ast)$ and $j\in\mathcal{S}(i^\ast)$) only if the remaining items (in the sum of $d_i$ for $i\in\mathcal{D}(j^\ast)$) are above a specific threshold, denoted by $t$ (with $t\in \mathbb{N}\cup\infty$).
\end{enumerate}
i.e, $u_x$ is such that $\sum_{i\in\mathcal{D}(j^\ast)}(u_x)_{\edge{i}{j^\ast}}=\min(\sum_{i\in\mathcal{D}(j^\ast)}x_{d_i},x_{s_{j^\ast}})$,
\\
$\sum_{j\in\mathcal{S}(i^\ast)}(u_x)_{\edge{i^\ast}{j}}=\min(x_{d_{i^\ast}},\sum_{j\in\mathcal{S}(i^\ast)}x_{s_j})$ and $\sum_{i\in\mathcal{D}(j^\ast)}\sum_{j\in\mathcal{S}(i^\ast)}(u_x)_{\edge{i}{j}}=k_t (x)$ where 
\[ k_t (x)=
\left\{\begin{array}{lr}
0 & if\; \sum_{i\in\mathcal{D}(j^\ast)}x_{d_i}-x_{s_{j^\ast}}\leq t \\
\sum_{i\in\mathcal{D}(j^\ast)}x_{d_i}-x_{s_{j^\ast}}-t & otherwise
\end{array}\right.
.\]

We define $D^\ast$ as the set of decision rules that are of threshold type with priority to $i^\ast$ and $j^\ast$ for any $t\in \mathbb{N}\cup\infty$.

\end{definition}

The next theorem shows that the stationary policy based on this decision rule of threshold-type with priority to $i^\ast$ and $j^\ast$ is optimal based on the results of Section~\ref{sec:N-optimal-structure} and some lemmas to bridge the gap between $\mathcal{G}$ and $\mathcal{G}^N$.

\begin{theorem}\label{thm:CMO-optimal-policy}
The optimal control for the discounted cost problem and the average cost problem is of threshold type with priority to $i^\ast$ and $j^\ast$.
\end{theorem}
\begin{proof}
Let $0\leq \theta <1$, let $\pi$ be a stationary policy on $Y$ and $\pi^\ast=(u(X(n)))_{n\geq 0}$ (with $u\in D^\ast$ as defined in Definition~\ref{def:CMO-threshold-policy}) be a threshold-type policy with priority in $i^\ast$ and $j^\ast$. Let $y_0=(q_0,a_0)\in\mathcal{Q}\times\mathcal{A}$ and $y^N_0=(p^N_{\mathcal{Q}}(q_0),p^N_{\mathcal{A}}(a_0))$. Using Lemma~\ref{lem:CMO-policy-construction}, there exists a policy $\pi^N$ such that:
\[v^{\pi}_\theta (y_0)=v^{\pi^N}_\theta (y^N_0)\geq v^{(\pi^N)^\ast}_\theta (y^N_0)=v^{\pi^\ast}_\theta (y_0)\]
where $(\pi^N)^\ast=(u^N(X^N(n)))_{n\geq 0}$ (with $u^N\in D^\sigma$ as defined in Definition~\ref{def:threshold_policy}) is a threshold-type policy with priority in $\edge{1}{1}$ and $\edge{2}{2}$. The inequality comes from our optimality result on $\mathcal{G}^N$: Theorem~\ref{thm:opt-control-thr-priority}. The last equality comes from Lemma~\ref{lem:CMO-optimal-policy}. This was done for any $y_0=(q_0,a_0) \in\mathcal{Q}\times\mathcal{A}$ and the average cost problem follows easily (with Theorem~\ref{thm:opt-control-thr-priority-average}), giving us the final result.
\end{proof}

\begin{lemma}\label{lem:CMO-optimal-policy}
Let $0\leq \theta <1$, let $\pi^\ast=(u(X(n)))_{n\geq 0}$ (with $u\in D^\ast$ as defined in Definition~\ref{def:CMO-threshold-policy}) be a threshold-type policy with priority in $i^\ast$ and $j^\ast$ and let $(\pi^N)^\ast=(u^N(X^N(n)))_{n\geq 0}$ (with $u^N\in D^\sigma$ as defined in Definition~\ref{def:threshold_policy}) be a threshold-type policy with priority in $\edge{1}{1}$ and $\edge{2}{2}$. Given Assumption~\ref{ass:equal-costs}, we have $v^{\pi^\ast}_\theta (y)=v^{(\pi^N)^\ast}_\theta (y^N)$ for all $y=(q,a)\in\mathcal{Q}\times\mathcal{A}$, with $y^N=(p^N_{\mathcal{Q}}(q),p^N_{\mathcal{A}}(a))$. The result remains true for the average cost problem with $g^{\pi^\ast}$ and $g^{(\pi^N)^\ast}$.
\end{lemma}
\begin{proof}
Let $0\leq \theta <1$, let $\pi^\ast=(u(X(n)))_{n\geq 0}$ (with $u\in D^\ast$ as defined in Definition~\ref{def:CMO-threshold-policy}) be a threshold-type policy with priority in $i^\ast$ and $j^\ast$ and let $(\pi^N)^\ast=(u^N(X^N(n)))_{n\geq 0}$ (with $u^N\in D^\sigma$ as defined in Definition~\ref{def:threshold_policy}) be a threshold-type policy with priority in $\edge{1}{1}$ and $\edge{2}{2}$. Let $y_0=(q_0,a_0)\in\mathcal{Q}\times\mathcal{A}$, with $y^N_0=(p^N_{\mathcal{Q}}(q_0),p^N_{\mathcal{A}}(a_0))$.
\medbreak
First, let us note that $(\pi^N)^\ast$ is the "projection" of $\pi^\ast$ on $\mathcal{G}^N$ in the following sense. Let $x\in\mathcal{Q}$, $u(x)\in D^\ast$ the matching of state $x$ ($u(x)\in U_x$) under the policy $\pi^\ast$ and $u^N(p^N_{\mathcal{Q}}(x))\in D^\sigma$ the matching of state $p^N_{\mathcal{Q}}(x)$ ($u^N(p^N_{\mathcal{Q}}(x))\in U^N_{p^N_{\mathcal{Q}}(x)}$) under the policy $(\pi^N)^\ast$. We can easily see that 
\\
$u^N_{\edge{1}{1}}(p^N_{\mathcal{Q}}(x))=\sum_{i\in\mathcal{D}(j^\ast)}u_{\edge{i}{j^\ast}}(x)$, $u^N_{\edge{2}{2}}(p^N_{\mathcal{Q}}(x))=\sum_{j\in\mathcal{S}(i^\ast)}u_{\edge{i^\ast}{j}}(x)$ and $u^N_{\edge{1}{2}}(p^N_{\mathcal{Q}}(x))=\sum_{i\in\mathcal{D}(j^\ast)}\sum_{j\in\mathcal{S}(i^\ast)}u_{\edge{i}{j}}(x)$.
\medbreak
Then, we are going to prove by induction that for any $a^N_1\in\mathcal{A}^N,\cdots,a^N_n\in\mathcal{A}^N$ and any $a_1\in (p^N_{\mathcal{A}})^{-1}(a^N_1),\cdots, a_n\in (p^N_{\mathcal{A}})^{-1}(a^N_n)$, we have $p^N_{\mathcal{Q}}(x_n)=x^N_n$.
\\
We already specifically chose $y^N_0$ to verify this property: $p^N_{\mathcal{Q}}(x_0)=p^N_{\mathcal{Q}}(q_0)+p^N_{\mathcal{A}}(a_0)=q^N_0 +a^N_0=x^N_0$. Now, assume that $p^N_{\mathcal{Q}}(x_{n-1})=x^N_{n-1}$. Then,
\begin{align*}
p^N_{\mathcal{Q}}(q_n) &= \left(\sum_{i\in\mathcal{D}(j^\ast)}(q_n)_{d_i},(q_n)_{d_{i^\ast}},(q_n)_{s_{j^\ast}},\sum_{j\in\mathcal{S}(i^\ast)}(q_n)_{s_j}\right)\\
&= \left(\sum_{i\in\mathcal{D}(j^\ast)}(x_{n-1}-u(x_{n-1}))_{d_i},(x_{n-1}-u(x_{n-1}))_{d_{i^\ast}},\right.\\
&\quad\left. (x_{n-1}-u(x_{n-1}))_{s_{j^\ast}},\sum_{j\in\mathcal{S}(i^\ast)}(x_{n-1}-u(x_{n-1}))_{s_j}\right)\\
&= \left(\sum_{i\in\mathcal{D}(j^\ast)}(x_{n-1})_{d_i}-\sum_{j\in\mathcal{S}(i)}u(x_{n-1})_{\edge{i}{j}},\right. \\
&\quad \left.(x_{n-1})_{d_{i^\ast}}-\sum_{j\in\mathcal{S}(i^\ast)}u(x_{n-1})_{\edge{i^\ast}{j}}, (x_{n-1})_{s_{j^\ast}}-\sum_{i\in\mathcal{D}(j^\ast)}u(x_{n-1})_{\edge{i}{j^\ast}},\right. \\
&\quad \left.\sum_{j\in\mathcal{S}(i^\ast)}(x_{n-1})_{s_j}-\sum_{i\in\mathcal{D}(j)}u(x_{n-1})_{\edge{i}{j}}\right)\\
&= p^N_{\mathcal{Q}}(x_{n-1})-\sum_{i\in\mathcal{D}(j^\ast)}u(x_{n-1})_{\edge{i}{j^\ast}}e_{\edge{1}{1}}-\sum_{j\in\mathcal{S}(i^\ast)}u(x_{n-1})_{\edge{i^\ast}{j}}e_{\edge{2}{2}}\\
&\quad -\sum_{i\in\mathcal{D}(j^\ast)}\sum_{j\in\mathcal{S}(i^\ast)}u(x_{n-1})_{\edge{i}{j}}e_{\edge{1}{2}}\\
&= p^N_{\mathcal{Q}}(x_{n-1})-u^N(p^N_{\mathcal{Q}}(x_{n-1}))_{\edge{1}{1}}e_{\edge{1}{1}}-u^N(p^N_{\mathcal{Q}}(x_{n-1}))_{\edge{2}{2}}e_{\edge{2}{2}}\\
&\quad  -u^N(p^N_{\mathcal{Q}}(x_{n-1}))_{\edge{1}{2}}e_{\edge{1}{2}}\\
&= p^N_{\mathcal{Q}}(x_{n-1}) -u^N(p^N_{\mathcal{Q}}(x_{n-1}))\\
&= x^N_{n-1} -u^N(x^N_{n-1}) \\
&=q^N_n
\end{align*}
and $p^N_{\mathcal{A}}(a_n)=a^N_n$ (because $p^N_{\mathcal{A}}((p^N_{\mathcal{A}})^{-1}(a^N))=a^N$ for all $a^N \in\mathcal{A}^N$). Thus, $p^N_{\mathcal{Q}}(x_{n})=p^N_{\mathcal{Q}}(q_{n})+p^N_{\mathcal{A}}(a_{n})=q^N_n+a^N_n=x^N_n$.
\\
Finally, using this property and Assumption~\ref{ass:equal-costs}, we have
\begin{align*}
\mathbb{E}^{\pi^\ast}_{y_0} [c(Y(n))]&=\sum_{a_1\in\mathcal{A},\cdots,a_n \in\mathcal{A}}c(x_{n})\prod_{k=1}^n \mathbb{P}(A=a_k) \\
&=\sum_{a^N_1\in\mathcal{A}^N,\cdots,a^N_n \in\mathcal{A}^N}\sum_{a_1\in (p^N_{\mathcal{A}})^{-1}(a^N_1)}\cdots \sum_{a_n\in (p^N_{\mathcal{A}})^{-1}(a^N_n)}c(x_{n})\\
&\quad \times\prod_{k=1}^n \mathbb{P}(A=a_k) \\
&=\sum_{a^N_1\in\mathcal{A}^N,\cdots,a^N_n \in\mathcal{A}^N}\sum_{a_1\in (p^N_{\mathcal{A}})^{-1}(a^N_1)}\cdots \sum_{a_n\in (p^N_{\mathcal{A}})^{-1}(a^N_n)}c^N(p^N_{\mathcal{Q}}(x_{n}))\\
&\quad \times\prod_{k=1}^n \mathbb{P}(A=a_k) \\
&=\sum_{a^N_1\in\mathcal{A}^N,\cdots,a^N_n \in\mathcal{A}^N}\sum_{a_1\in (p^N_{\mathcal{A}})^{-1}(a^N_1)}\cdots \sum_{a_n\in (p^N_{\mathcal{A}})^{-1}(a^N_n)}c^N(x^N_{n})\\
&\quad \times\prod_{k=1}^n \mathbb{P}(A=a_k) \\
&=\sum_{a^N_1\in\mathcal{A}^N,\cdots,a^N_n \in\mathcal{A}^N}c^N(x^N_{n})\prod_{k=1}^n \sum_{a_k\in (p^N_{\mathcal{A}})^{-1}(a^N_k)}\mathbb{P}(A=a_k) \\
&=\sum_{a^N_1\in\mathcal{A}^N,\cdots,a^N_n \in\mathcal{A}^N}c^N(x^N_{n})\prod_{k=1}^n \mathbb{P}(A^N=a^N_k) \\
&=\mathbb{E}^{(\pi^N)^\ast}_{y^N_0} [c^N(Y^N(n))].
\end{align*}
This equality is true for any $n\in\mathbb{N}$. Therefore, $v^{\pi^\ast}_\theta (y_0)=v^{(\pi^N)^\ast}_\theta (y^N_0)$ and $g^{\pi^\ast}(y_0)=g^{(\pi^N)^\ast} (y^N_0)$. This was done for any $y_0=(q_0,a_0) \in\mathcal{Q}\times\mathcal{A}$ with $y^N_0=(p^N_{\mathcal{Q}}(q_0),p^N_{\mathcal{A}}(a_0))$, giving the final result.
\end{proof}

\begin{lemma}\label{lem:CMO-policy-construction}
Let $0\leq \theta <1$, let $\pi$ be a stationary policy on $Y$, $y=(q,a)\in\mathcal{Q}\times\mathcal{A}$ and $y^N=(p^N_{\mathcal{Q}}(q),p^N_{\mathcal{A}}(a))$. Given Assumption~\ref{ass:equal-costs}, there exists a policy $\pi^N$ on $Y^N$ such that $v^\pi_\theta (y)=v^{\pi^N}_\theta (y^N)$. The result remains true for the average cost problem with $g^\pi$ and $g^{\pi^N}$.
\end{lemma}
\begin{proof}
The proof is very similar to the one of Lemma~\ref{lem:CMO-optimal-policy}. We just have to introduce new independent random variables $\hat{A}(n)$ which help bridge the gap between $Y$ and $Y^N$. See more details in Appendix~\ref{app:CMO}
\end{proof}

\subsection{Acyclic Graphs}
\label{sec:extreme-edges}

In Section~\ref{sec:N}, we fully characterizes the optimal matching control of 
the $N$-shaped matching graph, which is an acyclic graph. Therefore, in this section,
we study the optimal matching control of an arbitrary acyclic matching graph.
We show that, under certain assumptions on the costs of the nodes, the optimal matching
policy consists of prioritizing the matching of the extreme edges.



%
For an acyclic matching graph, we say that $\edge{i}{j}$ is an extreme edge if the unique adjacent node of $d_i$ is $s_j$
or if the unique adjacent node of $s_j$ is $d_i$. We denote by $\mathcal E^\ast$ the set of
extreme nodes. 
We say that an edge $\edge{i_1}{j_1}$ belongs to the neighborhood of an edge $\edge{i_2}{j_2}$ if
$i_1=i_2$ or $j_1=j_2$. We denote by $N(\edge{i}{j})$ the neighborhood of an edge 
$\edge{i}{j}$. 

\begin{figure}[htbp]
     \centering
    \begin{tikzpicture}[]
     \node[style={circle,draw}] at (1,0) (s1) {$s_1$};
     \node[style={circle,draw}] at (2,0) (s2) {$s_2$};
     \node[style={circle,draw}] at (3,0) (s3) {$s_3$};
     \node[style={circle,draw}] at (4,0) (s4) {$s_4$};
     \node[style={circle,draw}] at (5,0) (s5) {$s_5$};
     \node[style={circle,draw}] at (1,2) (d1) {$d_1$};
     \node[style={circle,draw}] at (2,2) (d2) {$d_2$};
     \node[style={circle,draw}] at (3,2) (d3) {$d_3$};
     \node[style={circle,draw}] at (4,2) (d4) {$d_4$};
     \node[style={circle,draw}] at (5,2) (d5) {$d_5$};
     \node[style={circle,draw}] at (6,2) (d6) {$d_6$};
     \draw[<-] (s1) -- (1,-1)  node[below] {$\beta_1$} ;
     \draw[<-] (s2) -- (2,-1)  node[below] {$\beta_2$} ;
     \draw[<-] (s3) -- (3,-1)  node[below] {$\beta_3$} ;
     \draw[<-] (d1) -- (1,3)  node[above] {$\alpha_1$} ;
     \draw[<-] (d2) -- (2,3)  node[above] {$\alpha_2$} ;
     \draw[<-] (d3) -- (3,3)  node[above] {$\alpha_3$} ;
     \draw[<-] (s4) -- (4,-1)  node[below] {$\beta_4$} ;
     \draw[<-] (s5) -- (5,-1)  node[below] {$\beta_5$} ;
     \draw[<-] (d4) -- (4,3)  node[above] {$\alpha_4$} ;
     \draw[<-] (d5) -- (5,3)  node[above] {$\alpha_5$} ;
     \draw[<-] (d6) -- (6,3)  node[above] {$\alpha_6$} ;
     \draw (s1) -- (d1);
     \draw (s2) -- (d1);
     \draw (s2) -- (d2);
     \draw (s3) -- (d2);
     \draw (s3) -- (d3);
     \draw (s3) -- (d4);
     \draw (s4) -- (d4);
     \draw (d5) -- (s4);
     \draw (s5) -- (d5);
     \draw (s5) -- (d6);
    \end{tikzpicture}  
   \caption{An example of an acyclic matching graph.}
   \label{fig:general}
  \end{figure}
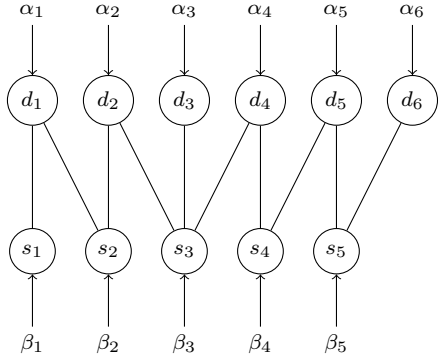

We assume that the neighborhood of all the edges is not empty, i.e., the matching graph is connected, and that in the neighborhood of
an extreme edge there are not extreme edges. An example of the matching graphs under study 
is provided in Figure~\ref{fig:general}. As it can be observed, the set of extreme edges is
$\mathcal E^\ast=\{\edge{1}{1},\edge{3}{3},\edge{6}{5}\}$ and the neighborhood of $\edge{1}{1}$
is $N(\edge{1}{1})=\{\edge{1}{2}\}$, whereas that of $\edge{3}{3}$ is
$N(\edge{3}{3})=\{\edge{2}{3},\edge{4}{3}\}$.

\subsubsection{Value Function Properties}

We start by using the increasing property in a given edge $\edge{i}{j}$ as defined in Definition~\ref{def:CG-increasing} that we will note $\mathcal{I}_{\edge{i}{j}}$.
\\
We also define the undesirability property for a given edge.

\begin{definition}[Undesirability property]
Let $\edge{i_1}{j_1}\in\mathcal E$. 
We say that a function $v$ is undesirable in $\edge{i_1}{j_1}$ or 
$v\in\mathcal U_{\edge{i_1}{j_1}}$ 
if for all $\edge{i_2}{j_2}\in N(\edge{i_1}{j_1})$
$$
v(q+e_{\edge{i_1}{j_1}}-e_{\edge{i_2}{j_2}},a)\geq v(q,a),
$$
for all $a\in\mathcal A$ and $q\in\mathcal Q$ such that $q_{s_{j_2}}\geq 1$ if $i_1=i_2$ and $q_{d_{i_2}}\geq 1$ if $j_1=j_2$.
\label{def:undesirability}
\end{definition}

\begin{remark}
We note that the undesirable property is the same as saying that, for all $\edge{i_2}{j_2}\in N(\edge{i_1}{j_1})$
$$
v(q+e_{\edge{i_1}{j_1}},q)\geq v(q+e_{\edge{i_2}{j_2}},a),
$$
that is, it is preferable to match the items in $\edge{i_1}{j_1}$ than in $\edge{i_2}{j_2}$.
\end{remark}

In the remainder of this section, we denote
\begin{equation}
V^\sigma=\bigcap_{\edge{i}{j}\in\mathcal E^\ast}\left(\mathcal I_{\edge{i}{j}}\cap\mathcal U_{\edge{i}{j}}\right)
\label{eq:v-sigma-general}
\end{equation}
and we assume that $c$ is a linear function with higher holding costs on the extreme edges rather than its neighbors, i.e., $c_{d_i}\geq c_{d_{i^\prime}}$ for all $\edge{i^\prime}{j}\in N(\edge{i}{j})$ (or $c_{s_j}\geq c_{s_{j^\prime}}$ for all $\edge{i}{j^\prime}\in N(\edge{i}{j})$). This additional assumption is needed for the cost function $c$ to satisfy the property of Definition~\ref{def:undesirability} and thus Assumption~\ref{ass:cost}. 

\subsubsection{Optimal decision rule}

In this section, we show that, for any $v\in V^\sigma$, there is a matching policy that prioritizes
the matchings in the extreme edges and minimizes $L_u^\theta v$.

\begin{definition}[Prioritizing the extreme edges]
We say that a matching policy prioritizes the extreme edges if it matches all the items in 
the extreme edges. This means that, for all $\edge{i}{j}\in\mathcal E^\ast$, $u_{\edge{i}{j}}=\min(x_{d_i},x_{s_j}). $
\end{definition}

We consider that $D^\sigma$ is the decision rule that prioritizes the extreme edges. We also 
consider that $\mathcal E^\ast=\{\edge{i_1}{j_1},\edge{i_2}{j_2},\dots,\edge{i_m}{j_m}\}$, which
means that the number of extreme edges is equal to $m$.

\begin{remark}
In the example of Figure~\ref{fig:general}, we have that $m=3$ and $i_1=j_1=1$, $i_2=j_2=3$
as well as $i_3=6$ and $j_3=5$.
\end{remark} 

We now show that, if $v\in V^\sigma$, there exists a decision rule that prioritizes the extreme edges.

\begin{proposition}
Let $a\in\mathcal A$, $q\in\mathcal Q$, $x=q+a,$ $v\in V^\sigma$, $0\leq \theta\leq 1$. 
There exists $u^\ast\in U_x$ such that 
$u^\ast\in \argmin_{u\in U_x}L_u^\theta v(q,a)$ and $u^\ast_{\edge{i}{j}}=\min(x_{d_i},x_{s_j})$ for all $\edge{i}{j}\in \mathcal E^\ast$. In particular, this result holds for the average cost operator $T_u.$
\label{prop:prioritize-extreme-general}
\end{proposition}

\begin{proof}
Let $q\in\mathcal{Q}$, $a\in\mathcal{A}$, $x=q+a$ and $u\in U_x$. We define $x_{\edge{i}{j}}=\min(x_{d_i},x_{s_j})$ for all $\edge{i}{j}\in\mathcal{E}$ to ease the notations. We 
recall that $\mathcal E^\ast=\{\edge{i_1}{j_1},\edge{i_2}{j_2},\dots,\edge{i_m}{j_m}\}$. For 
$k=1,\dots,m$, we define $$p_k=\min(x_{\edge{i_k}{j_k}}-u_{\edge{i_k}{j_k}},
\sum_{\edge{i}{j}\in N(\edge{i_k}{j_k})}u_{\edge{i}{j}}).$$

We now observe that, for all $\edge{i}{j}\in N(\edge{i_k}{j_k})$, we can define 
$0 \leq p_{i,j}\leq u_{\edge{i}{j}}$ such that $$p_k=\sum_{\edge{i}{j}\in N(\edge{i_k}{j_k})}p_{i,j}.$$  

We define 
$$
u^\prime=u+\sum_{k=1}^m\left(p_ke_{\edge{i_k}{j_k}}-\sum_{\edge{i}{j}\in N(\edge{i_k}{j_k})}p_{i,j}e_{\edge{i}{j}}\right).
$$
We assume that $u^\prime\in U_x$. Since $v\in V^\sigma,$ it follows that
\begin{equation}
L_{u^\prime}^\theta v(q,a)\leq L_{u}^\theta v(q,a).
\label{eq:general-proof-eq1}
\end{equation}

We now define $m_k=x_{\edge{i_k}{j_k}}-u^\prime_{\edge{i_k}{j_k}}$ for all $k=1,\dots,m$ and
$$
u^\ast=u^\prime+\sum_{k=1}^m m_ke_{\edge{i_k}{j_k}}.
$$

We assume that $u^\ast\in U_x$. Using that $v\in V^\sigma$
$$
L^\theta_{u^\ast} v(q,a)\leq L^\theta_{u^\prime} v(q,a),
$$
and, taking into account \eqref{eq:general-proof-eq1}, it follows that
$$
L^\theta_{u^\ast} v(q,a)\leq L^\theta_{u} v(q,a),
$$
which holds for any $u\in U_x$ and therefore also for $u\in \argmin L^\theta v(q,a)$. As a result,
$u^\ast\in\argmin L^\theta v(q,a)$. Besides,
for any extreme edge $\edge{i_k}{j_k}$, we have that
$$
u^\ast_{\edge{i_k}{j_k}}=u^\prime_{\edge{i_k}{j_k}}+x_{\edge{i_k}{j_k}}-u^\prime_{\edge{i_k}{j_k}}=x_{\edge{i_k}{j_k}},
$$
and the desired result follows if we show that $u^\prime,u^\ast\in U_x$.

\begin{itemize}
\item 
We aim to show that $u^\prime\in U_x$. To prove this, we only need to show that, 
for any $u\in U_x$ and any $k\in \{1,\dots,m\}$, $u_0\in U_x$ where
$$
u_0=u+p_k e_{\edge{i_k}{j_k}}-\sum_{\edge{i}{j}\in N(\edge{i_k}{j_k})}p_{i,j}e_{\edge{i}{j}}.
$$


We start by showing that $u_0\in\mathbb{N}^{n_{\mathcal{D}}+n_{\mathcal{S}}}$.
First, $(u_0)_{\edge{i_k}{j_k}}\geq 0$ because $u\in U_x$ and $p_k\geq 0$.
Then, for all $\edge{i}{j}\in N(\edge{i_k}{j_k})$, we have $u_{\edge{i}{j}}\geq p_{i,j}$ by definition. Thus, it follows that $(u_0)_{\edge{i}{j}}\geq 0$.

Now, we show \textit{(a)} and \textit{(b)} assuming that $d_{i_k}$ is of degree one (the proof for the case 
that $s_{j_k}$ is of degree one is symmetric and therefore we omitted it). We first show \textit{(a)} for
$d_{i_k}$ as follows:
\[
(u_0)_{\edge{i_k}{j_k}}=u_{\edge{i_k}{j_k}}+p_k\leq x_{\edge{i_k}{j_k}}\leq x_{d_{i_k}}
\]
where the first inequality holds by definition of $p_k$ and the second by definition of $x_{\edge{i_k}{j_k}}$.
We also show \textit{(a)} for all $i\in \mathcal{D}(j_k)$ as follows:
\begin{align*}
\sum_{r\in \mathcal{S}(i)}(u_0)_{\edge{i}{r}}&=(u_0)_{\edge{i}{j_k}}+\sum_{r\in \mathcal{S}(i)\setminus\{j_k\}}
(u_0)_{\edge{i}{r}}\\
&=u_{\edge{i}{j_k}}-p_{i,j_k}+\sum_{r\in \mathcal{S}(i)\setminus\{j_k\}}u_{\edge{i}{r}}\\
&\leq \sum_{r\in \mathcal{S}(i)}u_{\edge{i}{r}}\\
&\leq x_{d_i},
\end{align*}
where the first inequality holds since $p_{i,j_k}\geq 0$ and the second since $u\in U_x$.

We now show \textit{(b)} for $j_k$ as follows:
$$
\sum_{r\in \mathcal{D}(j_k)}(u_0)_{\edge{r}{j_k}}=\sum_{r\in \mathcal{D}(j_k)}u_{\edge{r}{j_k}}\leq x_{s_{j_k}},
$$
where the inequality holds since $u\in U_x$.

\item
We now aim to show that $u^\ast\in U_x$ and we observe that it is enough to show that, for 
any $k\in\{1,\dots,m\}$, 
$u_0^\prime=u^\prime+m_k e_{\edge{i_k}{j_k}}\in U_x$.

We first observe that, if $m_k=0$, then $u_0=u^\prime$ and therefore $u_0^\prime \in U_x$. 
Therefore, we now consider than $m_k>0$. For this case, 
$$
m_k>0\iff x_{\edge{i_k}{j_k}}>u^\prime_{\edge{i_k}{j_k}}=u_{\edge{i_k}{j_k}}+p_k.
$$

We observe that $p_k= x_{\edge{i_k}{j_k}}-u_{\edge{i_k}{j_k}}$ cannot be given since 
the above expression gives a contradiction. Therefore, we have that 
\\
$p_k=\sum_{\edge{i}{j}\in N(\edge{i_k}{j_k})}u_{\edge{i}{j}}$. For this case, $p_{i,j}=u_{\edge{i}{j}}$ for all $\edge{i}{j}\in N(\edge{i_k}{j_k})$ and therefore 
\begin{equation}
u^\prime_{\edge{i}{j}}=0.
\label{eq:proof-optimal-uprime}
\end{equation}

First, we have $u^\prime_0 \in \mathbb{N}^{n_{\mathcal{D}}+n_{\mathcal{S}}}$ because $m_k \geq 0$ and $u^\prime\in U_x$. 
Then, we show \textit{(a)} and \textit{(b)} assuming that $d_{i_k}$ is of degree one (the proof for the case that $s_{j_k}$ is of degree one is symmetric and therefore we omitted it). 
We first show \textit{(a)} for $i_k$ as follows:
$$
(u_0^\prime)_{\edge{i_k}{j_k}}=u^\prime_{\edge{i_k}{j_k}}+m_k=x_{\edge{i_k}{j_k}}\leq x_{d_{i_k}}.
$$

Finally, we show \textit{(b)} for $j_k$ as follows:
\begin{align*}
\sum_{r\in \mathcal{D}(j_k)}(u_0^\prime)_{\edge{r}{j_k}}&=(u_0^\prime)_{\edge{i_k}{j_k}}+\sum_{r\in \mathcal{D}(j_k)\setminus\{i_k\}}(u_0^\prime)_{\edge{r}{j_k}}\\
&=(u_0^\prime)_{\edge{i_k}{j_k}}+ \sum_{r\in \mathcal{D}(j_k)\setminus\{i_k\}}(u^\prime)_{\edge{r}{j_k}}\\
&=(u_0^\prime)_{\edge{i_k}{j_k}}\\
&=x_{\edge{i_k}{j_k}}\\
&\leq x_{s_{j_k}},
\end{align*}
where the third equality holds by \eqref{eq:proof-optimal-uprime}.
\end{itemize}
\end{proof}

 \subsubsection{Value Function Property Preservation}

We now show that the properties of the value function that belong to 
$V^\sigma$ are preserved by the dynamic programming operator. That is., we consider the value
function $v\in V^\sigma$ and we aim to show that $Lv\in V^\sigma$. 

We first show that the increasing property is preserved by the dynamic operator property. 
The proof uses the same arguments as for the preservation of the increasing 
property in $\edge{1}{1}$ and $\edge{2}{2}$ of Lemma~\ref{lem:increasing} and therefore we omit it.

\begin{lemma}
Let $\edge{i}{j}\in\mathcal E^\ast$. If $v\in\mathcal I_{\edge{i}{j}}\cap \mathcal U_{\edge{i}{j}}$, then $Lv\in\mathcal I_{\edge{i}{j}}$.
\label{lem:increasing-general}
\end{lemma}

We also show that the undesirability property is preserved by the dynamic programming operator. 

\begin{lemma}
Let $\edge{i}{j}\in\mathcal E^\ast$. If $v\in\mathcal I_{\edge{i}{j}}\cap \mathcal U_{\edge{i}{j}}$, then 
$Lv\in\mathcal U_{\edge{i}{j}}$.
\label{lem:undesirable}
\end{lemma}

\begin{proof}
Let $\edge{i_1}{j_1}\in\mathcal E^\ast$ and 
$\edge{i_2}{j_2}\in N(\edge{i_1}{j_1})$. We consider that $i_1=i_2$ (the case $j_1=j_2$ is symmetric and therefore it can be proven analogously). 
Let $\underline{a}\in\mathcal A$, $\underline{q}\in\mathcal Q$ such that $\underline{q}_{j_2}\geq 1$ and let $\underline{x}=\underline{q}+a$. 
We define $\overline {q}=\underline{q}+e_{\edge{i_1}{j_1}}-e_{\edge{i_2}{j_2}}$ and $\overline x=\overline q +a$. 
Since $v\in \mathcal U_{\edge{i}{j}}$, we know that $v(\overline{q},a)\geq v(\underline q,a)$. We aim to show
that $L^\theta v(\overline{q},a)\geq L^\theta v(\underline{q},a)$.

Let $u_{\overline x}\in\argmin_{u\in U_x}L^\theta_u v(\overline q,a)$, using Proposition~\ref{prop:prioritize-extreme-general}, we can choose $u_{\overline x}$ such that $(u_{\overline{x}})_{\edge{i_1}{j_1}}=\min(\overline{x}_{d_{i_1}},\overline{x}_{s_{j_1}})$. Assume that $\overline{x}_{d_{i_1}}\geq 1$, then $(u_{\overline{x}})_{\edge{i_1}{j_1}}\geq 1$ because $\overline{x}_{j_1}=\underline{x}_{j_1}+1\geq 1$. We define 
$u_{\underline{x}}=u_{\overline{x}}+e_{\edge{i_2}{j_2}}-e_{\edge{i_1}{j_1}}$. We know that
$u_{\underline{x}}\in U_{\underline{x}}$ since $0\leq (u_{\overline{x}})_{\edge{i_1}{j_1}}-1 \leq \overline{x}_{s_{j_1}}-1=\underline{x}_{s_{j_1}}$, $\sum_{r\in\mathcal{D}(j_2)}(u_{\underline{x}})_{\edge{r}{j_2}}=\left(\sum_{r\in\mathcal{D}(j_2)}(u_{\overline{x}})_{\edge{r}{j_2}}\right)+1\leq \overline{x}_{s_{j_2}}+1= \underline{x}_{s_{j_2}}$
and $u_{\overline{x}}\in U_{\overline{x}}$. Besides, 
$\overline{x}-u_{\overline x}=\underline{x}-u_{\underline x}$ and the desired result follows since
\begin{align*}
L^\theta v(\underline q,a)&\leq L^\theta_{u_{\underline x}}v(\underline q,a)\\
&=c(\underline x)+\theta\mathbb E[v(\underline x-u_{\underline x},A)] \\
&=c(\underline x)+\theta\mathbb E[v(\overline x-u_{\overline x},A)] \\
&=c(\underline x)-c(\overline x)+L^\theta v(\overline q,A) \\
&\leq L^\theta v(\overline q,A),
\end{align*}
where the last inequality holds since $c\in V^\sigma$. Assume now that $\overline{x}_{d_{i_1}}=0$, then we cannot do more matchings in $\overline{x}$ than we could do in $\underline{x}$, i.e $u_{\overline x}\in U_{\underline{x}}$. Indeed, we have $ (u_{\overline{x}})_{\edge{i_1}{j_1}} \leq \overline{x}_{d_{i_1}}=0\leq\underline{x}_{s_{j_1}}$ and $\sum_{r\in\mathcal{D}(j_2)}(u_{\overline{x}})_{\edge{r}{j_2}}\leq \overline{x}_{s_{j_2}}\leq \underline{x}_{s_{j_2}}$. Therefore,
\begin{align*}
L^\theta v(\underline q,a)&\leq L^\theta_{u_{\overline x}}v(\underline q,a)\\
&=c(\underline x)+\theta\mathbb E[v(\underline x-u_{\overline x},A)] \\
&\leq c(\overline x)+\theta\mathbb E[v(\overline x-u_{\overline x},A)] \\
&= L^\theta v(\overline q,A),
\end{align*}
where the last inequality holds since $c\in V^\sigma$ and $v\in V^\sigma$, giving us the final result.
\end{proof}

\subsubsection{Structure of the Optimal Policy}

In this section, we use the result of Theorem~\ref{thm:puterman_discounted} and of Theorem~\ref{thm:puterman_average} to show that
there exists an optimal policy that consists of a sequence of decision rules that belongs to
$D^\sigma$, i.e., that prioritizes the extreme edges. 

\begin{theorem}
The optimal control for the discounted cost problem prioritizes the extreme edges.
\label{thm:opt-control-general}
\end{theorem}

\begin{proof}
We apply Theorem~\ref{thm:puterman_discounted} where $V^\sigma$ is the set of functions 
defined in \eqref{eq:v-sigma-general} and 
$D^\sigma$ the set defined in Definition~\ref{prop:prioritize-extreme-general}.

From Lemma~\ref{lem:increasing-general} and Lemma~\ref{lem:undesirable}, it follows $(a)$. 
The result of Proposition~\ref{prop:prioritize-extreme-general} shows (b). 
And, since limits preserve inequalities, 
the point-wise convergence of functions of $V^\sigma$ belong to this set, which shows (c).
\end{proof}

The following theorem shows that the previous result is also verified 
for the average cost problem.

\begin{theorem}
The optimal control for the average cost problem 
prioritizes the extreme edges. 
\label{thm:opt-control-average-general}
\end{theorem}
\begin{proof}
We want to apply Theorem~\ref{thm:puterman_average} using the same value function set $V^\sigma$ and the same decision rule set $D^\sigma$ as in the proof of the previous proposition. 
%

Let $(\theta_n)_{n\in\mathbb N}$ be a sequence such that $0\leq \theta_n < 1$ for all $n\in\mathbb N$ and $\underset{n\to +\infty}{lim} \theta_n = 1$. Let $n\in\mathbb N$.  We know that $v_{\theta_n}^\ast \in V^\sigma$ (see the proof of Theorem~\ref{thm:opt-control-general}). The inequalities in the definitions of the properties used in $V^\sigma$ still hold if we add a constant to $v$, thus $v_{\theta_n}^\ast - v_{\theta_n}^\ast(0)e \in V^\sigma$. Using Assumption~\textit{(A3)} and Assumption~\textit{(A4)}, we have $H\leq v_{\theta_n}^\ast - v_{\theta_n}^\ast(0)e\leq M$, so $v_{\theta_n}^\ast - v_{\theta_n}^\ast(0)e \in V_H^\sigma$. This last result holds for each $n\in\mathbb N$ and since limits preserve inequalities $ V_H^\sigma$ is a closed set, $\underset{n\to +\infty}{lim} [v_{\theta_n}^\ast - v_{\theta_n}^\ast(0)e] \in V_H^\sigma$ which shows (a). The resulf of 
Proposition~\ref{prop:prioritize-extreme-general} shows (b) because the policy that belongs to $D^\sigma$ minimizes $L_u^1 v=T_u v$ 
if $v\in V_H^\sigma \subset V^\sigma$.
\end{proof}

\subsubsection{Neighborhood of Extreme edges}

For the above results, we have assumed that the extreme edges do not have other extreme
edges in their neighborhoods. We now explain how the results of this section also hold 
when in the neighborhood of an extreme edge there are other extreme edges. 
An example the matching models we now study consists of the matching graph of 
Figure~\ref{fig:general} with an additional demand node $d_7$ and an edge $\edge{7}{5}$.

We first note that, if the cost of the extreme edges that are neighbors is the same, these 
edges can be merged and seen as a single edge whose arrival probability is equal to the sum of 
their arrival probabilities. Otherwise, we require that the undesirability and 
increasing properties to be satisfied by the extreme edge with the highest cost and the 
above arguments can be used to show that the optimal policy prioritizes the most expensive
extreme edges.

\section{$W$-shaped graph}\label{sec:W}

\begin{figure}[htbp]
     \centering
    \begin{tikzpicture}[]

     \node[style={circle,draw}] at (1,0) (s1) {$s_1$};
     \node[style={circle,draw}] at (3,0) (s2) {$s_2$};
     \node[style={circle,draw}] at (0,2) (d1) {$d_1$};
     \node[style={circle,draw}] at (2,2) (d2) {$d_2$};
     \node[style={circle,draw}] at (4,2) (d3) {$d_3$};
     \draw[<-] (s1) -- (1,-1)  node[below] {$\beta_1$} ;
     \draw[<-] (s2) -- (3,-1)  node[below] {$\beta_2$} ;
     \draw[<-] (d1) -- (0,3)  node[above] {$\alpha_1$} ;
     \draw[<-] (d2) -- (2,3)  node[above] {$\alpha_2$} ;
     \draw[<-] (d3) -- (4,3)  node[above] {$\alpha_3$} ;
     \draw (s1) -- (d1) node[pos=0.5,left] {};
     \draw (s1) -- (d2) node[pos=0.5,right] {};
     \draw (d2) -- (s2) node[pos=0.75,right] {};
     \draw (s2) -- (d3) node[pos=0.5,right] {};
    \end{tikzpicture}  
   \caption{The $W$-shaped graph.}
   \label{fig:W-graph}
  \end{figure}
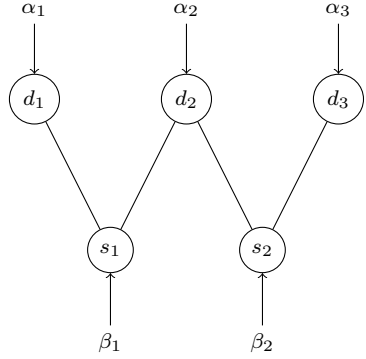

We now focus on a $W$-shaped matching graph. As it can be observed in Figure~\ref{fig:W-graph}, this matching
graph is formed by three demand nodes, two supply nodes and the following set of edge: $\mathcal{E}=\{\edge{1}{1},\edge{2}{1},\edge{2}{2},\edge{3}{2}\}$. Let us also define $\edge{1}{2}$ ($\edge{3}{1}$) as the imaginary edge ($\edge{1}{2},\edge{3}{1} \notin \mathcal E$) between $d_1$ and $s_2$ ($d_3$ and $s_1$) that we introduce to ease the notations.

We assume that $\alpha_1<\beta_1$ and  $\alpha_3<\beta_2$, which is required to ensure the stability 
of the system.

Proving the optimal policy on the $W$-shaped graph is much more difficult than on the $N$-shaped graph. We will study the former under different assumptions on the costs. First, we will present a conjecture of the optimal policy when there is a higher cost on the extreme nodes. Then, we will show through numerical evidence that it is not optimal to always match in priority the extreme edges when there is a higher cost on the middle node.

\subsection{Higher cost on the extreme nodes}

We define an extreme node as the node with degree one in an extreme edge (as defined in Section~\ref{sec:extreme-edges}). 
In the $W$-shaped graph, there are two extreme nodes: $d_1$ and $d_3$. 
We assume for the remainder of this section that there is a higher cost on those nodes rather than on $d_2$, i.e the cost function $c$ satisfy $\mathcal{U}_{\edge{1}{1}}\cap\mathcal{U}_{\edge{3}{2}}$ (as defined in Definition~\ref{def:undesirability}). 
For a linear cost function, this comes down to $c_{d_1}\geq c_{d_2}$ and $c_{d_3}\geq c_{d_2}$.
\\
Our goal is to extend the results of Section~\ref{sec:N} to this case. 
However, this $W$-shaped matching graph is much more difficult than than the $N$-shaped one.
We were not able to prove what is the optimal policy but we have a strong conjecture.
First, we will present our conjecture and explain why it should be the optimal policy.
Then, we will show which properties are needed to prove this conjecture and the work that has been done in that way.
Finally, we will explain what are the remaining challenges to solve.

\subsubsection{Conjecture of the optimal policy}

Using the insight gained with the $N$-shaped graph, we want to have some intuition about which policy should be optimal in the $W$-shaped graph.
We already know that matching in priority the extreme edges $\edge{1}{1}$ and $\edge{3}{2}$ is optimal because we are in the context of Section~\ref{sec:extreme-edges}.
Therefore, the question left is how do we want to match $\edge{2}{1}$ and $\edge{2}{2}$ ?
In the $N$-shaped graph, the non-extreme edge is matched only if it is above a threshold (based on the remaining items after matching the extreme edges) in order to have a buffer for the extreme nodes. 
Thus, translating the optimal policy of the $N$-shaped graph to the $W$-shaped graph would be to match in priority the edges $\edge{1}{1}$ and $\edge{3}{2}$, then only match $\edge{2}{1}$ if it is above a threshold (based on the remaining items in $s_1$ after the matchings in $\edge{1}{1}$) and only match $\edge{2}{2}$ if it is above a threshold (based on the remaining items in $s_2$ after the matchings in $\edge{3}{2}$).
After having matched everything we can in the edges $\edge{1}{1}$ and $\edge{3}{2}$, we obtain a state such that doing any possible number of matchings in either $\edge{2}{1}$ or $\edge{2}{2}$ does not prevent us from doing any possible number of matchings in the other edge.
So it seems reasonable to assume that the thresholds are independent.
Let us now define more formaly our conjecture of the optimal policy: 

\begin{definition}[Threshold-type decision rule]\label{def:threshold_policy_W}
A decision rule $u_x$ is said to be of threshold type in $\edge{2}{1}$ and $\edge{2}{2}$ with priority to $\edge{1}{1}$ and $\edge{3}{2}$ if:
\begin{enumerate}
\item it matches all of $\edge{1}{1}$ and $\edge{3}{2}$.
\item it matches $\edge{2}{1}$ ($\edge{2}{2}$) only if the remaining jobs in $s_1$ ($s_2$) are above a specific threshold, noted $t_{\edge{2}{1}}$ ($t_{\edge{2}{2}}$) with $t_{\edge{2}{1}}\in \mathbb{N}\cup\infty$ ($t_{\edge{2}{2}}\in \mathbb{N}\cup\infty$).
\end{enumerate}
i.e,
$ u_x =\min(x_{d_1},x_{s_1})e_{\edge{1}{1}}+\min(x_{d_3},x_{s_2})e_{\edge{3}{2}}+\min(k_{t_{\edge{2}{1}}} (x),x_{d_2})e_{\edge{2}{1}}+\min(j_{t_{\edge{2}{2}}} (x),x_{d_2})e_{\edge{2}{2}}$
where 
\\
$ k_{t_{\edge{2}{1}}}(x)=
\left\{\begin{array}{lr}
0 & if\; x_{s_1}-x_{d_1}\leq t_{\edge{2}{1}} \\
x_{s_1}-x_{d_1}-t_{\edge{2}{1}} & otherwise
\end{array}\right.
$
\\
and $ j_{t_{\edge{2}{2}}}(x)=
\left\{\begin{array}{lr}
0 & if\; x_{s_2}-x_{d_3}\leq t_{\edge{2}{2}} \\
x_{s_2}-x_{d_3}-t_{\edge{2}{2}} & otherwise
\end{array}\right.
$ 
\end{definition}

We will now present the work that has already been done in order to prove the optimality of this policy.

\subsubsection{Value Functions Properties}

We start by defining the properties that are needed to prove the optimality of our conjecture. 
\\
First, we consider the properties needed to prove that we match everything (with priority) in the extreme edges $\edge{1}{1}$ and $\edge{3}{2}$ using our results for acyclic graphs.
Thus, we assume the increasing and undesirability properties in the extreme edges $\edge{1}{1}$ and $\edge{3}{2}$ as defined in Definition~\ref{def:CG-increasing} and Definition~\ref{def:undesirability} that we will note $\mathcal{I}_{\edge{1}{1}}$, $\mathcal{I}_{\edge{3}{2}}$, $\mathcal{U}_{\edge{1}{1}}$ and $\mathcal{U}_{\edge{3}{2}}$. 
\\
Then, we use properties inspired from the $N$-shaped graph to prove that we have thresholds in $\edge{2}{1}$ and $\edge{2}{2}$ such as the convexity and boundery properties.
The convexity property is defined in $\edge{2}{1}$, $\edge{2}{2}$, $\edge{1}{2}$ and $\edge{3}{1}$ in a similar way as in Definition~\ref{def:convex}, that we will note $\mathcal{C}_{\edge{2}{1}}$, $\mathcal{C}_{\edge{2}{2}}$, $\mathcal{C}_{\edge{1}{2}}$ and $\mathcal{C}_{\edge{3}{1}}$. 
The property $\mathcal{C}_{\edge{2}{1}}$ (resp. $\mathcal{C}_{\edge{1}{2}}$) is defined for $q\in\mathcal{Q}$ such that $q_{s_1}\geq q_{d_1}$ (resp. $q_{d_1}\geq q_{s_1}$) and the property $\mathcal{C}_{\edge{2}{2}}$ (resp. $\mathcal{C}_{\edge{3}{1}}$) is defined for $q\in\mathcal{Q}$ such that $q_{s_2}\geq q_{d_3}$ (resp. $q_{d_3}\geq q_{s_2}$).
The boundary property is defined in $\edge{2}{1}$ and $\edge{2}{2}$ in a similar way as in Definition~\ref{def:boundary}, that we will note $\mathcal{B}_{\edge{2}{1}}$ and $\mathcal{B}_{\edge{2}{2}}$. 
For the property $\mathcal{B}_{\edge{2}{1}}$, $e_{\edge{1}{2}}$ is used in the left hand side and $e_{\edge{2}{1}}$ is used in the right hand side. 
For the property $\mathcal{B}_{\edge{2}{2}}$, $e_{\edge{3}{1}}$ is used in the left hand side and $e_{\edge{2}{2}}$ is used in the right hand side.
\\
Finally, we will need new properties to prove that we have a threshold in $\edge{2}{1}$ and $\edge{2}{2}$. The exchangeable property is there to be able to define the threshold based on the remaining items in $s_1$ or $s_2$ after having matched the extreme edges. The modularity property is there to prove the independence between the two thresholds.

\begin{definition}[Exchangeable property]\label{def:W_exchange}
A function $v$ is exchangeable in $\edge{2}{1}$ and $\edge{3}{1}$ or $v\in\mathcal{H}_{\edge{2}{1},\edge{3}{1}}$ if $\forall a\in\mathcal{A}$, $\forall q\in\mathcal Q$,
$$v(q+e_{\edge{2}{1}},a)-v(q,a)= v(q+e_{\edge{3}{1}},a)-v(q-e_{\edge{2}{1}}+e_{\edge{3}{1}},a).$$
Likewise, $v$ is exchangeable in $\edge{2}{2}$ and $\edge{1}{2}$ or $v\in\mathcal{H}_{\edge{2}{2},\edge{1}{2}}$ if $\forall a\in\mathcal{A}$, $\forall q\in\mathcal Q$,
$$v(q+e_{\edge{2}{2}},a)-v(q,a)= v(q+e_{\edge{1}{2}},a)-v(q-e_{\edge{2}{2}}+e_{\edge{1}{2}},a).$$
\end{definition} 

\begin{definition}[Modularity property]\label{def:W_modular}
A function $v$ is modular in $\edge{2}{1}$ and $\edge{2}{2}$ or $v\in\mathcal{M}_{\edge{2}{1},\edge{2}{2}}$ if $\forall a\in\mathcal{A}$, $\forall q\in\mathcal Q$,
$$v(q+e_{\edge{2}{1}}+e_{\edge{2}{2}},a)-v(q+e_{\edge{2}{1}},a)= v(q+e_{\edge{2}{2}},a)-v(q,a).$$
\end{definition} 
 
The properties $\mathcal I_{\edge{1}{1}}$, $\mathcal I_{\edge{3}{2}}$, $\mathcal U_{\edge{1}{1}}$, $\mathcal U_{\edge{3}{2}}$, $\mathcal{C}_{\edge{2}{1}}$, $\mathcal{C}_{\edge{2}{2}}$, $\mathcal{H}_{\edge{2}{1},\edge{3}{1}}$, 
\\
$\mathcal{H}_{\edge{2}{2},\edge{1}{2}}$ and $\mathcal{M}_{\edge{2}{1},\edge{2}{2}}$ will be used in Proposition~\ref{prop:W-optimal-control} which shows that if the value function satisfy them, then our conjecture is the optimal decision rule.
Using our insight from the $N$-shaped graph, $\mathcal B_{\edge{2}{1}}$, $\mathcal B_{\edge{2}{2}}$, $\mathcal C_{\edge{1}{2}}$ and $\mathcal C_{\edge{3}{1}}$ should be required to show that $\mathcal C_{\edge{2}{1}}$ and $\mathcal C_{\edge{2}{2}}$ are preserved by the operator $L^\theta$.

\subsubsection{Optimal decision rule}

First, the optimality of matching in priority the extreme edges $\edge{1}{1}$ and $\edge{3}{2}$ is comes immediately from our work in Section~\ref{sec:extreme-edges}.   

\begin{proposition}\label{prop:W_d1-d3-optimal-control}
Let $v\in\mathcal{I}_{\edge{1}{1}}\cap \mathcal{I}_{\edge{3}{2}}\cap \mathcal{U}_{\edge{1}{1}}\cap \mathcal{U}_{\edge{3}{2}}$, let $0\leq \theta\leq 1$. For
any $q\in\mathcal{Q}$ and any $a\in\mathcal A$, $x=q+a$, there exists $u^\ast\in U_x$ such that $u^*\in \argmin_{u\in U_x}L^\theta_u v(q,a)$ and 
$u^\ast_{\edge{1}{1}}=\min(x_{d_1},x_{s_1})$ and $u^\ast_{\edge{3}{2}}=\min(x_{d_3},x_{s_2})$.
\end{proposition}
\begin{proof}
This is a corollary of Proposition~\ref{prop:prioritize-extreme-general}. 
\end{proof}

Then, we show that there is a control of threshold-type in $\edge{2}{1}$ and in $\edge{2}{2}$ with priority to $\edge{1}{1}$ and $\edge{3}{2}$ that minimizes $L^\theta_u v$.

\begin{proposition}\label{prop:W-optimal-control}
Let $v\in \mathcal I_{\edge{1}{1}}\cap\mathcal I_{\edge{3}{2}}\cap\mathcal U_{\edge{1}{1}}\cap\mathcal U_{\edge{3}{2}}\cap\mathcal{C}_{\edge{2}{1}}\cap\mathcal{C}_{\edge{2}{2}}\cap\mathcal{H}_{\edge{2}{1},\edge{3}{1}}\cap\mathcal{H}_{\edge{2}{2},\edge{1}{2}}\cap\mathcal{M}_{\edge{2}{1},\edge{2}{2}}$. Let $0\leq \theta\leq 1$. There exists $u^\ast\in U_x$ such that $u^\ast$ is a matching policy of threshold type in $\edge{2}{1}$ and $\edge{2}{2}$ with priority to $\edge{1}{1}$ and $\edge{3}{2}$ (as defined in Definition~\ref{def:threshold_policy_W}) and $u^\ast\in\argmin_{u\in U_x} L_u^\theta v(x)$. In particular, this result holds for the average operator: $L_u^1=T_u$.
\end{proposition}
\begin{proof}
The idea of the proof is similar to the one of Proposition~\ref{prop:u-min-Lv}. However, you have to use the property $\mathcal{H}_{\edge{2}{1},\edge{3}{1}}$ to handle $\edge{3}{1}$ left over items when defining the threshold in $\edge{2}{1}$ and you have to use the property $\mathcal{H}_{\edge{2}{2},\edge{1}{2}}$ to handle $\edge{1}{2}$ left over items when defining the threshold in $\edge{2}{2}$. You also need the property $\mathcal{M}_{\edge{2}{1},\edge{2}{2}}$ to prove the independence between the two thresholds. See details in Appendix~\ref{app:W-optimal-control}.
\end{proof}

\subsubsection{Value Function Property Preservation}
\label{sec:W_propagation}

The preservation of all the properties by the dynamic programming operator is still a work in progress.
The properties related to matching in priority the extreme edges, i.e $\mathcal{I}_{\edge{1}{1}}$, $\mathcal{I}_{\edge{3}{2}}$, $\mathcal{U}_{\edge{1}{1}}$ and $\mathcal{U}_{\edge{3}{2}}$ are preserved as corollary of Lemma~\ref{lem:increasing-general} and Lemma~\ref{lem:undesirable}.
\\
However, it is very difficult to prove the preservation of the convexity properties, i.e $\mathcal{C}_{\edge{2}{1}}$ and $\mathcal{C}_{\edge{2}{2}}$. 
One challenge is that if we use a similar proof as for Lemma~\ref{lem:convex}, then we need an increasing property in the imaginary edges $\edge{1}{2}$ and $\edge{3}{1}$ (the undesirability property in $\edge{1}{1}$ and $\edge{3}{2}$ is not enough). 
The problem is that $\mathcal{I}_{\edge{1}{2}}$ and $\mathcal{I}_{\edge{3}{1}}$ cannot be preserved for all $q\in\mathcal{Q}$. 
We have to constrain both properties on subsets of $\mathcal{Q}$ and their preservation is still difficult at the boundary of these subsets. 

\subsection{Higher cost on the middle node}

Now, we want to study the $W$-shaped graph when our assumption of higher cost on the extreme nodes is not valid. 
For example, let us take the case where there is a higher cost on $d_2$ than on $d_3$.
Given that the optimal policy on the $N$-shaped graph always match in priority the extreme edges for any cost, we would expect to have the same result for the $W$-shaped graph.

\begin{figure}[h]

\centering
\hspace*{-1.5cm}
\includegraphics[scale=0.45]{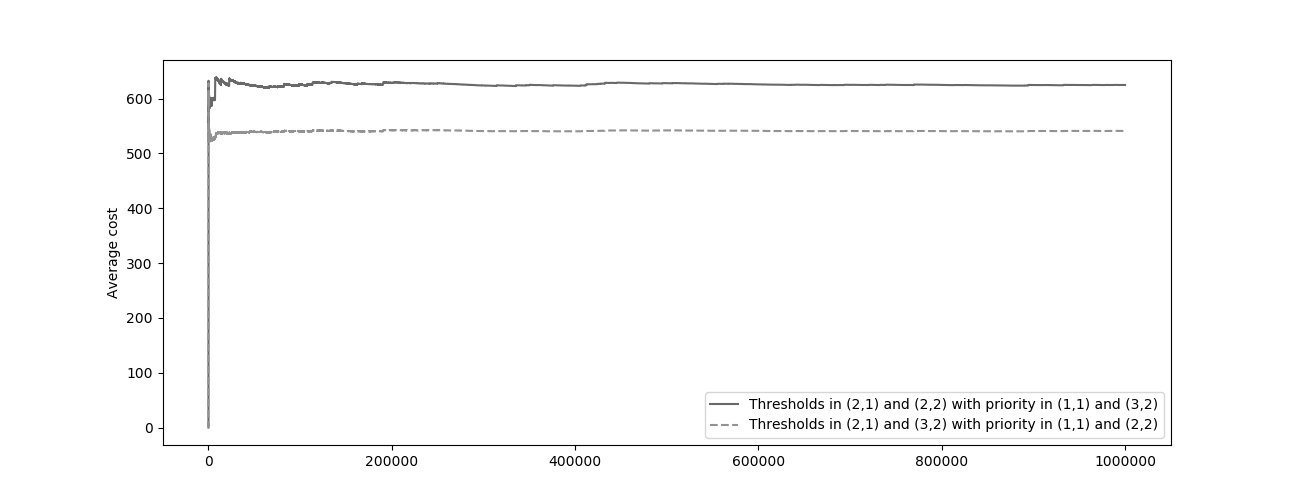}
\caption{W-shaped model. Average cost of our conjecture and a similar policy but with a priority in $\edge{2}{2}$ instead of $\edge{3}{2}$.}
\label{fig:W-extreme-edges-not-optimal}
\end{figure}

However, we have numerical evidence that always matching in priority $\edge{1}{1}$ and $\edge{3}{2}$ is not optimal. Indeed, let $c_{d_1}=10$, $c_{d_2}=10$, $c_{d_3}=1$, $c_{s_1}=1$, $c_{s_2}=1000$, $\alpha_1=0.5-\epsilon$, $\alpha_2=0.25+\epsilon$, $\alpha_3=0.25$, $\beta_1=0.5$, $\beta_2=0.5$ and $\epsilon=0.1$. For these parameters, we observed that a policy of threshold type in $\edge{2}{1}$ and $\edge{3}{2}$ with priority in $\edge{1}{1}$ and $\edge{2}{2}$ (using a similar definition as Definition~\ref{def:threshold_policy_W} but now $k_{t_{\edge{2}{1}}}(x)$ depends on the remaining items in $s_1$ and $d_2$ plus $d_3$ and $j_{t_{\edge{3}{2}}}(x)$ only depends on the remaining items in $d_3$) performed better than our conjecture, i.e a policy of threshold type in $\edge{2}{1}$ and $\edge{2}{2}$ with priority in $\edge{1}{1}$ and $\edge{3}{2}$ (see Figure~\ref{fig:W-extreme-edges-not-optimal}). For both policies, we selected the best thresholds by numerical experiments which are $t_{\edge{2}{1}}=14$ and $t_{\edge{3}{2}}=0$ for the former policy and $t_{\edge{2}{1}}=11$ and $t_{\edge{2}{2}}=0$ for the latter policy.

This numerical experiment was based on the concept of workload and results in heavy-traffic regime found in \cite{BoundedRegret}. 
We chose these parameters in order for the nodes $d_1$ and $s_2$ to be critical in term of stability. 
By that, we mean that the stability conditions for $D=\{d_1\}$ and $S=\{s_2\}$ are the closest to be violated among all subsets (see \eqref{eq:stability}). 
Heavy-traffic results suggest, in that case, that we should have some buffer in $\mathcal{D}(s_2)=\{d_2,d_3\}$. 
In addition, because we have higher cost in $d_2$ rather than $d_3$, this implies that we prefer to have the buffer in $d_3$ rather than $d_2$. 
All these conditions lead to matching in priority $\edge{2}{2}$ rather than $\edge{3}{2}$ a better performing policy.

To conclude, the optimal policy on the $W$-shaped graph should take various forms depending on the cost structure and results from heavy-traffic regime could lead us to the right ones.

\section{Conclusion}\label{sec:Conclusion}

We consider bipartite matching graphs with linear costs in the buffer size. We model this problem
as a Markov Decision problem where the goal is to find the optimal matching control, that is, how to match items so as to minimize the cost of the system. We study the derived optimal control problem for the discounted cost and the average cost problems. In both cases, we 
fully characterize the optimal policy in a wide range of matching models. For a complete
graph, the optimal policy consists of matching all items. For the $N$-shaped matching graph, we show that there exists an optimal policy that is of threshold type for the diagonal edge with priority to the end edges of the matching graph.
We also fully characterize the optimal threshold for the average cost problem. Then, we show that, under some assumptions in the costs of the nodes, this threshold-based structure of the optimal matching policy extends to more complex models such as a complete graph minus one edge. In an arbitrary acyclic graph, we show that, when the cost of the extreme nodes is larger or equal than the cost of
its neighbor nodes, the optimal policy always matches all in the extreme edges. Finally, we
investigate the $W$-shaped matching model. For this case, we conjecture that the optimal matching policy is also of threshold type with priority to the extreme edges when the cost of
the extreme nodes is large and we provide numerical evidence that, when the cost of the 
middle node is large, the optimal matching control does not prioritize the extreme edges.
Therefore, the optimal matching control seems to depend on the cost of the nodes for the $W$-shaped matching model.

The characterization of the optimal matching policy in an arbitrary bipartite matching graph
remains as an open question.
For future work, it would be interesting to prove our conjecture of the optimal matching policy 
on the $W$-shaped matching graph when the cost of the extreme nodes is large and characterize the optimal policy when 
the cost of the intermediate node is large. We believe that this result will allow us to characterize
the optimal matching policy of more complex matching graphs, such as of graphs that can be projected on $W$-shaped graphs or of an arbitrary acyclic graph. Another possible future research is the 
study of the optimal matching policy for general (non-bipartite) graphs. In this case, items arrive
one by one to the system and there are not two different sets of nodes. In spite of these
differences with respect to bipartite matching graphs, we think that the techniques 
applied in this article can be used to characterize the optimal matching policy for non-bipartite 
matching graphs. 

\bibliographystyle{spmpsci} 
\bibliography{matching}

\begin{thebibliography}{10}
\providecommand{\url}[1]{{#1}}
\providecommand{\urlprefix}{URL }
\expandafter\ifx\csname urlstyle\endcsname\relax
  \providecommand{\doi}[1]{DOI~\discretionary{}{}{}#1}\else
  \providecommand{\doi}{DOI~\discretionary{}{}{}\begingroup
  \urlstyle{rm}\Url}\fi

\bibitem{ExactFCFS}
Adan, I., Weiss, G.: Exact fcfs matching rates for two infinite multitype
  sequences.
\newblock Operations Research \textbf{60}(2), 475--489 (2012).
\newblock \doi{10.1287/opre.1110.1027}.
\newblock \urlprefix\url{https://doi.org/10.1287/opre.1110.1027}

\bibitem{ABMW18}
Adan, I.J.B.F., Busic, A., Mairesse, J., Weiss, G.: Reversibility and further
  properties of {FCFS} infinite bipartite matching.
\newblock Math. Oper. Res. \textbf{43}(2), 598--621 (2018).
\newblock \doi{10.1287/moor.2017.0874}.
\newblock \urlprefix\url{https://doi.org/10.1287/moor.2017.0874}

\bibitem{Kidney}
Ashlagi, I., Jaillet, P., Manshadi, V.H.: Kidney exchange in dynamic sparse
  heterogenous pools.
\newblock In: Proceedings of the Fourteenth ACM Conference on Electronic
  Commerce, EC '13, pp. 25--26. ACM, New York, NY, USA (2013).
\newblock \doi{10.1145/2482540.2482565}.
\newblock \urlprefix\url{http://doi.acm.org/10.1145/2482540.2482565}

\bibitem{BKQ}
Banerjee, S., Kanoria, Y., Qian, P.: {The Value of State Dependent Control in
  Ridesharing Systems}.
\newblock ArXiv e-prints  (2018)

\bibitem{Stability}
Busic, A., Gupta, V., Mairesse, J.: Stability of the bipartite matching model.
\newblock Advances in Applied Probability \textbf{45}(2), 351--378 (2013).
\newblock \urlprefix\url{http://arxiv.org/abs/1003.3477}

\bibitem{BoundedRegret}
Busic, A., Meyn, S.P.: Approximate optimality with bounded regret in dynamic
  matching models.
\newblock ArXiv e-prints  (2014).
\newblock \urlprefix\url{http://arxiv.org/abs/1411.1044}

\bibitem{CDB}
Cadas, A., Bu\v{s}i\'{c}, A., Doncel, J.: Optimal control of dynamic bipartite
  matching models.
\newblock In: Proceedings of the 12th EAI International Conference on
  Performance Evaluation Methodologies and Tools, VALUETOOLS 2019, pp. 39--46.
  ACM, New York, NY, USA (2019).
\newblock \doi{10.1145/3306309.3306317}.
\newblock \urlprefix\url{http://doi.acm.org/10.1145/3306309.3306317}

\bibitem{FCFSModel}
Caldentey, R., Kaplan, E.H., Weiss, G.: Fcfs infinite bipartite matching of
  servers and customers.
\newblock Advances in Applied Probability \textbf{41}(3), 695--730 (2009).
\newblock \urlprefix\url{http://www.jstor.org/stable/27793900}

\bibitem{FMMM}
Feldman, J., Mehta, A., Mirrokni, V.S., Muthukrishnan, S.: Online stochastic
  matching: Beating 1-1/e.
\newblock In: I.C. Society (ed.) Proceedings of the 2009 50th Annual IEEE
  Symposium on Foundations of Computer Science, pp. 117--126. ACM Press (2009)

\bibitem{GW}
Gurvich, I., Ward, A.: On the dynamic control of matching queues.
\newblock Stoch. Syst. \textbf{4}(2), 479--523 (2014).
\newblock \doi{10.1214/13-SSY097}.
\newblock \urlprefix\url{https://doi.org/10.1214/13-SSY097}

\bibitem{HJM12}
Hyon, E., Jean-Marie, A.: Scheduling services in a queuing system with
  impatience and setup costs.
\newblock The Computer Journal \textbf{55}(5), 553--563 (2012).
\newblock \doi{10.1093/comjnl/bxq096}

\bibitem{JL}
Jaillet, P., Lu, X.: Online stochastic matching: New algorithms with better
  bounds.
\newblock Mathematics of Operations Research \textbf{39}(3), 624--646 (2014).
\newblock \doi{10.1287/moor.2013.0621}

\bibitem{Mairesse_Stability}
Mairesse, J., Moyal, P.: Stability of the stochastic matching model.
\newblock Journal of Applied Probability \textbf{53}(4), 1064--1077 (2016).
\newblock \doi{10.1017/jpr.2016.65}

\bibitem{VSA}
Manshadi, V.H., Gharan, S.O., Saberi, A.: Online stochastic matching: Online
  actions based on offline statistics.
\newblock Mathematics of Operations Research \textbf{37}(4), 559--573 (2012).
\newblock \doi{10.1287/moor.1120.0551}

\bibitem{NS}
Nazari, M., Stolyar, A.L.: Reward maximization in general dynamic matching
  systems.
\newblock Queueing Systems  (2018).
\newblock \doi{10.1007/s11134-018-9593-y}.
\newblock \urlprefix\url{https://doi.org/10.1007/s11134-018-9593-y}

\bibitem{KidneySite}
for Organ~Sharing, U.N.: Online.
\newblock
  \urlprefix\url{https://unos.org/wp-content/uploads/unos/Living_Donation_KidneyPaired.pdf}

\bibitem{puterman2005markov}
Puterman, M.L.: Markov decision processes: discrete stochastic dynamic
  programming.
\newblock Wiley series in probability and statistics. Wiley-Interscience (2005)

\bibitem{SR}
Schalkoff, R.J.: Pattern Recognition: Statistical, Structural and Neural
  Approaches.
\newblock John Wiley \& Sons, Inc., New York, NY, USA (1991)

\bibitem{ZDC}
Zdeborov\'a, L., Decelle, A., Chertkov, M.: Message passing for optimization
  and control of a power grid: Model of a distribution system with redundancy.
\newblock Phys. Rev. E \textbf{80}, 046112 (2009).
\newblock \doi{10.1103/PhysRevE.80.046112}.
\newblock \urlprefix\url{https://link.aps.org/doi/10.1103/PhysRevE.80.046112}

\end{thebibliography}

\begin{appendices}

\section{Proof of Lemma~\ref{lem:B-convex}}\label{app:convex}

\subsection{Preservation of $\mathcal{B}$}
\begin{proof}
Let $a\in\mathcal{A}$. Since $v\in \mathcal B$, we have $v(0,a)-v(e_{\edge{2}{1}},a)\leq v(e_{\edge{1}{2}},a)-v(0,a)$ (this inequality also holds for the cost function $c$ because of Assumption~\ref{ass:cost}). We aim to show that $L^\theta v(0,a)-L^\theta v(e_{\edge{2}{1}},a)\leq L^\theta v(e_{\edge{1}{2}},a)-L^\theta v(0,a)$. For any $x\in\mathcal{Q}$, we choose $u_x \in \argmin_u L_u^\theta v(x)$ such that $u_{x}=\min(x_{d_1},x_{s_1})e_{\edge{1}{1}}+\min(x_{d_2},x_{s_2})e_{\edge{2}{2}}+k_t(x)e_{\edge{1}{2}}$, using Proposition~\ref{prop:u-min-Lv}.

We are going to show the preservation for each possible value of $a$:
\begin{itemize}
\item If $a=e_{\edge{1}{1}}$ or $a=e_{\edge{2}{2}}$. Suppose that $k_t(e_{\edge{1}{2}})=0$. Then,
\begin{align*}
L^\theta v(0,a)-L^\theta v(e_{\edge{2}{1}},a)&=c(a)-c(a+e_{\edge{2}{1}})+\theta\mathbb E[v(0,A)-v(e_{\edge{2}{1}},A)] \\
&\leq c(a+e_{\edge{1}{2}})-c(a)+\theta\mathbb E[v(e_{\edge{1}{2}},A)-v(0,A)] \\
&=L^\theta v(e_{\edge{1}{2}},a)-L^\theta v(0,a)
\end{align*}
because $c\in\mathcal{B}$ and $v\in\mathcal{B}$. Suppose now that $k_t(e_{\edge{1}{2}})>0$. Then,
\begin{align*}
L^\theta v(0,a)-L^\theta v(e_{\edge{2}{1}},a)&=c(a)-c(a+e_{\edge{2}{1}})+\theta\mathbb E[v(0,A)-v(e_{\edge{2}{1}},A)] \\
&\leq c(a)-c(a+e_{\edge{2}{1}}) \\
&\leq c(a+e_{\edge{1}{2}})-c(a)+\theta\mathbb E[v(0,A)-v(0,A)] \\
&=L^\theta v(e_{\edge{1}{2}},a)-L^\theta v(0,a)
\end{align*}
because $v\in\mathcal{I}_{\edge{2}{1}}$ and $c\in\mathcal{B}$.
\item If $a=e_{\edge{1}{2}}$. Suppose that $k_t(2e_{\edge{1}{2}})=0$. Then,
\begin{align*}
L^\theta v(0,a)-L^\theta v(e_{\edge{2}{1}},a)&=c(a)-c(a+e_{\edge{2}{1}})+\theta\mathbb E[v(e_{\edge{1}{2}},A)-v(0,A)] \\
&\leq c(a+e_{\edge{1}{2}})-c(a)+\theta\mathbb E[v(2e_{\edge{1}{2}},A)\\
&\quad -v(e_{\edge{1}{2}},A)] \\
&=L^\theta v(e_{\edge{1}{2}},a)-L^\theta v(0,a)
\end{align*}
because $c\in\mathcal{B}$ and $v\in\mathcal{C}_{\edge{1}{2}}$. Suppose now that $k_t(2e_{\edge{1}{2}})>0$ and $k_t(e_{\edge{1}{2}})=0$. Then,
\begin{align*}
L^\theta v(0,a)-L^\theta v(e_{\edge{2}{1}},a)&=c(a)-c(a+e_{\edge{2}{1}})+\theta\mathbb E[v(e_{\edge{1}{2}},A)-v(0,A)] \\
&=c(a)-c(a+e_{\edge{2}{1}})+L^\theta v(0,a)-L_{u_{a}+e_{\edge{1}{2}}}^\theta v(0,a) \\
&\leq c(a)-c(a+e_{\edge{2}{1}}) \\
&\leq c(a+e_{\edge{1}{2}})-c(a)+\theta\mathbb E[v(e_{\edge{1}{2}},A)-v(e_{\edge{1}{2}},A)] \\
&=L^\theta v(e_{\edge{1}{2}},a)-L^\theta v(0,a)
\end{align*}
because $c\in\mathcal{B}$. Finally, suppose that $k_t(2e_{\edge{1}{2}})>0$ and $k_t(e_{\edge{1}{2}})>0$. Then,
\begin{align*}
L^\theta v(0,a)-L^\theta v(e_{\edge{2}{1}},a)&=c(a)-c(a+e_{\edge{2}{1}})+\theta\mathbb E[v(0,A)-v(0,A)] \\
&\leq c(a+e_{\edge{1}{2}})-c(a)+\theta\mathbb E[v(0,A)-v(0,A)] \\
&=L^\theta v(e_{\edge{1}{2}},a)-L^\theta v(0,a)
\end{align*}
because $c\in\mathcal{B}$.
\item If $x=e_{\edge{2}{1}}$. Then,
\begin{align*}
L^\theta v(0,a)-L^\theta v(e_{\edge{2}{1}},a)&=c(a)-c(a+e_{\edge{2}{1}})+\theta\mathbb E[v(e_{\edge{2}{1}},A)\\
&\quad -v(2e_{\edge{2}{1}},A)] \\
&\leq c(a+e_{\edge{1}{2}})-c(a)+\theta\mathbb E[v(0,A)-v(e_{\edge{2}{1}},A)] \\
&=L^\theta v(e_{\edge{1}{2}},a)-L^\theta v(0,a)
\end{align*}
because $c\in\mathcal{B}$ and $v\in\mathcal{C}_{\edge{2}{1}}$. 
\end{itemize}
\end{proof}

\subsection{Preservation of $\mathcal{C}_{\edge{2}{1}}$}
\begin{proof}
Let $a\in\mathcal{A}$ and $\underline{q}\in\mathcal Q$ such that $\underline{q}_{s_1}\geq \underline{q}_{d_1}$, $\underline{x}=\underline{q}+a$. We define $\overline{q}=\underline{q}+e_{\edge{2}{1}}$, $\overline{x}=\overline{q}+a$, $\overline{\overline{q}}=\overline{q}+e_{\edge{2}{1}}$ and $\overline{\overline{x}}=\overline{\overline{q}}+a$. Since $v$ is convex in $\edge{2}{1}$, we have $v(\overline{q},a)-v(\underline{q},a)\leq v(\overline{\overline{q}},a)-v(\overline{q},a)$ (this inequality also holds for the cost function $c$ because of Assumption~\ref{ass:cost}). We aim to show that $L^\theta v(\overline{q},a)-L^\theta v(\underline{q},a)\leq L^\theta v(\overline{\overline{q}},a)-L^\theta v(\overline{q},a)$. For $y\in\left\{\underline{x},\overline{x},\overline{\overline{x}}\right\}$, let $u_{y}\in \argmin_u L_u^\theta v(y)$. From Proposition~\ref{prop:u-min-Lv}, we can choose $u_y$ such that $ u_{y}=\min(y_{d_1},y_{s_1})e_{\edge{1}{1}}+\min(y_{d_2},y_{s_2})e_{\edge{2}{2}}+k_t(y)e_{\edge{1}{2}}$.
Let us also define $m=\underline{x}-u_{\underline{x}}$. We can distinguish 2 cases: (a) $\underline{q}_{s_1}\geq \underline{q}_{d_1}+1$ or $a\in\mathcal{A}\setminus\{e_{\edge{1}{2}}\}$ and (b) $\underline{q}_{s_1}= \underline{q}_{d_1}$ and $a=e_{\edge{1}{2}}$:
\begin{enumerate}
\item[(a)] If $\underline{q}_{s_1}\geq \underline{q}_{d_1}+1$ or $a\in\mathcal{A}\setminus\{e_{\edge{1}{2}}\}$. Then,
\begin{align*}
L^\theta v(\overline{q},a)-L^\theta v(\underline{q},a)&=c(\overline{x})-c(\underline{x})+\theta\mathbb E[v(m+e_{\edge{2}{1}},A)-v(m,A)] \\
&\leq c(\overline{\overline{x}})-c(\overline{x})+\theta\mathbb E[v(m+2e_{\edge{2}{1}},A)-v(m+e_{\edge{2}{1}},A)]\\
&=L^\theta v(\overline{\overline{q}},a)-L^\theta v(\overline{q},a)
\end{align*}
because $c\in\mathcal{C}_{\edge{2}{1}}$ and $v\in\mathcal{C}_{\edge{2}{1}}$.
\item[(b)] If $\underline{q}_{s_1}= \underline{q}_{d_1}$ and $a=e_{\edge{1}{2}}$. Suppose that $k_t(\underline{x})=0$. Then, 
\begin{align*}
L^\theta v(\overline{q},a)-L^\theta v(\underline{q},a)&=c(\overline{x})-c(\underline{x})+\theta\mathbb E[v(m-e_{\edge{1}{2}},A)-v(m,A)] \\
&\leq c(\overline{\overline{x}})-c(\overline{x})+\theta\mathbb E[v(m-e_{\edge{1}{2}}+e_{\edge{2}{1}},A)\\
&\quad -v(m-e_{\edge{1}{2}},A)]\\
&=L^\theta v(\overline{\overline{q}},a)-L^\theta v(\overline{q},a)
\end{align*}
because $c\in\mathcal{C}_{\edge{2}{1}}$ and $v\in\mathcal{B}$. Suppose now that $k_t(\underline{x})>0$. Then, 
\begin{align*}
L^\theta v(\overline{q},a)-L^\theta v(\underline{q},a)&=c(\overline{x})-c(\underline{x})+\theta\mathbb E[v(m+e_{\edge{2}{1}},A)-v(m,A)] \\
&\leq c(\overline{\overline{x}})-c(\overline{x})+\theta\mathbb E[v(m+2e_{\edge{2}{1}},A)-v(m+e_{\edge{2}{1}},A)]\\
&=L^\theta v(\overline{\overline{q}},a)-L^\theta v(\overline{q},a)
\end{align*}
because $c\in\mathcal{C}_{\edge{2}{1}}$ and $v\in\mathcal{C}_{\edge{2}{1}}$.
\end{enumerate}
\end{proof}

\section{Proof of Proposition~\ref{prop:opt-threshold}}\label{app:opt-threshold}
\begin{proof}

Let $u_t^\infty \in \Pi^{T_{\edge{1}{2}}}$. Let us look at the Markov chain derived from this policy. The set of possible states (except for $Y_0$) is $\mathcal{S}^{\mathcal{A}}=\{ (s_i,a) : i\in\mathbb{N},a\in\mathcal{A}\}$ with 
\[ s_i = \left\{ 
\begin{array}{lr}
(t-i,0,0,t-i) & if\; i\leq t \\
(0,i-t,i-t,0) &
\end{array}
\right. \]
\smallbreak
The $s_i$ are all possible states after the matchings (using $u_t^\infty$). In order to see more clearly the behavior of the Markov chain, we group some states together. Let us define $\mathcal{S}=\{ S_i : i\in\mathbb{N}\}$ with 
\\
$S_0 = \{(s_0 ,e_{\edge{1}{2}}),(s_0,e_{\edge{1}{1}}),(s_0,e_{\edge{2}{2}}),(s_1,e_{\edge{1}{2}})\}$ and 
\\
$S_i = \{(s_{i-1},e_{\edge{2}{1}}),(s_i,e_{\edge{1}{1}}),(s_i,e_{\edge{2}{2}}),(s_{i+1},e_{\edge{1}{2}})\}$ for all $i\in\mathbb{N}^\ast$. Figure~\ref{fig:MarkovChain} shows that this Markov chain defined on $\mathcal S$ is clearly irreducible.
\smallbreak
\begin{figure*}[t!]
     \centering
    \begin{tikzpicture}[->,>=stealth',shorten >=1pt,auto,node distance=2cm,
                    semithick]
	\node[randomVariable,minimum size=1.1cm]        (S0)                      {$S_{0}$};
	\node[param,minimum size=1.1cm]         (Dot1) [right=1cm of S0]  {$\cdots$};
	\node[randomVariable,minimum size=1.1cm]         (St) [right=1cm of Dot1]  {$S_{t}$};
	\node[randomVariable,minimum size=1.1cm]         (St+1) [right=1cm of St]  {$S_{t+1}$};
	\node[param,minimum size=1.1cm]        (Dot2) [right=1cm of St+1]  {$\cdots$};
  \path 
	      (S0) edge  [loop left]   node[text width=1.2cm]  {$p_{e_{\edge{1}{1}}}$\\$ +p_{e_{\edge{2}{2}}}$\\$ +p_{e_{\edge{1}{2}}}$} (S0)
	      (S0) edge  [bend left]   node  {$p_{e_{\edge{2}{1}}}$} (Dot1)
	      
	      (Dot1) edge  [bend left]   node  {$p_{e_{\edge{1}{2}}}$} (S0)
	      (Dot1) edge  [bend left]   node  {$p_{e_{\edge{2}{1}}}$} (St)
	      (St) edge  [loop below]   node  {$p_{e_{\edge{1}{1}}} +p_{e_{\edge{2}{2}}}$} (St)
	      (St) edge  [bend left]   node  {$p_{e_{\edge{1}{2}}}$} (Dot1)
	      (St) edge  [bend left]   node  {$p_{e_{\edge{2}{1}}}$} (St+1)
	      (St+1) edge  [loop below]   node  {$p_{e_{\edge{1}{1}}} +p_{e_{\edge{2}{2}}}$} (St+1)
	      (St+1) edge  [bend left]   node  {$p_{e_{\edge{1}{2}}}$} (St)
	      (St+1) edge  [bend left]   node  {$p_{e_{\edge{2}{1}}}$} (Dot2)
	      (Dot2) edge  [bend left]   node  {$p_{e_{\edge{1}{2}}}$} (St+1);
\end{tikzpicture} 
   \caption{The graph associated to the Markov chain derived from $u_t^\infty$ and defined on the state space $\mathcal S$. $p_{e_{\edge{1}{1}}}=\alpha\beta$, $p_{e_{\edge{2}{2}}}=(1-\alpha)(1-\beta)$, $p_{e_{\edge{1}{2}}}=\alpha(1-\beta)$ and $p_{e_{\edge{2}{1}}}=\beta(1-\alpha)$.}
   \label{fig:MarkovChain}
  \end{figure*}
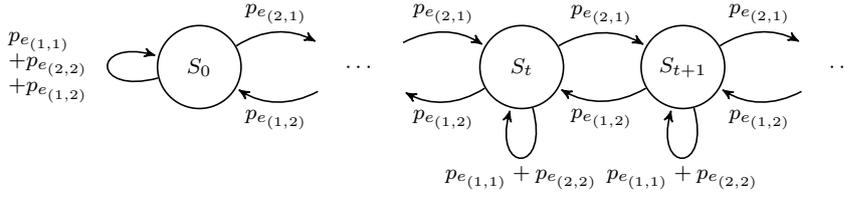
\smallbreak
The detailed balance equations are the following:
\[ \beta(1-\alpha)\pi_{S_i}=\alpha(1-\beta)\pi_{S_{i+1}}\quad\quad i=0,1,\dots\]
Solving these equations under the constraint that $\sum_{i=0}^\infty\pi_{S_i}=1$ give:
\begin{equation}\label{eq:balance}
\pi_{S_i}=\rho^{i}(1-\rho)\quad\quad i=0,1,\dots
\end{equation}
with $\rho=\frac{\beta(1-\alpha)}{\alpha(1-\beta)}\in(0,1)$. So \eqref{eq:balance} is the stationary distribution of our Markov chain (on $\mathcal S$) and the latter is positive recurrent. Using these results, we can derive caracteristics of the Markov chain on $\mathcal{S}^{\mathcal{A}}$. First of all, it is irreducible too because of the irreducibility of the chain on $\mathcal S$ and the fact that within a group $S_i$, there exists a path from each state to each other states either by looping in the group $S_i$ or by going to a neighboor ($S_{i-1}$ or $S_{i+1}$) and coming back to the initial group $S_i$. Then, since the arrival process is independant of the state and because we must have $\pi_{S_i}=\pi_{(s_{i-1},e_{\edge{2}{1}})}+\pi_{(s_{i},e_{\edge{1}{1}})}+\pi_{(s_{i},e_{\edge{2}{2}})}+\pi_{(s_{i+1},e_{\edge{1}{2}})}$, we can think of the following as the stationary distribution:
\begin{equation}\label{eq:stationary-dist}
\pi_{(s_{i},a)} = \rho^{i}(1-\rho)p_a 
\end{equation}
with $p_a$ as defined in Figure~\ref{fig:MarkovChain} (i.e $p_{e_{\edge{1}{1}}}=\alpha\beta$, $p_{e_{\edge{2}{2}}}=(1-\alpha)(1-\beta)$, $p_{e_{\edge{1}{2}}}=\alpha(1-\beta)$ and $p_{e_{\edge{2}{1}}}=\beta(1-\alpha)$), for all $i\in\mathbb N$ and $a\in\mathcal{A}$. Let us verify that \eqref{eq:stationary-dist} is indeed a stationary distribution:
\begin{align*}
\sum_{k\in\mathbb N}\sum_{a\in\mathcal{A}}\pi_{(s_{k},a)}p((s_{k},a),(s_{i},a^\prime))&=p_{a^\prime}(\pi_{(s_{i-1},e_{\edge{2}{1}})}+\pi_{(s_{i},e_{\edge{1}{1}})}+\pi_{(s_{i},e_{\edge{2}{2}})}\\
&\quad +\pi_{(s_{i+1},e_{\edge{1}{2}})})\\
&=p_{a^\prime}(\rho^{i-1}(1-\rho)p_{e_{\edge{2}{1}}}+\rho^{i}(1-\rho)p_{e_{\edge{1}{1}}}\\
&\quad +\rho^{i}(1-\rho)p_{e_{\edge{2}{2}}}+\rho^{i+1}(1-\rho)p_{e_{\edge{1}{2}}}) \\
&=p_{a^\prime} \rho^{i}(1-\rho)(\frac{\beta(1-\alpha)}{\rho}+\alpha\beta\\
&\quad +(1-\alpha)(1-\beta)+\rho\alpha(1-\beta)) \\
&=p_{a^\prime} \rho^{i}(1-\rho)\\
& =\pi_{(s_{i},a^\prime)}
\end{align*}
\[ \sum_{i\in\mathbb N}\sum_{a\in\mathcal{A}}\pi_{(s_{i},a)}=\sum_{i\in\mathbb N}\rho^{i}(1-\rho) \sum_{a\in\mathcal{A}} p_a = \sum_{i\in\mathbb N}\rho^{i}(1-\rho)=1 \]
Therefore, \eqref{eq:stationary-dist} is the stationary distribution of the Markov chain derived from the policy $u_t^\infty$ and the latter is positive recurrent. Using the monotone convergence theorem and the strong law of large number for Markov chains, we can compute the average cost $g^{u_t^\infty}$:
\[ g^{u_t^\infty}(y)=\lim_{N\to\infty}\dfrac1N\sum_{n=0}^{N-1}\mathbb E_y^{u_t^\infty}[c(Y(n))]=\mathbb E_{\pi}[c(Y)]\]
where $\mathbb E_{\pi}$ means the expectation over the stationary distribution $\pi$ defined as \eqref{eq:stationary-dist}. Finally, we can compute this value:

\begin{align}
\mathbb E_{\pi}[c(Y)]&=\sum_{i=1}^t \sum_{a\in\mathcal{A}}c(s_{t-i}+a)\pi_{(s_{t-i},a)}+\sum_{i\in\mathbb N}\sum_{a\in\mathcal{A}}c(s_{t+i}+a)\pi_{(s_{t+i},a)} \nonumber\\
&=\sum_{i=1}^t((c_{d_1}+c_{s_2})i+c_{d_1}+c_{s_1})\rho^{t-i}(1-\rho)\alpha\beta\nonumber\\
& \quad\quad\quad+((c_{d_1}+c_{s_2})i+c_{d_2}+c_{s_1})\rho^{t-i}(1-\rho)(1-\alpha)\beta\nonumber\\
& \quad\quad\quad+((c_{d_1}+c_{s_2})i+c_{d_1}+c_{s_2})\rho^{t-i}(1-\rho)\alpha(1-\beta)\nonumber\\
& \quad\quad\quad+((c_{d_1}+c_{s_2})i+c_{d_2}+c_{s_2})\rho^{t-i}(1-\rho)(1-\alpha)(1-\beta)\nonumber\\
& \quad+ \sum_{i\in\mathbb{N}}((c_{d_2}+c_{s_1})i+c_{d_1}+c_{s_1})\rho^{t+i}(1-\rho)\alpha\beta\nonumber\\
& \quad\quad\quad+((c_{d_2}+c_{s_1})i+c_{d_2}+c_{s_1})\rho^{t+i}(1-\rho)(1-\alpha)\beta\nonumber\\
& \quad\quad\quad+((c_{d_2}+c_{s_1})i+c_{d_1}+c_{s_2})\rho^{t+i}(1-\rho)\alpha(1-\beta)\nonumber\\
& \quad\quad\quad+((c_{d_2}+c_{s_1})i+c_{d_2}+c_{s_2})\rho^{t+i}(1-\rho)(1-\alpha)(1-\beta)\nonumber\\\
&=\sum_{i=1}^t(c_{d_1}+c_{s_2})i\rho^{t-i}(1-\rho)\nonumber\\
&\quad\quad\quad+(c_{d_1}+c_{s_1})\rho^{t-i}(1-\rho)\alpha\beta\nonumber\\
&\quad\quad\quad+(c_{d_2}+c_{s_1})\rho^{t-i}(1-\rho)(1-\alpha)\beta\nonumber\\
& \quad\quad\quad+(c_{d_1}+c_{s_2})\rho^{t-i}(1-\rho)\alpha(1-\beta)\nonumber\\
&\quad\quad\quad+(c_{d_2}+c_{s_2})\rho^{t-i}(1-\rho)(1-\alpha)(1-\beta)\nonumber\\
& \quad+ \sum_{i\in\mathbb{N}}(c_{d_2}+c_{s_1})i\rho^{t+i}(1-\rho)\nonumber\\
& \quad\quad\quad+ (c_{d_2}+c_{s_1})\rho^{t+i}(1-\rho)\alpha\beta\nonumber\\
&\quad\quad\quad+(c_{d_2}+c_{s_1})\rho^{t+i}(1-\rho)(1-\alpha)\beta\nonumber\\
& \quad\quad\quad+(c_{d_2}+c_{s_1})\rho^{t+i}(1-\rho)\alpha(1-\beta)\nonumber\\
&\quad\quad\quad+(c_{d_2}+c_{s_2})\rho^{t+i}(1-\rho)(1-\alpha)\label{eq:unequalcosts-exp}
\end{align}

It is easy to see that, for any $c$, the following properties hold:
$\sum_{i=1}^tc\cdot i\cdot \rho^{t-i}= c\left(t-\rho\frac{1-\rho^{t}}{1-\rho}\right)$ and 
$\sum_{i\in\mathbb N}c\cdot i\cdot \rho^{t-i}= c\frac{\rho^{t+1}}{1-\rho}.
$
Also, for any $c$ and $q$, we have that
$q \cdot c \cdot (1-\rho)\sum_{i=1}^t\rho^{t-i}=q\cdot c\cdot (1-\rho^t)$ and 
$q \cdot c \cdot (1-\rho)\sum_{i\in\mathbb N}\rho^{t+i}=q\cdot c\cdot \rho^t.$
Using these properties in \eqref{eq:unequalcosts-exp}, we obtain that
\begin{align}
\mathbb E_{\pi}[c(Y)]&=(c_{d_1}+c_{s_2})\left(t-\rho\frac{1-\rho^{t}}{1-\rho}\right)\nonumber\\
&\quad+(1-\rho^t)
((c_{d_1}+c_{s_1})\alpha\beta+(c_{d_1}+c_{s_2})\alpha(1-\beta)\nonumber\\
&\quad +(c_{d_2}+c_{s_1})\alpha(1-\beta)+(c_{d_2}+c_{s_2})(1-\alpha)(1-\beta))\nonumber\\
&\quad +(c_{d_2}+c_{s_1})\frac{\rho^{t+1}}{1-\rho}+\rho^t
((c_{d_1}+c_{s_1})\alpha\beta\nonumber\\
&\quad+(c_{d_1}+c_{s_2})\alpha(1-\beta) +(c_{d_2}+c_{s_1})\alpha(1-\beta)+(c_{d_2}\nonumber\\
&\quad+c_{s_2})(1-\alpha)(1-\beta))\nonumber\\
& = (c_{d_1}+c_{s_2})t+(c_{d_1}+c_{d_2}+c_{s_1}+c_{s_2})\frac{\rho^{t+1}}{1-\rho}
\nonumber\\
&\quad-(c_{d_1}+c_{d_2})\frac{\rho}{1-\rho} +((c_{d_1}+c_{s_1})\alpha\beta\nonumber\\
&\quad+(c_{d_1}+c_{s_2})\alpha(1-\beta) +(c_{d_2}+c_{s_1})\alpha(1-\beta)\nonumber\\
&\quad+(c_{d_2}+c_{s_2})(1-\alpha)(1-\beta)) \label{eq:average-cost}.
\end{align}
We aim to obtain the value of $t$ that minimize \eqref{eq:average-cost}. Thus, we show that \eqref{eq:average-cost} 
is convex in $t$ since its second derivative with respect to a $t$ is positive:
$$
(c_{d_1}+c_{d_2}+c_{s_1}+c_{s_2})\frac{\rho^{t+1}}{1-\rho}(\log{\rho})^2,
$$
which positive for $\rho\in (0,1)$.  Hence, the minimum of \eqref{eq:average-cost} is given when its derivative 
with respect to $t$ is equal to zero:
\begin{align*}
c_{d_1}+c_{s_2}+(c_{d_1}+c_{d_2}+c_{s_1}+c_{s_2})\frac{\rho^{t+1}}{1-\rho}(\log{\rho})=0 \iff \\
1+(1+R)\frac{\rho^{t+1}}{1-\rho}(\log{\rho})=0,
\end{align*}
where $R=\frac{c_{s_1}+c_{d_2}}{c_{d_1}+c_{s_2}}$. The root of the previous equation is
$$
t=\frac{1}{\log{\rho}}\log{\left(\frac{\rho-1}{(R+1)\log{\rho}}\right)}-1.
$$
%
%
%
%

Since this value is not necessarily integer, the optimal threshold $t^\ast$ is obtained in the minimum of \eqref{eq:average-cost} between the ceil and the floor.
\end{proof}

We now aim to show that $t^*$ is always positive. First, we show that the value of $k$, as defined in the previous
result, is increasing with $\rho$.

\begin{lemma}
The function $w(\rho,R)=\frac{1}{\log\rho}\log{\frac{\rho-1}{(R+1)\log\rho}}$ is increasing with $\rho$, 
where $\rho\in(0,1)$.
\end{lemma}

\begin{proof}
Let $w(\rho,R)=\frac{1}{\log\rho}\log{\frac{\rho-1}{(R+1)\log\rho}}$. We compute the derivative of $w$ with
respect to $\rho$ and we obtain
\begin{align*}
\frac{\partial w}{\partial \rho}&=\frac{-1/\rho}{(\log{\rho})^2}\log{\left(\frac{\rho-1}{(R+1)\log{\rho}}\right)}\nonumber\\
&\quad+
\frac{1}{\log{\rho}}\frac{(R+1)\log{\rho}}{\rho-1}\frac{(R+1)\log{\rho}-\frac{(R+1)(\rho-1)}{\rho}}{(R+1)^2(\log{\rho})^2}\\
&=\frac{-1/\rho}{(\log{\rho})^2}\left(\log{\left(\frac{\rho-1}{(R+1)\log{\rho}}\right)}\right.\\&\quad-\left.\frac{1}{\rho-1}
\left(\frac{\rho(R+1)\log{\rho}-(R+1)(\rho-1)}{R+1}\right)\right)\\
&=\frac{-1/\rho}{(\log{\rho})^2}\left(\log{\left(\frac{\rho-1}{(R+1)\log{\rho}}\right)}\right.\\&-\left.\frac{1}{\rho-1}
\left(\frac{\rho(R+1)\log{\rho}-(R+1)}{R+1}\right)\right)\\
&=\frac{-1/\rho}{(\log{\rho})^2}\left(\log{\left(\frac{\rho-1}{(R+1)\log{\rho}}\right)}-\frac{\rho\log{\rho}-(\rho-1)}{\rho-1}\right)\\
&=\frac{-1/\rho}{(\log{\rho})^2}\left(\log{\left(\frac{\rho-1}{(R+1)\log{\rho}}\right)}-\frac{\rho\log{\rho}}{\rho-1}+1\right)
\end{align*}
The last expression is positive if and only if
\begin{equation}
\log{\left(\frac{\rho-1}{(R+1)\log{\rho}}\right)}-\frac{\rho\log{\rho}}{\rho-1}+1<0.
\label{eq:inequality-proof-1}
\end{equation}
Let $B_1(\rho,R)=\log{\left(\frac{\rho-1}{(R+1)\log{\rho}}\right)}-\frac{\rho\log{\rho}}{\rho-1}+1$. We now study the 
value of $B_1(\rho,R)$ when $\rho\to 1$. First, we use L'Hopital's rule to prove the following properties:
$$
\lim_{\rho\to 1}\frac{\rho-1}{(R+1)\log{\rho}}=\lim_{\rho\to 1}\frac{1}{(R+1)/\rho}=\log{\frac{1}{(R+1)}},
$$
$$
\lim_{\rho\to 1}\frac{\log{\rho}}{(\rho-1)/\rho}=\lim_{\rho\to 1}\frac{1/\rho}{1/\rho^2}=1.
$$
We now observe that when $\rho\to 1$, it follows that $B_1(\rho,R)=\log{1/(R+1)}-1+1$, 
which is negative since $R\geq 0$ (note that the costs are assumed to be all positive). 
As a result, $\eqref{eq:inequality-proof-1}$ holds if $B_1(\rho,R)$ is increasing with $\rho$,
with $\rho\in (0,1)$. We compute the derivative of $B_1(\rho,R)$ with respect to $\rho$.
\begin{align*}
\frac{\partial B_1}{\partial \rho}&=\frac{(R+1)\log{\rho}}{\rho-1}\cdot
\frac{(R+1)\log{\rho}-\frac{(\rho-1)(R+1)}{\rho}}{(R+1)^2(\log{\rho}^2)}
\\&\quad-\frac{(\log{\rho}+1)(\rho-1)-\rho\log{\rho}}{(\rho-1)^2}\\
& =\frac{(R+1)\log{\rho}-\frac{(\rho-1)(R+1)}{\rho}}{(\rho-1)(R+1)(\log{\rho})}\\
&\quad-\frac{(\rho\log{\rho}+\rho-\log{\rho}-1-\rho\log{\rho})}{(\rho-1)^2}\\
& =\frac{(R+1)\log{\rho}-\frac{(\rho-1)(R+1)}{\rho}}{(\rho-1)(R+1)(\log{\rho})}
-\frac{(\rho-\log{\rho}-1)}{(\rho-1)^2}\\
&=\frac{(\rho-1)\left(-\frac{(\rho-1)(R+1)}{\rho}\right)-(R+1)(\log\rho)(-\log{\rho})}{(\rho-1)^2(R+1)(\log{\rho})}\\
&=\frac{(\rho-1)\left(-\frac{(\rho-1)(R+1)}{\rho}\right)-(R+1)(\log\rho)(-\log{\rho})}{(\rho-1)^2(R+1)(\log{\rho})}\\
&=\frac{(\rho-1)\left(-\frac{(\rho-1)}{\rho}\right)-(\log\rho)(-\log{\rho})}{(\rho-1)^2(\log{\rho})}\\
&=\frac{-(\rho-1)^2/\rho+(\log\rho)^2}{(\rho-1)^2(\log{\rho})}
\end{align*}

And the last expression is positive for $\rho\in(0,1)$ if and only if $-(\rho-1)^2/\rho+(\log\rho)^2<0$.
Furthermore, since $(\rho-1)/\rho$ and $\log{\rho}$ are negative when $\rho\in(0,1)$, we have that 
$$
\frac{-(\rho-1)^2}{\rho}+(\log\rho)^2<0\iff \frac{-(\rho-1)}{\sqrt\rho}+(\log\rho)>0.
$$

Let $B_2(\rho,R)=\frac{-(\rho-1)}{\sqrt\rho}+(\log\rho)$. From the previous reasoning, the desired result follows
if $B_2(\rho,R)>0$. We now observe that $B_2(\rho,R)$ tends to zero when $\rho\to 1$. Therefore, to show 
that $B_2(\rho,R)$ is positive it is enough to prove that it is decreasing with $\rho$, where $\rho\in(0,1).$
We now calculate the derivative of $B_2(\rho,R)$ with respect to $\rho$.
\begin{align*}
\frac{\partial B_2(\rho,R)}{\partial \rho}&=\frac{1}{\rho}-\frac{\frac{(1-\rho)}{2\sqrt\rho}+\sqrt\rho}{\rho}=
\frac{1}{\rho}-\frac{(1-\rho)+2\rho}{2\sqrt{\rho}\rho}=\frac{2\sqrt{\rho}-1-\rho}{2\sqrt{\rho}\rho}\\&=
-\frac{(1-\sqrt{\rho})^2}{2\sqrt{\rho}\rho},
\end{align*}
which is always negative and the proof ends.
\end{proof}

From the previous result, it follows that the smallest value of $k$ is given when $\rho\to 0$. In the following
result, we give the value of $f(\rho,R)$ when tends $\rho\to 0$.

\begin{lemma}
We have that
$$
\lim_{\rho\to 0}w(\rho,R)=0.
$$
\end{lemma}

\begin{proof}
We aim to compute the value of $\frac{\log{\frac{\rho-1}{(R+1)\log\rho}}}{\log\rho}$ when $\rho\to 0$. Since
the numerator and the denominator of this expression tend to $\log{0}$ when $\rho\to 0$, we use L'Hopital's
rule as follows
\begin{align*}
\lim_{\rho\to 0}\frac{\log{\frac{\rho-1}{(R+1)\log\rho}}}{\log\rho}&=
\lim_{\rho\to 0}\frac{\frac{(\log{\rho})-\frac{1}{\rho}(\rho-1)}{(R+1)(\log{\rho})^2}}{1/\rho}\\&=
\lim_{\rho\to 0}\frac{\rho\left(\log{\rho}-\frac{1}{\rho}(\rho-1)\right)}{(R+1)(\log{\rho})^2}\\&=
\lim_{\rho\to 0}\frac{\rho}{(R+1)\log{\rho}}-\frac{\rho-1}{(R+1)\log\rho},
\end{align*}
and both terms tend to zero when $\rho\to 0$.
\end{proof}

From this result, we have that the smallest value of $k$ is $-1$. Therefore, 
if $f(-1)<f(0)$, where $f$ is defined as in Proposition~\ref{prop:opt-threshold}, $t^*$ gets negative values.
In the following result, we show that this condition is never satisfied.

\begin{lemma}
Let $f(\cdot)$ as defined in Proposition~\ref{prop:opt-threshold}. For all $\rho\in(0,1)$, $f(-1)>f(0)$.
\end{lemma}

\begin{proof}
To show this result, we first compute the value of $f(0)$ and of $f(1)$.
\begin{align*}
f(0)&=(c_{d_1}+c_{d_2}+c_{s_1}+c_{s_2})\frac{\rho}{1-\rho}
-(c_{d_1}+c_{s_2})\frac{\rho}{1-\rho}\\&\quad +((c_{d_1}+c_{s_1})\alpha\beta+(c_{d_2}+c_{s_2})(1-\alpha)(1-\beta)\\&\quad+
(c_{d_2}+c_{s_1})(1-\alpha)\beta+(c_{d_1}+c_{s_2})\alpha(1-\beta))\\
f(-1)&=-(c_{d_1}+c_{s_2})+\frac{c_{d_1}+c_{d_2}+c_{s_1}+c_{s_2}}{1-\rho}
-(c_{d_1}+c_{s_2})\frac{\rho}{1-\rho}\\&\quad+((c_{d_1}+c_{s_1})\alpha\beta+(c_{d_2}+c_{s_2})(1-\alpha)(1-\beta)\\&\quad+
(c_{d_2}+c_{s_1})(1-\alpha)\beta+(c_{d_1}+c_{s_2})\alpha(1-\beta))
\end{align*}
From this expressions, it follows that $f(0)<f(-1)$ if and only if
\begin{multline*}
(c_{d_1}+c_{d_2}+c_{s_1}+c_{s_2})\frac{\rho}{1-\rho}
-(c_{d_1}+c_{s_2})\frac{\rho}{1-\rho}<\\-(c_{d_1}+c_{s_2})+(c_{d_1}+c_{d_2}+c_{s_1}+c_{s_2})\frac{1}{1-\rho}
-\frac{(c_{d_1}+c_{s_2})\rho}{1-\rho}
\end{multline*}
This is equivalent to show that
$$
(c_{d_2}+c_{s_1})\frac{\rho}{1-\rho}<-(c_{d_1}+c_{s_2})+\frac{c_{d_1}+c_{d_2}+c_{s_1}+c_{s_2}}{1-\rho}
-\frac{(c_{d_1}+c_{s_2})\rho}{1-\rho}
$$
$$
\iff(c_{d_1}+c_{d_2}+c_{s_1}+c_{s_2})\frac{\rho}{1-\rho}<-(c_{d_1}+c_{s_2})+\frac{c_{d_1}+c_{d_2}+c_{s_1}+c_{s_2}}{1-\rho}
$$
$$
\iff(c_{d_1}+c_{d_2}+c_{s_1}+c_{s_2})\frac{1-\rho}{1-\rho}>(c_{d_1}+c_{s_2})\iff
$$
$$
 c_{d_1}+c_{d_2}+c_{s_1}+c_{s_2}>c_{d_1}+c_{s_2},
$$
where the last expression always holds for positive values of the costs and the proof ends.
\end{proof}

From the last three lemmas, we conclude that the value of the threshold $t^*$ of the optimal matching 
policy is positive.

\section{Proof of Lemma~\ref{lem:CMO-policy-construction}}\label{app:CMO}

\begin{proof}
Let $0\leq \theta <1$, let $\pi$ be a stationary policy on $Y$, $y_0=(q_0,a_0)\in\mathcal{Q}\times\mathcal{A}$ and $y^N_0=(p^N_{\mathcal{Q}}(q_0),p^N_{\mathcal{A}}(a_0))$.
We start by constructing a history dependent policy $\pi^N=(u^N_n)_{n\geq 0}$ on $Y^N$ that will make $Y^N$ "follow" (in some sense) the projection of $Y$ on $\mathcal{G}^N$.
First, let us introduce new independent random variables $\hat{A}(n)$ that we sample just after the arrivals on $\mathcal{G}^N$. $\hat{A}(n)$ is defined on $\mathcal{A}$ with the following distribution: $\forall n\in\mathbb{N}^\ast$, $\forall a\in\mathcal{A}$, $\forall a^N\in\mathcal{A}^N$,
\[ \mathbb{P}(\hat{A}(n)=a|A^N(n)=a^N)=\left\{\begin{array}{lr}
\frac{\mathbb{P}(A(n)=a)}{\mathbb{P}(A^N(n)=a^N)}\quad & if\; a\in (p_{\mathcal{A}}^N)^{-1}(a^N)\\
0 & Otherwise
\end{array}\right. .\]
Then, we define $u_n^N$ the decision rule of $\pi^N$ at time $n$ based on the history of the trajectory, i.e $(A^N(1),\hat{A}(1),\cdots,A^N(n),\hat{A}(n))=(a^N_1,\hat{a}^N_1,\cdots,a^N_{n},\hat{a}^N_{n})$, the initial state $y_0=(q_0,a_0)$ and the stationary policy $\pi$.
Let $\hat{x}_n \in\mathcal{Q}$ be the state we end up by starting in $x_0=q_0 +a_0$ and following the dynamics \eqref{eq:x-evolution} with the sequence of arrivals $\hat{a}^N_1,\cdots,\hat{a}^N_n$ and under the policy $\pi$.
Let $u\in U_{\hat{x}_n}$ be the decision rule applied for the state $\hat{x}_n$ under the policy $\pi$. We construct $u^N_n \in  U_{p^N_{\mathcal{Q}}(\hat{x}_n)}$ such that $(u^N_n)_{\edge{1}{1}}=\sum_{i\in\mathcal{D}(j^\ast)}u_{\edge{i}{j^\ast}}$, $(u^N_n)_{\edge{2}{2}}=\sum_{j\in\mathcal{S}(i^\ast)}u_{\edge{i^\ast}{j}}$ and $(u^N_n)_{\edge{1}{2}}=\sum_{i\in\mathcal{D}(j^\ast)}\sum_{j\in\mathcal{S}(i^\ast)}u_{\edge{i}{j}}$.
\medbreak
Now, let us show by induction that, under $\pi^N$, we have $p^N_{\mathcal{Q}}(x_n)=x^N_n$ for any $a^N_1\in\mathcal{A}^N,\cdots,a^N_n\in\mathcal{A}^N$ and any $\hat{a}_1\in (p^N_{\mathcal{A}})^{-1}(a^N_1),\cdots, \hat{a}_n\in (p^N_{\mathcal{A}})^{-1}(a^N_n)$ such that $\hat{a}_1=a_1,\cdots,\hat{a}_n=a_n$.
\\
We already specifically chose $y^N_0$ to verify this property: $p^N_{\mathcal{Q}}(x_0)=p^N_{\mathcal{Q}}(q_0)+p^N_{\mathcal{A}}(a_0)=q^N_0 +a^N_0=x^N_0$. Now, assume that $p^N_{\mathcal{Q}}(x_{n-1})=x^N_{n-1}$.
First, let us note that $x_{n-1}=\hat{x}_{n-1}$ because $\hat{a}_1=a_1,\cdots,\hat{a}_{n-1}=a_{n-1}$ and they both follow the same dynamics under the same policy $\pi$. 
Then,
\begin{align*}
p^N_{\mathcal{Q}}(q_n) &= \left(\sum_{i\in\mathcal{D}(j^\ast)}(q_n)_{d_i},(q_n)_{d_{i^\ast}},(q_n)_{s_{j^\ast}},\sum_{j\in\mathcal{S}(i^\ast)}(q_n)_{s_j}\right)\\
&= \left(\sum_{i\in\mathcal{D}(j^\ast)}(x_{n-1}-u(x_{n-1}))_{d_i},(x_{n-1}-u(x_{n-1}))_{d_{i^\ast}}\right.\\
&\quad\left.,(x_{n-1}-u(x_{n-1}))_{s_{j^\ast}},\sum_{j\in\mathcal{S}(i^\ast)}(x_{n-1}-u(x_{n-1}))_{s_j}\right)\\
&= \left(\sum_{i\in\mathcal{D}(j^\ast)}(x_{n-1})_{d_i}-\sum_{j\in\mathcal{S}(i)}u(x_{n-1})_{\edge{i}{j}},\right.\\
&\quad\left.(x_{n-1})_{d_{i^\ast}}-\sum_{j\in\mathcal{S}(i^\ast)}u(x_{n-1})_{\edge{i^\ast}{j}},(x_{n-1})_{s_{j^\ast}}-\sum_{i\in\mathcal{D}(j^\ast)}u(x_{n-1})_{\edge{i}{j^\ast}},\right.\\
&\quad\left.\sum_{j\in\mathcal{S}(i^\ast)}(x_{n-1})_{s_j}-\sum_{i\in\mathcal{D}(j)}u(x_{n-1})_{\edge{i}{j}}\right)\\
&= p^N_{\mathcal{Q}}(x_{n-1})-\sum_{i\in\mathcal{D}(j^\ast)}u(x_{n-1})_{\edge{i}{j^\ast}}e_{\edge{1}{1}} -\sum_{j\in\mathcal{S}(i^\ast)}u(x_{n-1})_{\edge{i^\ast}{j}}e_{\edge{2}{2}}\\
&\quad -\sum_{i\in\mathcal{D}(j^\ast)}\sum_{j\in\mathcal{S}(i^\ast)}u(x_{n-1})_{\edge{i}{j}}e_{\edge{1}{2}}\\
&= p^N_{\mathcal{Q}}(x_{n-1})-\sum_{i\in\mathcal{D}(j^\ast)}u(\hat{x}_{n-1})_{\edge{i}{j^\ast}}e_{\edge{1}{1}} -\sum_{j\in\mathcal{S}(i^\ast)}u(\hat{x}_{n-1})_{\edge{i^\ast}{j}}e_{\edge{2}{2}}\\
&\quad -\sum_{i\in\mathcal{D}(j^\ast)}\sum_{j\in\mathcal{S}(i^\ast)}u(\hat{x}_{n-1})_{\edge{i}{j}}e_{\edge{1}{2}}\\
&= p^N_{\mathcal{Q}}(x_{n-1})-(u^N_n)_{\edge{1}{1}}e_{\edge{1}{1}}-(u^N_n)_{\edge{2}{2}}e_{\edge{2}{2}} -(u^N_n)_{\edge{1}{2}}e_{\edge{1}{2}}\\
&= x^N_{n-1} -u^N_n \\
&=q^N_n
\end{align*}
and $p^N_{\mathcal{A}}(a_n)=a^N_n$ (because $p^N_{\mathcal{A}}((p^N_{\mathcal{A}})^{-1}(a^N))=a^N$ for all $a^N \in\mathcal{A}^N$). Thus, $p^N_{\mathcal{Q}}(x_{n})=p^N_{\mathcal{Q}}(q_{n})+p^N_{\mathcal{A}}(a_{n})=q^N_n+a^N_n=x^N_n$.
\\
Then, using this property and Assumption~\ref{ass:equal-costs}, we have
\begin{align*}
\mathbb{E}^\pi_{y_0} [c(Y(n))]&=\sum_{a_1\in\mathcal{A},\cdots,a_n \in\mathcal{A}}c(x_{n})\prod_{k=1}^n \mathbb{P}(A(k)=a_k) \\
&=\sum_{a^N_1\in\mathcal{A}^N,\cdots,a^N_n \in\mathcal{A}^N}\sum_{a_1\in (p^N_{\mathcal{A}})^{-1}(a^N_1)}\cdots \sum_{a_n\in (p^N_{\mathcal{A}})^{-1}(a^N_n)}c(x_{n})\\
&\quad \times\prod_{k=1}^n \mathbb{P}(A(k)=a_k) \\
&=\sum_{a^N_1\in\mathcal{A}^N,\cdots,a^N_n \in\mathcal{A}^N}\sum_{a_1\in (p^N_{\mathcal{A}})^{-1}(a^N_1)}\cdots \sum_{a_n\in (p^N_{\mathcal{A}})^{-1}(a^N_n)}c^N(p^N_{\mathcal{Q}}(x_{n}))\\
&\quad \times\prod_{k=1}^n \mathbb{P}(A(k)=a_k) \\
&=\sum_{a^N_1\in\mathcal{A}^N,\cdots,a^N_n \in\mathcal{A}^N}\sum_{\hat{a}_1\in (p^N_{\mathcal{A}})^{-1}(a^N_1)}\cdots \sum_{\hat{a}_n\in (p^N_{\mathcal{A}})^{-1}(a^N_n)}c^N(x^N_{n})\\
&\quad \times\prod_{k=1}^n \mathbb{P}(A(k)=\hat{a}_k ) \\
&=\sum_{a^N_1\in\mathcal{A}^N,\cdots,a^N_n \in\mathcal{A}^N}\sum_{\hat{a}_1\in (p^N_{\mathcal{A}})^{-1}(a^N_1)}\cdots \sum_{\hat{a}_n\in (p^N_{\mathcal{A}})^{-1}(a^N_n)}c^N(x^N_{n})\\
&\quad \times\prod_{k=1}^n \mathbb{P}(\hat{A}(k)=\hat{a}_k |A^N(k)=a^N_k)\mathbb{P}(A^N(k)=a^N_k) \\
&=\sum_{a^N_1\in\mathcal{A}^N,\cdots,a^N_n \in\mathcal{A}^N}\sum_{\hat{a}_1\in (p^N_{\mathcal{A}})^{-1}(a^N_1)}\cdots \sum_{\hat{a}_n\in (p^N_{\mathcal{A}})^{-1}(a^N_n)}c^N(x^N_{n})\\
&\quad \times\prod_{k=1}^n \mathbb{P}(A^N(k)=a^N_k,\hat{A}(k)=\hat{a}_k) \\
&=\mathbb{E}^{\pi^N}_{y^N_0} [c^N(Y^N(n))].
\end{align*}
This equality is true for any $n\in\mathbb{N}$. Therefore, $v^\pi_\theta (y_0)=v^{\pi^N}_\theta (y^N_0)$ and $g^\pi(y_0)=g^{\pi^N}(y^N_0)$.
\end{proof}

\section{Proof of Proposition~\ref{prop:W-optimal-control}}\label{app:W-optimal-control}

\begin{proof}
Let $a\in\mathcal{A}$, $q\in\mathcal Q$, $x=q+a$ and $u\in U_x$. First of all, we need to introduce some notations. Let $m_{\edge{1}{1}}=\min(x_{s_1},x_{d_1})$ (resp. $m_{\edge{3}{2}}=\min(x_{s_2},x_{d_3})$) be the maximal number of matchings that can be done in $\edge{1}{1}$ (resp. $\edge{3}{2}$). Let 
\[ K_x =\left\{\begin{array}{lr}
\{0\} & if \; x_{s_1}\leq x_{d_1} \\
\{0,\cdots,\min(x_{s_1}-x_{d_1},x_{d_2})\} & else
\end{array}\right. \]
be the set of possible matching in $\edge{2}{1}$ after having matched every possible $\edge{1}{1}$ and $\edge{3}{2}$. Let 
\[ J_x =\left\{\begin{array}{lr}
\{0\} & if \; x_{s_2}\leq x_{d_3} \\
\{0,\cdots,\min(x_{s_2}-x_{d_3},x_{d_2})\} & else
\end{array}\right.\] 
be the set of possible matching in $\edge{2}{2}$ after having matched every possible $\edge{1}{1}$ and $\edge{3}{2}$.

We can note that $K_x$ does not depend on the number of matchings in $\edge{2}{2}$. 
Indeed, if $x_{s_1}\geq x_{d_1}$ and $x_{s_2}\geq x_{d_3}$, then $x_{d_2}=x_{s_1}-x_{d_1}+x_{s_2}-x_{d_3}$ because $x\in\mathcal Q$, thus $min(x_{s_1}-x_{d_1},x_{d_2})=x_{s_1}-x_{d_1}$ which can not be modified by any matching in $\edge{2}{2}$. 
If $x_{s_1}\geq x_{d_1}$ and $x_{s_2}\leq x_{d_3}$, then no matchings can be made in $\edge{2}{2}$ ($J_x=\{0\}$). 
If $x_{s_1}\leq x_{d_1}$, then no matchings can be made in $\edge{2}{1}$ ($K_x=\{0\}$) and this can not be changed with matchings in $\edge{2}{2}$. 
A symmetric argument can be made to show that $J_x$ does not depend on the number of matchings in $\edge{2}{1}$.

We assumed that $v\in \mathcal I_{\edge{1}{1}}\cap\mathcal I_{\edge{3}{2}}\cap\mathcal U_{\edge{1}{1}}\cap\mathcal U_{\edge{3}{2}}$, so we can use Proposition~\ref{prop:W_d1-d3-optimal-control} : $ \exists u^\prime\in U_x$ such that $L_{u^\prime}^\theta v(q,a)\leq L_u^\theta v(q,a)$ and $u^\prime = m_{\edge{1}{1}}e_{\edge{1}{1}}+m_{\edge{3}{2}}e_{\edge{3}{2}}+ke_{\edge{2}{1}}+je_{\edge{2}{2}}$ with $k\in K_x,j \in J_x$. We now have to prove that there exists $t_{\edge{2}{1}}\in\mathbb{N}\cap\infty$ and $t_{\edge{2}{2}}\in\mathbb{N}\cap\infty$ such that 
\begin{equation}\label{eq:optimal_pol_W}L_{u^\ast}^\theta v(x)\leq L_{u^\prime}^\theta v(x),\quad \forall k\in K_x,\forall j\in J_x
\end{equation}
where $u^\ast$ is defined as in Definition~\ref{def:threshold_policy_W}. Let us first define the threshold in $\edge{2}{1}$ as $t_{\edge{2}{1}} = \min\{k\in\mathbb{N}: \mathbb{E}[v((k+1)e_{\edge{2}{1}},A)-v(k e_{\edge{2}{1}},A)]\geq 0\}$ and the threshold in $\edge{2}{2}$ as $t_{\edge{2}{2}} = \min\{j\in\mathbb{N}: \mathbb{E}[v((j+1)e_{\edge{2}{2}},A)-v(j e_{\edge{2}{2}},A)]\geq 0\}$ (with the convention that $\min\{\emptyset\}=\infty$). Then, given $x$, there are four cases:
\begin{enumerate}
\item $x_{s_1}= x_{d_1}$ and $x_{s_2}= x_{d_3}$. In that case $K_x=\{0\}$ and $J_x=\{0\}$, $k_{t_{\edge{2}{1}}}=0$ and $j_{t_{\edge{2}{2}}}=0$. Thus we have $u^\ast = u^\prime$.
\item $x_{s_1}> x_{d_1}$ and $x_{s_2}> x_{d_3}$. In that case, we have $x_{d_2}=x_{s_1}- x_{d_1} + x_{s_2}- x_{d_3}$. So the number of matchings in $\edge{2}{1}$ ($\edge{2}{2}$) for $u^\ast$ is exactly $k_{t_{\edge{2}{1}}}$ ($j_{t_{\edge{2}{2}}}$). We define $u^2=u^\prime - (k-k_{t_{\edge{2}{1}}}(x))e_{\edge{2}{1}}$. Suppose that $k<k_{t_{\edge{2}{1}}}(x)$ (this is only possible if $k_{t_{\edge{2}{1}}}(x)=x_{s_1}-x_{d_1}-t_{\edge{2}{1}}>0$), then by definition of $t_{\edge{2}{1}}$ and convexity in $\edge{2}{1}$ ($v\in\mathcal{C}_{\edge{2}{1}}$) we have: $\forall p\in\{0,\cdots,x_{s_1}-x_{d_1}-k-t_{\edge{2}{1}}-1\}$,
\begin{align*}
 & &\mathbb{E}[v((t_{\edge{2}{1}}+p+1)e_{\edge{2}{1}},A)-v((t_{\edge{2}{1}}+p) e_{\edge{2}{1}},A)]&\geq 0 \\
 &\iff &L_{u^2 -(p+1) e_{\edge{2}{1}}- (j-x_{s_2}+x_{d_3})e_{\edge{2}{2}}}^\theta v(q,a)&\\
 & &\quad - L_{u^2 -p e_{\edge{2}{1}}- (j-x_{s_2}+x_{d_3})e_{\edge{2}{2}}}^\theta v(q,a)&\geq 0  \\
 &\iff & L_{u^2 -(p+1) e_{\edge{2}{1}}}^\theta v(q,a)- L_{u^2 -p e_{\edge{2}{1}}}^\theta v(q,a)&\geq 0
 \end{align*}
because $v\in\mathcal{M}_{\edge{2}{1},\edge{2}{2}}$. This means that
\[ L_{u^\prime}^\theta v(q,a)\geq L_{u^\prime +e_{\edge{2}{1}}}^\theta v(q,a) \geq \cdots \geq L_{u^2}^\theta v(q,a)  \]
Suppose now that $k>k_{t_{\edge{2}{1}}}(x)$ (this is only possible if $t_{\edge{2}{1}}>0$), then by definition of $t_{\edge{2}{1}}$ and convexity in $\edge{2}{1}$ ($v\in\mathcal{C}_{\edge{2}{1}}$) we have: $\forall p\in\{0,\cdots,\min\{x_{s_1}-x_{d_1},t_{\edge{2}{1}}\} -x_{s_1}+x_{d_1}+k-1\}$,
\begin{align*}
 & &\mathbb{E}[v((\min\{x_{s_1}-x_{d_1},t_{\edge{2}{1}}\}-p)e_{\edge{2}{1}},A)&\\
 & &\quad -v((\min\{x_{s_1}-x_{d_1},t_{\edge{2}{1}}\}-p-1) e_{\edge{2}{1}},A)]&\leq 0 \\
 &\iff &L_{u^2 +p e_{\edge{2}{1}}- (j-x_{s_2}+x_{d_3})e_{\edge{2}{2}}}^\theta v(q,a)&\\
 & &\quad - L_{u^2 +(p+1) e_{\edge{2}{1}}- (j-x_{s_2}+x_{d_3})e_{\edge{2}{2}}}^\theta v(q,a)&\leq 0  \\
 &\iff & L_{u^2 +p e_{\edge{2}{1}}}^\theta v(q,a)- L_{u^2 +(p+1) e_{\edge{2}{1}}}^\theta v(q,a)&\leq 0
 \end{align*}
because $v\in\mathcal{M}_{\edge{2}{1},\edge{2}{2}}$ which means that
\[L_{u^2}^\theta v(q,a)\leq \cdots \leq L_{u^\prime -e_{\edge{2}{1}}}^\theta v(q,a) \leq L_{u^\prime}^\theta v(q,a) \]
Thus, we showed that $L_{u^2}^\theta v(x)\leq  L_{u^\prime}^\theta v(x)$ for any $k\in K_x$ and any $j\in J_x$. Now, let us compare $u^2$ and $u^\ast$. We can do a similar proof as we just did but with $\edge{2}{2}$ instead of $\edge{2}{1}$. Suppose that $j<j_{t_{\edge{2}{2}}}(x)$ (this is only possible if $j_{t_{\edge{2}{2}}}(x)=x_{s_2}-x_{d_3}-t_{\edge{2}{2}}>0$), then by definition of $t_{\edge{2}{2}}$ and convexity in $\edge{2}{2}$ ($v\in\mathcal{C}_{\edge{2}{2}}$) we have: $\forall p\in\{0,\cdots,x_{s_2}-x_{d_3}-j-t_{\edge{2}{2}}-1\}$,
\begin{align*}
 & &\mathbb{E}[v((t_{\edge{2}{2}}+p+1)e_{\edge{2}{2}},A)-v((t_{\edge{2}{2}}+p) e_{\edge{2}{2}},A)]&\geq 0 \\
 &\iff &L_{u^\ast -(p+1) e_{\edge{2}{2}}- k_{t_{\edge{2}{1}}}e_{\edge{2}{1}}}^\theta v(q,a)- L_{u^\ast -p e_{\edge{2}{2}}- k_{t_{\edge{2}{1}}}e_{\edge{2}{1}}}^\theta v(q,a)&\geq 0  \\
 &\iff & L_{u^\ast -(p+1) e_{\edge{2}{2}}}^\theta v(q,a)- L_{u^\ast -p e_{\edge{2}{2}}}^\theta v(q,a)&\geq 0
 \end{align*}
because $v\in\mathcal{M}_{\edge{2}{1},\edge{2}{2}}$. This means that
\[ L_{u^2}^\theta v(q,a)\geq L_{u^2 +e_{\edge{2}{2}}}^\theta v(q,a) \geq \cdots \geq L_{u^\ast}^\theta v(q,a)  \]
Suppose now that $j>j_{t_{\edge{2}{2}}}(x)$ (this is only possible if $t_{\edge{2}{2}}>0$), then by definition of $t_{\edge{2}{2}}$ and convexity in $\edge{2}{2}$ ($v\in\mathcal{C}_{\edge{2}{2}}$) we have: $\forall p\in\{0,\cdots,\min\{x_{s_2}-x_{d_3},t_{\edge{2}{2}}\} -x_{s_2}+x_{d_3}+j-1\}$,
\begin{align*}
 & &\mathbb{E}[v((\min\{x_{s_2}-x_{d_3},t_{\edge{2}{2}}\}-p)e_{\edge{2}{2}},A)&\\
 & &\quad -v((\min\{x_{s_2}-x_{d_3},t_{\edge{2}{2}}\}-p-1) e_{\edge{2}{2}},A)]&\leq 0 \\
 &\iff &L_{u^\ast +p e_{\edge{2}{2}}- k_{t_{\edge{2}{1}}}e_{\edge{2}{1}}}^\theta v(q,a)- L_{u^\ast +(p+1) e_{\edge{2}{2}}- k_{t_{\edge{2}{1}}}e_{\edge{2}{1}}}^\theta v(q,a)&\leq 0  \\
 &\iff & L_{u^\ast +p e_{\edge{2}{2}}}^\theta v(q,a)- L_{u^\ast +(p+1) e_{\edge{2}{2}}}^\theta v(q,a)&\leq 0
 \end{align*}
because $v\in\mathcal{M}_{\edge{2}{1},\edge{2}{2}}$ which means that
\[L_{u^\ast}^\theta v(q,a)\leq \cdots \leq L_{u^2 -e_{\edge{2}{2}}}^\theta v(q,a) \leq L_{u^2}^\theta v(q,a) \]
Thus, we showed that $L_{u^\ast}^\theta v(q,a)\leq L_{u^2}^\theta v(q,a)\leq  L_{u^\prime}^\theta v(q,a)$ for any $k\in K_x$ and any $j\in J_x$.
\item $x_{s_1}\geq x_{d_1}$ and $x_{s_2}\leq x_{d_3}$. In that case, we have $J_x=\{0\}$ and $j_{t_{\edge{2}{2}}}=0$. If $x_{s_1}= x_{d_1}$, then $K_x=\{0\}$ and $k_{t_{\edge{2}{1}}}=0$. Thus, we have $u^\ast = u^\prime$. Otherwise, suppose that $k<\min\{k_{t_{\edge{2}{1}}}(x),x_{d_2}\}$ (this is only possible if $k_{t_{\edge{2}{1}}}(x)=x_{s_1}-x_{d_1}-t_{\edge{2}{1}}>0$ and $x_{d_2}>0$), then by definition of $t_{\edge{2}{1}}$ and because $v\in\mathcal{C}_{\edge{2}{1}}\cap \mathcal{H}_{\edge{2}{1},\edge{3}{1}}$, we have: $\forall p\in\{0,\cdots,x_{d_2}-k-\max\{t_{\edge{2}{1}}-x_{d_3}+x_{s_2},0\}-1\}$,
\begin{multline*}
 \mathbb{E}[v((\max\{t_{\edge{2}{1}},x_{d_3}-x_{s_2}\}+p+1)e_{\edge{2}{1}},A)\\
 -v((\max\{t_{\edge{2}{1}},x_{d_3}-x_{s_2}\}+p) e_{\edge{2}{1}},A)\geq 0 \\
 \iff \mathbb{E}[v((\max\{t_{\edge{2}{1}}-x_{d_3}+x_{s_2},0\}+p+1)e_{\edge{2}{1}}+(x_{d_3}-x_{s_2})e_{\edge{3}{1}},A) \\
  -v((\max\{t_{\edge{2}{1}}-x_{d_3}+x_{s_2},0\}+p) e_{\edge{2}{1}}+(x_{d_3}-x_{s_2})e_{\edge{3}{1}},A)]\geq 0 \\
 \iff L_{u^\ast -(p+1) e_{\edge{2}{1}}}^\theta v(q,a)- L_{u^\ast -p e_{\edge{2}{1}}}^\theta v(q,a)\geq 0
 \end{multline*}
which means that
\[ L_{u^\prime}^\theta v(q,a)\geq L_{u^\prime +e_{\edge{2}{1}}}^\theta v(q,a) \geq \cdots \geq L_{u^\ast}^\theta v(q,a)  \]
Suppose now that $k>\min\{k_{t_{\edge{2}{1}}}(x),x_{d_2}\}$ (this is only possible if
\\
 $k_{t_{\edge{2}{1}}}(x)<x_{d_2}$), then by definition of $t_{\edge{2}{1}}$ and because $v\in\mathcal{C}_{\edge{2}{1}}\cap \mathcal{H}_{\edge{2}{1},\edge{3}{1}}$, we have: $\forall p\in\{0,\cdots,\min\{x_{d_2},t_{\edge{2}{1}} -x_{d_3}+x_{s_2}\} -x_{d_2}+k-1\}$,
\begin{multline*}
 \mathbb{E}[v((\min\{x_{s_1}-x_{d_1},t_{\edge{2}{1}}\}-p)e_{\edge{2}{1}},A)\\
  -v((\min\{x_{s_1}-x_{d_1},t_{\edge{2}{1}}\}-p-1) e_{\edge{2}{1}},A)]\leq 0 \\
 \iff \mathbb{E}[v((\min\{x_{d_2},t_{\edge{2}{1}} -x_{d_3}+x_{s_2}\}-p)e_{\edge{2}{1}}+(x_{d_3}-x_{s_2})e_{\edge{3}{1}},A)  \\
  -v((\min\{x_{d_2},t_{\edge{2}{1}} -x_{d_3}+x_{s_2}\}-p-1) e_{\edge{2}{1}}+(x_{d_3}-x_{s_2})e_{\edge{3}{1}},A)]\leq 0 \\
 \iff L_{u^\ast +p e_{\edge{2}{1}}}^\theta v(q,a)- L_{u^\ast +(p+1) e_{\edge{2}{1}}}^\theta v(q,a)\leq 0
\end{multline*}
which means that
\[L_{u^\ast}^\theta v(q,a)\leq \cdots \leq L_{u^\prime -e_{\edge{2}{1}}}^\theta v(q,a) \leq L_{u^\prime}^\theta v(q,a) \]
Thus, we showed that $L_{u^\ast}^\theta v(q,a)\leq  L_{u^\prime}^\theta v(q,a)$ for any $k\in K_x$ and any $j\in J_x$.
\item $x_{s_1}\leq x_{d_1}$ and $x_{s_2}\geq x_{d_3}$. In that case, we have $K_x=\{0\}$ and $k_{t_{\edge{2}{1}}}=0$. If $x_{s_2}= x_{d_3}$, then $J_x=\{0\}$ and $j_{t_{\edge{2}{2}}}=0$. Thus, we have $u^\ast = u^\prime$. Otherwise, suppose that $j<\min\{j_{t_{\edge{2}{2}}}(x),x_{d_2}\}$ (this is only possible if $j_{t_{\edge{2}{2}}}(x)=x_{s_2}-x_{d_3}-t_{\edge{2}{2}}>0$ and $x_{d_2}>0$), then by definition of $t_{\edge{2}{2}}$ and because $v\in\mathcal{C}_{\edge{2}{2}}\cap \mathcal{H}_{\edge{2}{2},\edge{1}{2}}$, we have: $\forall p\in\{0,\cdots,x_{d_2}-j-\max\{t_{\edge{2}{2}}-x_{d_1}+x_{s_1},0\}-1\}$,
\begin{multline*}
 \mathbb{E}[v((\max\{t_{\edge{2}{2}},x_{d_1}-x_{s_1}\}+p+1)e_{\edge{2}{2}},A)\\
 -v((\max\{t_{\edge{2}{2}},x_{d_1}-x_{s_1}\}+p) e_{\edge{2}{2}},A)]\geq 0 \\
 \iff \mathbb{E}[v((\max\{t_{\edge{2}{2}}-x_{d_1}+x_{s_1},0\}+p+1)e_{\edge{2}{2}}+(x_{d_1}-x_{s_1})e_{\edge{1}{2}},A)  \\
  -v((\max\{t_{\edge{2}{2}}-x_{d_1}+x_{s_1},0\}+p) e_{\edge{2}{2}}+(x_{d_1}-x_{s_1})e_{\edge{1}{2}},A)]\geq 0 \\
 \iff  L_{u^\ast -(p+1) e_{\edge{2}{2}}}^\theta v(q,a)- L_{u^\ast -p e_{\edge{2}{2}}}^\theta v(q,a)\geq 0
 \end{multline*}
which means that
\[ L_{u^\prime}^\theta v(q,a)\geq L_{u^\prime +e_{\edge{2}{2}}}^\theta v(q,a) \geq \cdots \geq L_{u^\ast}^\theta v(q,a)  \]
Suppose now that $j>\min\{j_{t_{\edge{2}{2}}}(x),x_{d_2}\}$ (this is only possible if 
\\
$j_{t_{\edge{2}{2}}}(x)<x_{d_2}$), then by definition of $t_{\edge{2}{2}}$ and because $v\in\mathcal{C}_{\edge{2}{2}}\cap \mathcal{H}_{\edge{2}{2},\edge{1}{2}}$, we have: $\forall p\in\{0,\cdots,\min\{x_{d_2},t_{\edge{2}{2}} -x_{d_1}+x_{s_1}\} -x_{d_2}+j-1\}$,
\begin{multline*}
 \mathbb{E}[v((\min\{x_{s_2}-x_{d_3},t_{\edge{2}{2}}\}-p)e_{\edge{2}{2}},A)\\
-v((\min\{x_{s_2}-x_{d_3},t_{\edge{2}{2}}\}-p-1) e_{\edge{2}{2}},A)]\leq 0 \\
 \iff \mathbb{E}[v((\min\{x_{d_2},t_{\edge{2}{2}} -x_{d_1}+x_{s_1}\}-p)e_{\edge{2}{2}}+(x_{d_1}-x_{s_1})e_{\edge{1}{2}},A)  \\
 -v((\min\{x_{d_2},t_{\edge{2}{2}} -x_{d_1}+x_{s_1}\}-p-1) e_{\edge{2}{2}}+(x_{d_1}-x_{s_1})e_{\edge{1}{2}},A)]\leq 0 \\
 \iff  L_{u^\ast +p e_{\edge{2}{2}}}^\theta v(q,a)- L_{u^\ast +(p+1) e_{\edge{2}{2}}}^\theta v(q,a)\leq 0
\end{multline*}
which means that
\[L_{u^\ast}^\theta v(q,a)\leq \cdots \leq L_{u^\prime -e_{\edge{2}{2}}}^\theta v(q,a) \leq L_{u^\prime}^\theta v(q,a) \]
Thus, we showed that $L_{u^\ast}^\theta v(q,a)\leq  L_{u^\prime}^\theta v(q,a)$ for any $k\in K_x$ and any $j\in J_x$.
\end{enumerate}
\end{proof}

\end{appendices}

\end{document}